# Parenthood Penalties in Academia: Childcare Responsibilities, Gender Role Beliefs and Institutional Support

Xi Hong[1], Xiang Zheng[1], Haimiao Yuan[2], and Chaoqun Ni*[1]

*1 Information School, University of Wisconsin-Madison, Madison, Wisconsin, USA*

*2 College of Education, The University of Iowa, Iowa City, Iowa, USA*

Correspondence concerning this article should be addressed to Dr. Chaoqun Ni, Information School, University of Wisconsin-Madison, Madison, WI 53706, USA. Email: chaoqun.ni@wisc.edu

**Abstract:** Women remain persistently underrepresented and underrecognized across multiple dimensions of academic outcomes, highlighting the urgency of understanding the structural and social mechanisms that contribute to these disparities. This study investigates the role of parenthood in shaping gender disparities in academic careers, focusing on the complex interplay between gender, childcare responsibilities, gender role beliefs, institutional support, and scientists' career outcomes. Using a large-scale survey of 5,700 academics and bibliometric data from Web of Science, this study reveals that childcare responsibilities significantly mediate gender disparities in both subjective and objective academic outcomes. In particular, women shoulder a greater caregiving load than men when their partners are employed full-time outside academia. Strengthening egalitarian gender role beliefs among women academics has played an important role in shifting caregiving norms, as greater endorsement of such beliefs is associated with reduced childcare responsibilities—a pattern not observed among men. Institutional parental support policies show mixed mitigating gender disparities in academia: For fathers, a paused tenure clock can offset the negative effect of childcare responsibilities on research productivity and collaborations, whereas the paternity leave exacerbates the negative effect of childcare responsibilities on research satisfaction. For mothers, while flexible work schedules and childcare support can alleviate the negative association between childcare responsibilities and career outcomes, policies such as tenure clock extensions and paternity leave inadvertently intensify it. Our study demonstrates that addressing these disparities necessitates a comprehensive approach that combines policy reform with shifts in individual attitudes and broader sociocultural change.

## Introduction

Despite substantial progress toward gender parity in doctoral education, women remain underrepresented and underrecognized in academia. Globally, women make up 47% of doctoral students and exceed 50% in certain regions such as North America and Central Asia (UIS, 2022), and represent 43% of doctoral graduates worldwide (UNESCO, 2015). However, this educational progress has not translated into proportional representation in research careers. Women account for only 28% of researchers globally, and their presence further declines at senior academic ranks (AAUP, 2023; Spoon et al., 2023; UNESCO, 2015). Moreover, they remain underrepresented in key indicators of academic recognition and influence. Women publish less in prestigious journals (Ni et al., 2025), are underrepresented in lead authorship positions (Prozesky, 2006; Elsevier, 2017), are less likely to receive journal invitations to submit papers (Holman et al., 2018), and are less likely to receive project funding (Ley & Hamilton, 2008; Witteman et al., 2019) and prestigious awards (Lunnemann et al., 2019; Meho, 2021) than their male counterparts. These disparities are compounded by slower career advancement, lower salaries, more work-life conflict, and limited representation in academic leadership roles (Casad et al., 2022; Malkinson et al., 2023).

Gender disparities in academia have detrimental effects on both individual scholars and the broader scientific community. For women, obstacles in career advancement often lead to lower job satisfaction and a higher likelihood of leaving academia (Spoon et al., 2023; Xu, 2008), which further reinforce and exacerbate gender stereotypes, resulting in hiring discrimination, salary inequities, and reduced career support or opportunities for women scientists (Barbezat & Hughes, 2005; Casad et al., 2021). The underrepresentation of women in senior positions also means fewer female role models and mentors, which may negatively affect the recruitment, training, and retention of female students in certain disciplines (Moss-Racusin et al., 2012; Sheltzer & Smith, 2014). Moreover, given women's vital role in boosting collaboration and new discoveries in scientific teams (Nielsen et al., 2017), the underrepresentation of women may further impede the pace of science advancement and diminish the quality of research outcomes. Collectively, these patterns underscore the systemic and self-reinforcing nature of gender disparities in academia, highlighting the urgent need to identify their root causes and implement structural reforms to foster a more equitable, inclusive, and productive academic environment.

Prior research suggests that parenthood plays a pivotal role in observed gender disparities in academia. While the current gender gaps cannot be solely attributed to parenthood, as factors like conscious and unconscious bias against women's work and an unsupportive workplace climate also contribute (Oliveira et al., 2021; Zhang et al., 2022; Spoon et al., 2023), parenthood remains a powerful explanatory factor (Derrick et al., 2022; Liu et al., 2025; Misra et al., 2012). Research shows that gender differences in academic achievement are minimal among scientists without children, but pronounced disparities arise among those who are parents (Zheng et al., 2022). Furthermore, childless women in academia tend to experience faster promotion than male counterparts, while mothers are promoted more slowly than fathers (Takahashi & Takahashi, 2015). These patterns have been exacerbated since the onset of the COVID-19 pandemic, which intensified caregiving responsibilities and disproportionately impacted women's research productivity (Caldarulo et al., 2022; Madsen et al., 2022). Together, these findings underscore that parenthood plays a significant role in perpetuating gender disparities in academia, highlighting an urgent need for targeted interventions and policies to support academic parents, particularly women.

Studies have examined the underlying mechanism through which parenthood shapes the careers of women scientists. Key areas of focus include career breaks due to childbirth and challenges associated with reentry to academic work (Davies et al., 2022; Nielsen et al., 2024; Akram & Pflaeger Young, 2021), heightened work-family conflicts stemming from traditional gender role beliefs and unbalanced caregiving responsibilities, and the persistence of maternal discrimination against mothers in academia (Staniscuaski et al., 2023; Hosseini & Villanueva, 2023; Carter, 2024; Thébaud & Taylor, 2021). In addition, studies have identified a lack of adequate institutional support—or the failure of existing policies—to effectively meet the needs of academic parents (O'Laughlin & Bischoff, 2005; Stalford, 2005; Windsor & Crawford, 2020; Eren, 2022; Sougou et al., 2022).

Despite extensive discourse, caregiving, one of the most demanding aspects of parenthood, remains under-researched from a quantitative perspective in the context of gender gaps in academia. Prior research has consistently documented that women scholars shoulder a disproportionate share of caregiving responsibilities compared to men, largely due to entrenched gender norms and expectations (Moore, 1995; Santero & Cannito, 2024; Wilton & Ross, 2017). These caregiving pressures have been linked to heightened work–family conflict and an increased likelihood of women pursuing less favorable career paths or exiting academia altogether. However, most existing studies rely on theoretical frameworks, qualitative data, or small-scale samples (Deutsch & Yao, 2014; Townsley & Broadfoot, 2008; Anders, 2004). There is a pressing need for large-scale, quantitative investigations to rigorously assess the impact of caregiving—particularly childcare responsibilities—on academic career outcomes. Such analyses should include both objective indicators (e.g., publication productivity, citation impact, collaboration patterns) and subjective dimensions (e.g., career satisfaction, perceived recognition). Moreover, the influence of partners' employment status and occupational type on caregiving dynamics remains inconclusive and warrants further empirical exploration (Derrick et al., 2022; Vohlídalová, 2017).

In addition to childcare responsibilities, evolving gender role beliefs and the ambiguous effects of institutional parental support policies are critical factors in understanding how parenthood contributes to gender disparities in academia. Recent studies document a shift toward more egalitarian gender role beliefs, with women increasingly rejecting traditional gender norms and men becoming more engaged in family and childcare responsibilities (Bleske-Rechek & Gunseor, 2022; Boehnke, 2011; Meeussen et al., 2016; Sallee et al., 2016). However, it remains unclear whether similar shifts are occurring within the academic community. Are academics adopting more egalitarian gender role beliefs, and if so, are these changes translating into a more equitable division of caregiving responsibilities? At the same time, institutional support for parenting—often viewed as essential to reducing the parental penalty—has produced mixed outcomes. While some policies offer meaningful relief, others may inadvertently reinforce gender disparities (Antecol et al., 2018). These complexities raise an important question: What types of parenting support policies can positively mitigate gender gaps in academia, particularly by easing the burden of childcare responsibilities?

This study investigates the complex interplay among gender, gender role beliefs, childcare responsibilities, institutional support, and scientists' career outcomes. Specifically, it addresses the following research questions:

(1) To what extent do childcare responsibilities explain the relationship between gender and academic achievement as a mediator?

(2) Do academics' egalitarian gender role beliefs moderate the relationship between gender and childcare responsibilities?

(3) Do institutional parenting support policies moderate the association between childcare responsibilities and academic outcomes?

Based on prior research, we propose the following hypotheses:

***H1:*** Childcare responsibilities play a significant mediating role in explaining gender gaps in career outcomes in academia as a mediator.

***H2:*** Higher levels of egalitarian gender role beliefs will lead to greater involvement of male scholars in childcare responsibilities while reducing the caregiving involvement of female scholars.

***H3:*** Institutional parenting support policies could mitigate the negative impact of childcare responsibilities on academics' career outcomes.

## Data and Methods

To comprehensively examine the interplay among gender, childcare responsibilities, and academic outcomes, this study relies on two sets of data: a large-scale survey and bibliographic data from Clarivate's Web of Science (WoS).

## Survey data collection

Our analytical sample includes researchers affiliated with US and Canadian institutions and responded to a survey we distributed through Qualtrics. The survey was distributed to collect US and Canadian researchers' experiences and opinions regarding various aspects of parenting and their career trajectories. Prior to the survey distribution, we identified 396,674 scholars who published at least one paper between 2000 and 2019 and were associated with institutions in the US or Canada. We then randomly sampled 25% of the list (99,168 scholars) and distributed our survey via the Qualtrics survey platform between June 2019 and July 2019. We distributed the survey to the sampled list instead of the full list of scholars due to the weekly account limit (50,000 email addresses per week) set by Qualtrics. Of the 99,168 authors we contacted, 10,333 began the survey, and 9,105 completed it. After excluding cases with missing critical information, such as parenthood status and gender, as well as individuals not holding tenure-track faculty positions, the dataset includes 7,764 valid respondents. For the purposes of this study, the final analytic sample is further restricted to those who identified as parents, resulting in 5,700 scholars based at institutions in the United States or Canada. Among them, 2,540 are men (45%) and 3,160 are women (55%). To assess the representation of gender distribution of this study, the gender composition of the population, surveyed, respondents, and analytical sample is provided in Table S1 in the supplementary file. This survey was approved by the Institutional Review Board of the affiliated university, and all participants provided informed consent before taking part.

## Key variables

### *Variables of Academic Career Outcome*

Academic career achievement is measured from two dimensions: ***subjective*** and ***objective*** career outcomes. **Subjective career outcomes** are quantified based on the perceived career achievement reported by survey respondents, including **research satisfaction** (satisfaction about the progress made towards meeting research achievement goals), **career satisfaction** (satisfaction about the progress made towards meeting career achievement goals), and **community recognition** (being recognized for contributions to scholarly communities).

We measure **objective academic career outcomes** through three key dimensions of their research profiles: **productivity** (number of published papers), **citation** (number of citations received), and **collaboration** (number of coauthors), based on the Web of Science (WOS) database. To account for differences in publishing practices across disciplines and the cumulative nature of these metrics, we normalized each measure by discipline and time, as explained in supplementary materials (see "Objective career achievement measures"). The normalized indicators used included average relative publication (ARP) for productivity, average relative citation (ARC) for citation impact, and average relative coauthor (ARCo) for the extent of collaboration, following a previous study (Zheng et al., 2022). The descriptive statistics for these measures are presented in Supplementary Table S4.

### *Variables of Childcare Responsibilities*

We measured individuals' childcare responsibilities using a self-reported, reverse-scored survey item that assessed the level of caregiving support they received from others in their family or private network (e.g., partners, relatives, friends). This indirect approach was chosen to reduce potential social desirability bias, which often affects responses to sensitive questions. Research has shown that individuals—particularly men—may overreport their own caregiving involvement due to ego-defensive tendencies or impression management (Fisher, 1993; Derrick et al., 2022). To mitigate this bias, indirect questioning strategies that focus on external circumstances rather than self-reported behaviors are considered more reliable (Westfall et al., 1957, p138). Accordingly, participants were asked how much support they received with childcare, with response options ranging from "extremely" to "not at all." These responses were reverse-coded to construct a childcare responsibilities variable, such that individuals reporting "not at all" were assigned the highest level of childcare responsibilities, while those reporting "extremely high" support were assigned the lowest. While this single-item measure offers a practical and bias-reducing approach, we acknowledge its limitations and encourage future research to incorporate more detailed and multi-dimensional measures of caregiving responsibilities.

### *Variables of Egalitarian Gender Role Beliefs*

The variable "Egalitarian gender role beliefs" in this study was constructed to capture multiple dimensions of respondents' attitudes toward egalitarian gender role beliefs, particularly in relation to work and family responsibilities. Gender role beliefs are considered to "represent people's perceptions of men's and women's social roles in the society in which they live" (Eagly & Wood, 2012, p459), and are known to influence how individuals perceive their own and others' obligations within both the household and the workplace (Kerr & Holden, 1996). Traditional gender role beliefs typically assign primary responsibilities for family and childcare to women, shaping career and caregiving decisions for both genders. Women are often socialized to prioritize family over career—demonstrating greater "family centrality"—while men are generally less likely to do so (Snir et al., 2009). In contrast, egalitarian gender role beliefs advocate for a more equal distribution of domestic and caregiving responsibilities, and recognize women's identities not only as caregivers but also as professionals and contributors to the workforce (Lorber, 1994). This variable thus serves as a critical lens through which to examine how gendered expectations shape childcare responsibilities and, in turn, academic career outcomes.

To operationalize the concept of egalitarian gender role beliefs and examine their moderating effect on the relationship between gender and childcare responsibilities, this study measured respondents' attitudes across four key dimensions drawn from prior research (Davern et al., 2021; Prasad & Baron, 1996). These dimensions reflect widely studied indicators of gender role egalitarianism: (1) the belief that a working mother can develop a relationship with her children that is just as warm and secure as that of a non-working mother; (2) acceptance of women working while men stay home to care for the family; (3) disagreement with the notion that women should reduce their paid work for the sake of their family; and (4) disagreement with the claim that a young child is likely to suffer if both parents are employed (See Supplementary File for full survey wording). Respondents indicated their level of agreement with each statement on a seven-point Likert scale, and an average score across the four items was computed to represent each individual's overall level of egalitarian gender role beliefs. Higher average scores correspond to stronger endorsement of egalitarian views. Descriptive statistics for each item and the composite measure are reported in Supplementary Table S5.

### *Variables of Institutional Parental Support*

Institutional support can be a crucial factor in facilitating the professional development and career advancement of faculty members with children. Programs such as onsite and subsidized childcare, flexible work schedules, and family leave policies are designed to alleviate work-family conflict and help retain academic parents in the workforce. Our study estimated the level of institutional support for parents using self-reported survey data across four dimensions: availability of childcare support, paternity leave, tenure clock extension (paused tenure clock), and flexible work schedules. Each item was coded dichotomously (1 = "Yes," 0 = "No"). This composite measure is used to examine whether and how institutional support moderates the relationship between childcare responsibilities and academic career outcomes.

## Methods and Analytical Framework

This study investigates the mechanism through which gender, childcare responsibilities and academic career outcomes interact, with particular attention to how individual attitudes and institutional support shape these relationships. As illustrated in Figure 1, the analytical framework centers on three core relationships: (1) the mediating role of childcare responsibilities in explaining gender disparities in academic career outcomes, (2) the moderating effect of egalitarian gender role beliefs on the relationship between gender and childcare responsibilities, and (3) the moderating role of institutional parenting support in shaping the impact of childcare responsibilities on academic outcomes.

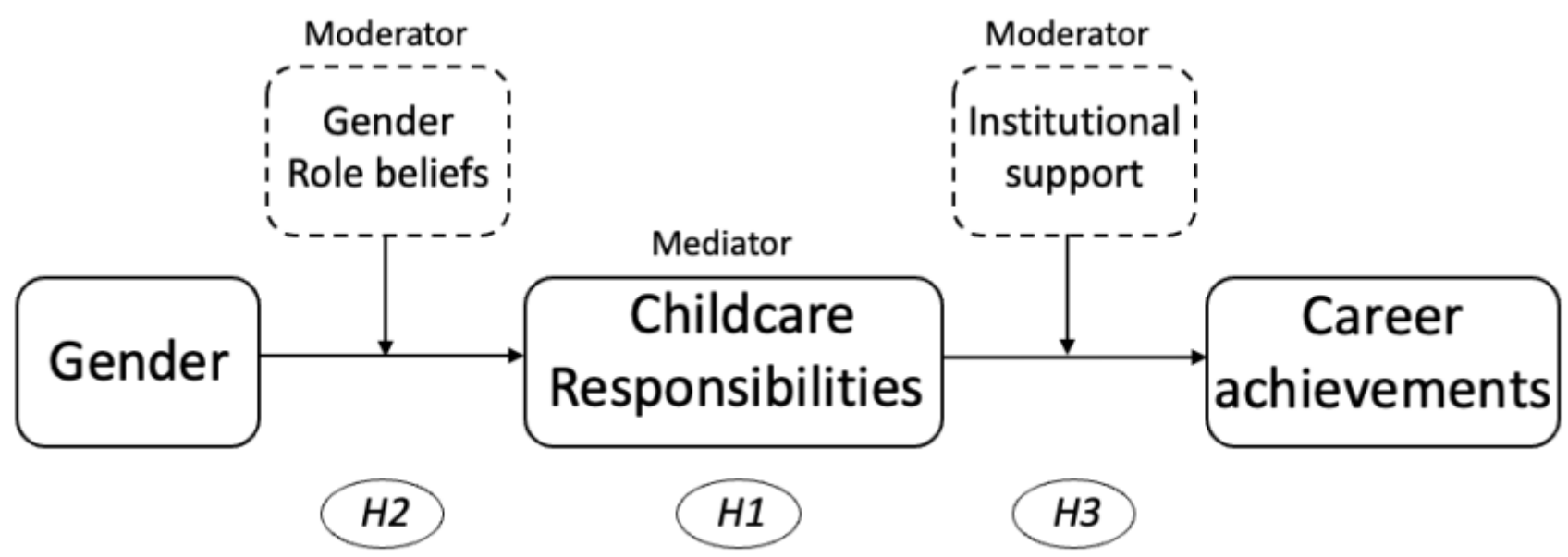

**Figure 1.** Analytical framework

To empirically test these relationships, we utilize Structural Equation Modeling (SEM) alongside moderated linear regressions. SEM is defined as "a collection of statistical techniques that allow a set of relationships between one or more independent variables and one or more dependent variables to be examined" (Ullman & Bentler, 2012, p661). In this study, SEM is employed to assess the mediating effect of childcare responsibilities on the relationship between gender and academic career outcomes. Mediation analysis allows us to evaluate whether and how an intermediary variable (i.e., childcare responsibilities) explains the relationship between the independent variable (i.e., gender) and a dependent variable (i.e., academic career outcomes). The indirect effect is based on the product of two coefficients ($a_1$ and $b_1$): the effect of independent variable on dependent variables ($a_1$) and the effect of the mediator on career outcome measures ($b_1$), following standard mediation modeling practices (Bollen & Stine, 1990; Pearl, 2022).

To assess moderation, we employ moderated regression analyses to examine whether egalitarian gender role beliefs and institutional support alter the strength or direction of the relationships described above. A moderator is a variable that affects the direction or magnitude of the relation between an independent and a dependent variable. A moderating effect is present when the interaction term between the moderator and the independent variable is statistically significant (Baron & Kenny, 1986). We further interpret these interactions using marginal effects, which quantify how the effect of an independent variable changes across different values of the moderator (Brambor et al., 2006). For example, this study examines how the effect of childcare responsibilities on academic achievement varies with the presence or absence of institutional parental support.

The models include several control variables to account for confounding factors, including disciplinary area, career stage, number of children, and race. Likert-scale items are treated as continuous variables, consistent with recommendations for large-sample studies using 5-7 response categories (Johnson & Creech, 1983; Norman, 2010). Moreover, to account for potential clustering effects and correlated errors within institutions, we cluster standard errors by respondents' institutional affiliation.

By incorporating both individual-level beliefs and institutional-level supports into our framework, this study seeks to provide a comprehensive understanding of how childcare responsibilities contribute to gender disparities in academic careers—and how these effects may be mitigated by more equitable social norms and supportive organizational structures. The full analytical framework is presented in Figure 1.

## Results

### Gender gaps in academic outcomes, egalitarian gender role beliefs, and childcare responsibilities

As established in prior work (Zheng et al., 2022), gender disparities persist in the academic career outcomes—particularly among researchers with children. These disparities are evident across both subjective and indicators of career outcomes. Among academics with children, females reported significantly lower levels of perceived career outcomes than males: A higher proportion of women reported dissatisfaction (including "somewhat dissatisfied," "dissatisfied," and "strongly dissatisfied") with their research achievement (29%), career achievement (19%), and community recognition (15%), compared with men (21%, 15%, and 10%, respectively). Conversely, fewer women reported satisfaction (including "somewhat satisfied," "satisfied," and "strongly satisfied") with their research achievement (68%), career achievement (78%), and community recognition (79%), whereas higher proportions of men expressed satisfaction in these areas (77%, 83%, and 84%, respectively). Objective career outcome measures based on bibliometric data reflect similar trends—mothers also had 20% fewer annual relative publications (women's mean ARP = 1.814, men's mean ARP = 2.272), 6% fewer annual relative citations (women's mean

ARC = 2.141, men's mean ARC = 2.278), and 9% fewer annual relative coauthors (women's mean ARCo =0.981, men's mean ARCo = 1.084) compared to fathers (Figure 2). In contrast, gender differences among academics without children were minimal across these same indicators. This divergence underscores the potential centrality of childcare responsibilities in driving gender-based disparities in academic careers.

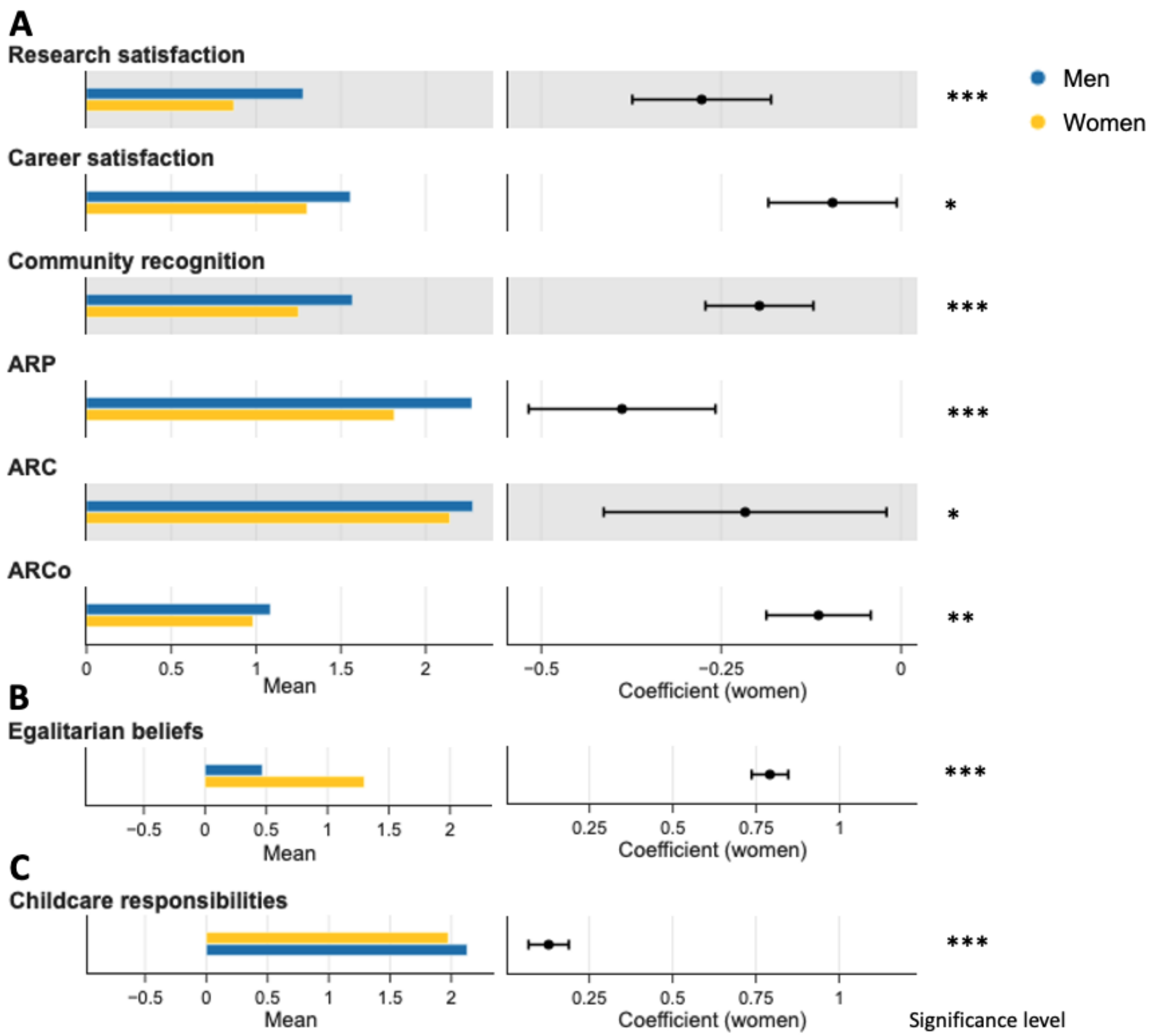

**Figure 2.** Gender differences in career outcomes, egalitarian gender role beliefs, and childcare responsibilities. (A) Means of subjective and objective career achievement measures by gender and the coefficients of gender on these measures in linear

regressions. (B) Mean of egalitarian gender role beliefs and the coefficient of gender on egalitarian gender role beliefs in linear regressions. (C) Means of childcare responsibilities by gender and the coefficient of gender on childcare responsibilities in linear regressions. Coefficient (women): the coefficient of the variable gender (0 denotes men and 1 denotes women) in linear regression models, with subjective and objective career achievement, egalitarian gender role beliefs, and childcare responsibilities measures as dependent variables, respectively. Control variables like disciplinary area, career stage, number of children, and race were incorporated into the models. Significance level: * $p < 0.05$, ** $p < 0.01$, *** $p < 0.001$.

Our results also reveal significant gender gaps in both egalitarian gender role beliefs and the division of childcare responsibilities (Figure 2). Overall, mothers displayed higher levels of egalitarian gender role beliefs compared to fathers, with overall scores being 1.8 times higher (See supplementary materials Table S5). Specifically, 83% of mothers agreed or strongly agreed that a working mother can form just as warm and secure a relationship with her children as a non-working mother, compared to only 60% of fathers. Similarly, 67% of mothers agreed or strongly agreed that it is acceptable for them to work while their spouse/partner stays home to care for the family, while only 45% of fathers agreed with the reverse scenario. Additional gaps were observed on items challenging traditional norms: 19% of mothers, compared to 11% of fathers, disagreed or strongly disagreed that a woman should reduce her paid work for the sake of family responsibilities. Similarly, 73% of mothers disagreed or strongly disagreed that a young child is likely to suffer if both parents work, in contrast to only 43% of fathers .

Despite holding more egalitarian views, mothers continue to bear a disproportionate share of childcare responsibilities. On average, mothers reported significantly lower levels of support from others in their caregiving network. Among mothers, 38% reported receiving an extreme level of childcare support from family, 48% reported receiving high or moderate support, and 14% reported slight or no support. In comparison, 45% of fathers reported receiving extreme support, 42% received high or moderate support, and only 13% reported receiving slight or no support. These patterns raise an important question: to what extent does the unequal division of childcare contribute to the broader gender disparities observed in academic careers?

## The mediating effect of childcare responsibilities on gender gaps in academic career outcomes

To address this question, we examined whether childcare responsibilities mediate the relationship between gender and academic career outcomes. Our findings suggest that childcare responsibilities play a meaningful explanatory role, particularly in subjective career assessments. Compared to fathers, mothers' greater caregiving burden accounts for 4% of the gender gap in research satisfaction responsibilities(Indirect Effect [IE] = -0.010, 95% CI [-0.017, -0.003]), 16% of the gap in career satisfaction (IE = -0.013, 95% CI [-0.021, -0.004]), and 6% of the gap in perceived community recognition (IE = -0.010, 95% CI [-0.017, -0.004]).

In terms of objective career outcomes, childcare responsibilities explain 4% of the gender gap in ARP (IE = -0.017, 95% CI [-0.026, -0.007]) and 6% of the gap in research collaborations measured by ARCo (IE = -0.007, 95% CI [-0.012, -0.002]) compared to men. These findings, illustrated in **Figure 3**, underscore the significant—albeit partial—role that unequal caregiving burdens play in sustaining gender disparities in both objective and subjective career outcomes.

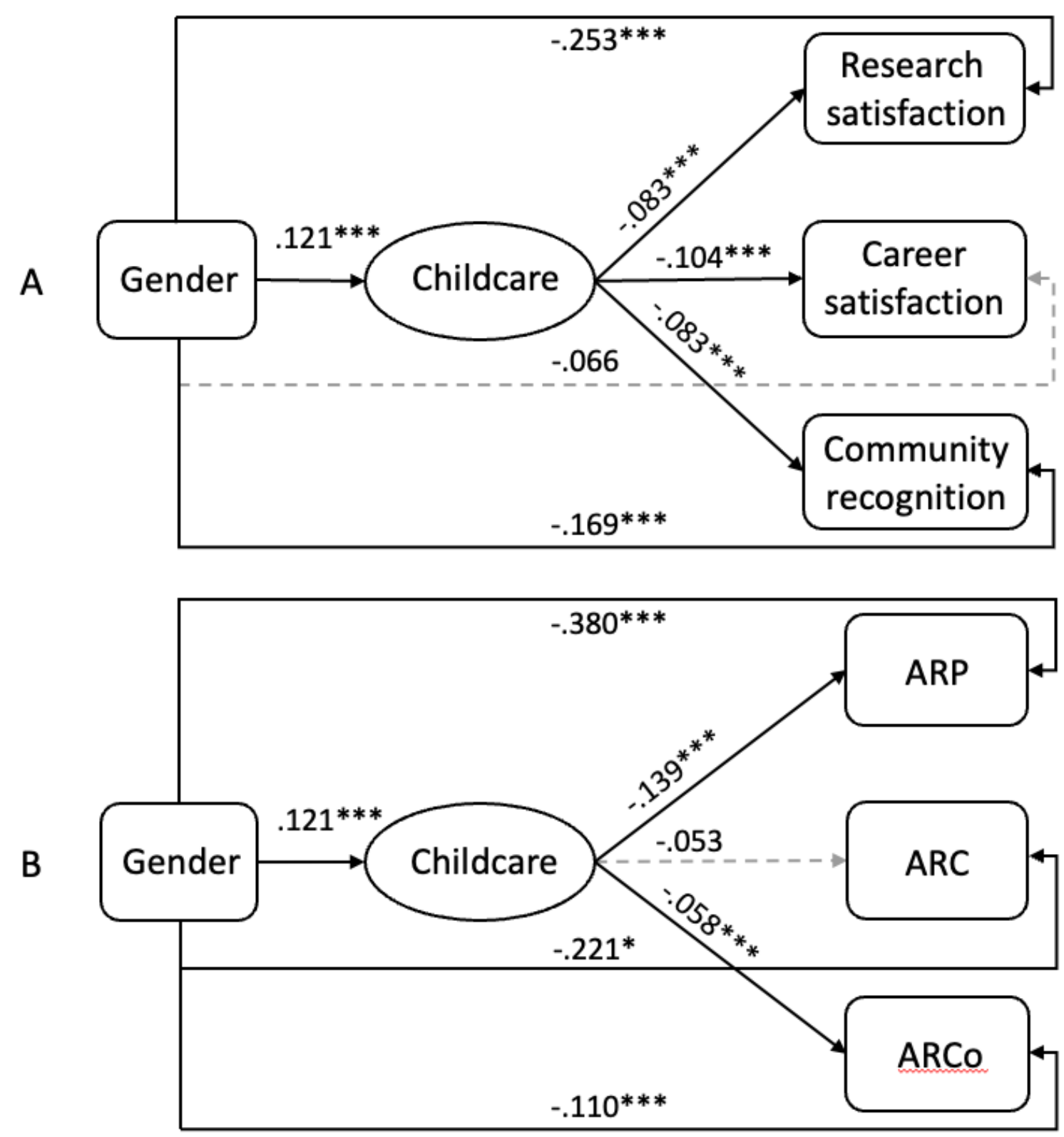

**Figure 3.** (A) Mediation effect analysis models of childcare responsibilities between gender and three subjective career achievement measures. (B) Mediation effect analysis models of childcare responsibilities between gender and three objective career achievement measures. Black lines and gray dashed lines denote significant and insignificant coefficients, respectively. Significance level: * P < 0.05, ** P < 0.01, *** P < 0.001.

We further examine whether the mediating role of childcare responsibilities in gender disparities varies by the employment status and job type of academics' partners (See supplementary materials Table S6.2-S6.3 for full model results). Our findings indicate that childcare responsibilities contribute more strongly to gender gaps in academic outcomes when scholars' partners hold wage-based, inflexible jobs. In these cases, mothers assume greater caregiving burdens, which significantly mediate differences in both subjective and objective career measures. Specifically, for academics with wage-employed partners, childcare responsibilities account for 4% of the gender gap in perceived research satisfaction (IE = -0.010, 95% CI [-0.019, -0.001]), 25% of the gap in career satisfaction (IE = -0.014, 95% CI [-0.025, -0.004]), 6% in community recognition (IE = -0.010, 95% CI [-0.019, -0.002]), 4% in research productivity (ARP) (IE = -0.017, 95% CI [-0.029, -0.005]), and 5% in research collaborations (ARCo) (IE = -0.007, 95% CI [-0.013, -0.000]). However, if partners are self-employed, students, or unemployed—conditions typically associated with greater schedule flexibility—these mediating effects are not observed. This suggests that childcare responsibilities may be more equitably shared in households where partners have flexible work arrangements.

We also explore whether the nature of the partner's job—specifically, whether it is research-related—affects the distribution of caregiving. Among academics whose partners do not work in research-related fields, mothers' higher involvement in childcare significantly mediates 3% of the gender gap in perceived research satisfaction (IE = -0.008, 95% CI [-0.017, 0.001]), 14% of the gap in career satisfaction (IE = -0.010, 95% CI [-0.019, -0.000]), 5% of the gap in community recognition (IE = -0.009, 95% CI [-0.017, -0.000]), 3% of the gap in ARP (IE = -0.012, 95% percentile CI [-0.022, -0.001]), and 4% of the gap in ARC (IE = -0.007, 95% CI [-0.013, -0.000]) than fathers. These mediating effects are not observed when both partners work in research, suggesting a more balanced division of caregiving responsibilities in dual-academic households. Together, these results underscore the importance of partner context in shaping the gendered impact of childcare on academic careers.

## The moderating effect of egalitarian gender role beliefs

Building on our findings that childcare responsibilities mediate gender disparities in academic achievement, we further examine whether egalitarian gender role beliefs moderate this relationship. Specifically, we test whether holding more egalitarian views weakens the association between gender and childcare responsibilities—potentially reducing the caregiving burden disproportionately placed on women in academia.

Our analysis reveals a significant moderating effect of egalitarian gender role beliefs on the relationship between gender and childcare responsibilities. As individuals' endorsement of egalitarian beliefs increases, the gender gap in reported childcare duties narrows. When egalitarian gender role beliefs are one standard deviation below the mean, the marginal effect of gender on childcare responsibilities is 0.321 (95% CI [0.219, 0.423]), reflecting a substantial 16% difference in caregiving responsibilities between male and female academics. However, at one standard deviation above the mean, this marginal effect drops to 0.077 (95% CI [-0.011, 0.165]), indicating that the gender gap is no longer statistically significant. These results suggest that more egalitarian views are associated with a more equitable distribution of caregiving responsibilities, and they hold across different partner employment statuses and job types (see Supplementary Tables S7.1–S7.2).

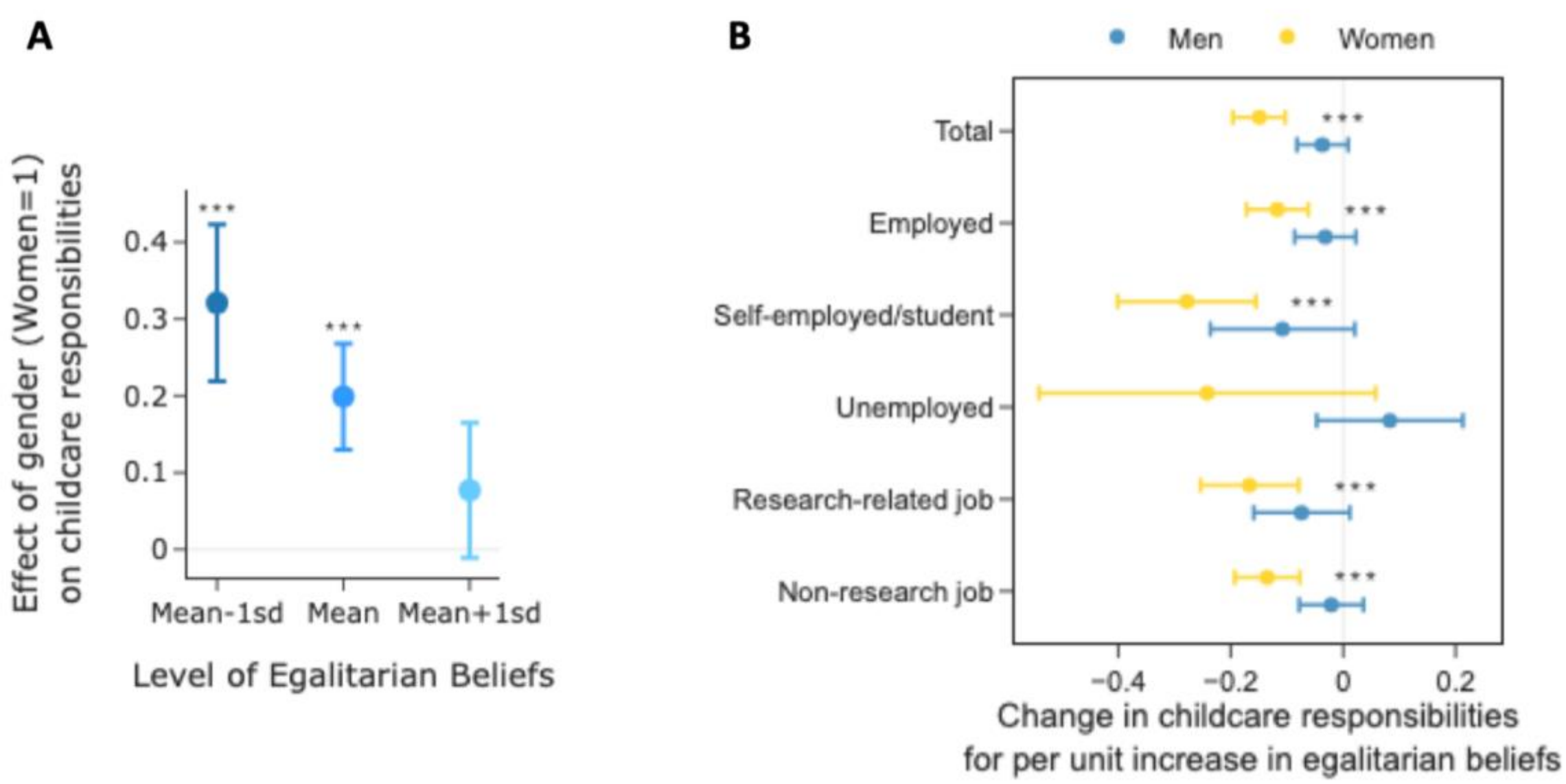

**Figure 4.** (A) Gender gap in childcare responsibilities at different levels of egalitarian gender role beliefs. The centers of the plots are the coefficients of gender on childcare responsibilities in the total sample, while the ends of the plots are the 95% confidence intervals of these coefficients. (B) Change in childcare responsibilities for per unit increase in egalitarian gender role beliefs, by gender and the job status and employment type of respondents' partners. The centers of the plots are the coefficients of egalitarian gender role beliefs on childcare responsibilities in different groups, while the ends of the plots

are the 95% confidence intervals of these coefficients. Significance level: * P < 0.05, ** P < 0.01, *** P < 0.001.

To further test how egalitarian gender role beliefs influence caregiving responsibilities within each gender group, we conducted linear regression analyses stratified by gender. Our findings suggest that among women, stronger egalitarian gender role beliefs are significantly associated with lower levels of childcare responsibilities—a reduction of 0.149 units in caregiving burden for each one-unit increase in egalitarian beliefs (95% CI [-0.196, -0.103]). In contrast, no significant relationship was observed among men; their level of egalitarian belief had no discernible effect on their self-reported childcare involvement. These patterns, visualized in Figure 4, are consistent across partner job types and employment statuses. This asymmetry suggests that while egalitarian beliefs may help women adjust their caregiving roles to better balance family and career, similar beliefs do not necessarily lead to increased caregiving participation among men.

### The moderating effect of institutional support

In addition to the influence of family-level dynamics and individual gender role beliefs, institutional parenting support may play a critical role in moderating the impact of childcare responsibilities on academic career outcomes. This section displays how different forms of institutional support—such as flexible schedules, paused tenure clocks, paternity leave, and childcare assistance—moderate the relationship between childcare responsibilities and academic outcomes. To assess potential gender differences in these effects, we conducted separate moderation analyses for men and women.

Our results show that institutional support moderates the impact of childcare responsibilities in gender-specific ways. For fathers, a paused tenure clock fully offsets the negative association between childcare responsibilities and research productivity measured by ARP (CE = 0.454, 95% CI [0.044, 0.863]). It also marginally mitigates the negative effect on research collaboration measured by ARCo (CE = 0.163, 95% percentile CI [-0.007, 0.333]). However, paternity leave appears to exacerbate the negative association between childcare responsibilities and men's research satisfaction (CE = -0.143, 95% percentile CI [-0.295, 0.010]). For mothers, flexible work schedules significantly alleviate the negative effect of childcare responsibilities on career satisfaction by 60% (CE = 0.114, 95% CI [0.003, 0.225]). Conversely, a paused tenure clock and paternity leave show marginal but negative moderating effects, intensifying the association between childcare duties and reduced career satisfaction (CE = -0.092, 95% percentile CI [-0.201, 0.017]), and research satisfaction (CE = -0.105, 95% percentile CI [-0.219, 0.009]), respectively. Full model estimates are reported in Table S8.1.

The effectiveness of institutional support also varies by the partner's employment status, particularly their work flexibility. For men whose partners are wage-employed or in the military—jobs typically associated with inflexible schedules—paternity leave worsens the impact of childcare responsibilities on research collaboration (CE = -0.427, 95% CI [-0.845, -0.008]). In contrast, when their partner is out of work (and not looking for work) or retired, paternity leave becomes beneficial, fully offsetting the negative effects on both career satisfaction (CE = 0.448, 95% CI [0.154, 0.742]) and research satisfaction (CE = 0.445, 95% CI [0.091, 0.798]). Similarly for men whose partners are self-employed, students, or seeking work, a paused tenure clock fully mitigates the impact of childcare on ARCo (CE = 0.126, 95% CI [0.001, 0.251]). However, this same policy has the opposite effect—exacerbating the negative relationship between childcare responsibilities and both career satisfaction (CE = -0.504, 95% CI [-0.982, -0.027]) and research satisfaction (CE = -0.473, 95% CI [-0.915, -0.031]) if their partner is out of work but not looking for work or retired.

For women, the moderating effects of institutional support are more complex and context-dependent. When their partners are self-employed, students, or actively seeking work, paternity leave—despite being positively associated with women's perceived community recognition overall—actually turns the effect of childcare responsibilities on recognition from neutral to negative (CE = -0.260, 95% CI [-0.516, -0.004]). It implies the presence of support does not always reduce caregiving burdens and may in some cases reinforce gendered expectations. On the other hand, direct childcare support proves effective: it fully offsets the negative effects of childcare responsibilities on both ARP (CE = 0.527, 95% CI [0.285, 0.770]) and ARCo (CE = 0.474, 95% CI [0.261, 0.687]). When women have partners who are out of the workforce but not seeking employment, flexible scheduling is particularly impactful—fully mitigating the negative effects on both career satisfaction (CE = 0.691, 95% CI [0.126, 1.256]) and research satisfaction (CE = 0.638, 95% CI [0.004, 1.273]). However, some forms of support, like childcare assistance, display mixed effects—positively moderating community recognition (CE = 1.110, 95% CI [0.053, 2.167]) but negatively affecting ARP(CE = -1.189, 95% CI [-2.120, -0.257]).

Partner job type—specifically whether it is research-related—further shapes the moderating effects of institutional support. For men with research-active partners, flexible scheduling and paternity leave each have negative moderating effects. Flexible schedules shift the impact of childcare responsibilities on community recognition from insignificant to negative (CE = -0.216, 95% CI [-0.405, -0.026]), , while paternity leave exacerbates the negative relationship with career satisfaction (CE = -0.321, 95% CI [-0.577, -0.066]). For women, the same flexible schedule policy substantially reduces the negative effect of childcare responsibilities on career satisfaction by 92% (CE = 0.222, 95% CI [0.019, 0.424]). However, when paternity leave is available, the relationship between childcare duties and research satisfaction shifts from neutral to negative (CE = -0.192, 95% CI [0.383, -0.001]). Additionally, for women whose partners are in non-research positions, a paused tenure clock policy exacerbates the negative association between childcare responsibilities and career satisfaction by 1.4 times (CE = -0.152, 95% CI [-0.287, -0.018]).

To visually illustrate the moderating effect of institutional support, **Figure 5** displays the marginal effects of childcare responsibilities on academic career outcomes, comparing scenarios with and without institutional parental support. Each subplot compares the marginal effect of childcare responsibilities on academic outcomes in the presence versus absence of support. For example, in the subplot illustrating the association between childcare responsibilities and career satisfaction, the left side displays the estimated coefficient of childcare responsibilities on career satisfaction in the absence of support (Coefficient = −0.192, 95% CI [−0.258, −0.126]), while the right bar shows the adjusted estimate when flexible scheduling is available( Coefficient = −0.078, 95% CI [−0.157, 0.001]), reflecting the moderating effect (CE = 0.114, 95% CI [0.003, 0.225]) to the original estimate. Full moderating model results by partners' employment status and job type are provided in the supplementary materials (See Table S8.2-S8.6).

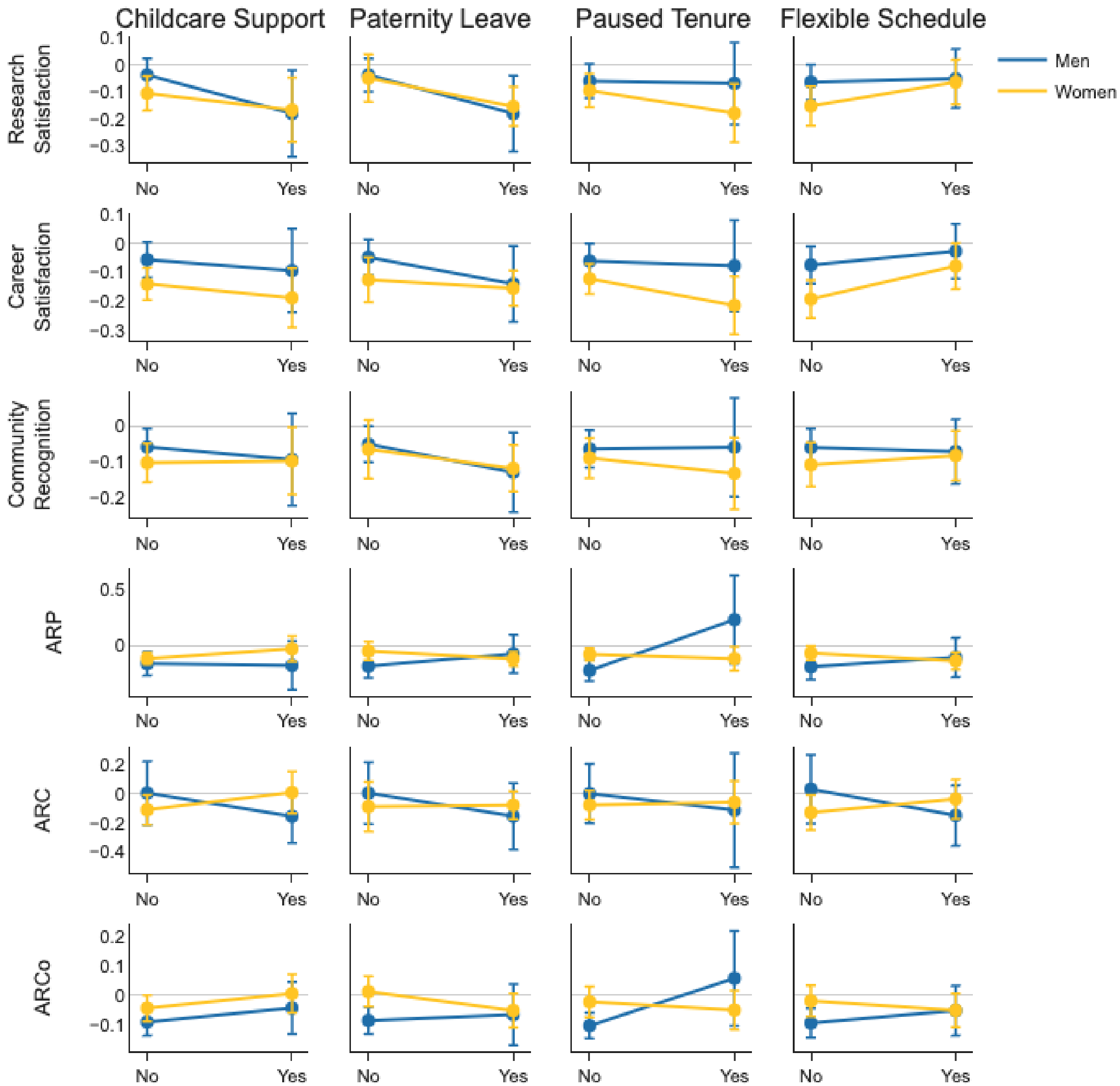

**Figure 5.** The marginal effects of childcare responsibilities on career outcomes with (Yes) or without (No) institutional parental support (moderators) by gender. Moderators include four types of institutional parental support. The centers of the plots are the marginal effect coefficients of childcare responsibilities on career outcomes, while the ends of the plots are the 95% confidence intervals of these coefficients.

## Discussions

### Evolving Beliefs, Persistent Burdens: The Disconnect Between Attitudes and Practice

A central discovery of this study is the misalignment between evolving gender role beliefs and the actual distribution of caregiving responsibilities. Our results show that both men and women in academic settings increasingly endorse egalitarian beliefs, with women expressing even stronger support for equitable gender roles. Yet, despite these progressive attitudes, women continue to shoulder a disproportionate share of childcare duties than their male counterparts.

One explanation for this disconnect lies in enduring societal and institutional norms that implicitly link masculinity to limited domestic involvement, regardless of men's stated intentions to share caregiving responsibilities (Haines & Stroessner, 2019; Sallee et al., 2016). These findings reinforce the notion that structural expectations and role prioritization models continue to channel women toward caregiving and away from professional advancement. To truly close the gender gap in academia, policy reform and cultural change must not only promote egalitarian values but also dismantle the structural barriers that prevent those values from being realized in practice.

Despite this misalignment, our study also affirms the importance of egalitarian gender role beliefs in shifting caregiving patterns—especially among women. We find that as women's egalitarian beliefs increase, their reported caregiving responsibilities decrease.

This effect is not observed among men, suggesting that women’s beliefs play a more active role in reshaping family labor dynamics in academic households. This pattern may reflect the phenomenon of "maternal gatekeeping"—where mothers, often unintentionally, regulate or limit paternal involvement in caregiving(Artis & Pavalko, 2003; De Luccie, 1995; Meteyer & Perry-Jenkins, 2010). While prior evidence has largely focused on general populations, our study adds to this literature by demonstrating that in academia, women’s own beliefs are a critical factor in how caregiving is distributed. These findings highlight the need for strategies that not only empower women to redefine their roles but also engage men more directly in redistributing caregiving responsibilities.

### The mediating roles of childcare responsibilities in explaining gender gaps in academia

Our research confirms that childcare responsibilities significantly mediate gender disparities in both objective and subjective academic career outcomes. Prior studies have predominantly focused on how parenthood affects gender disparities within academia (Zheng et al., 2022); however, they have not adequately quantified the effect of childcare as a significant component or reached a consensus about how partners’ job statuses and types might interact with it (Derrick et al., 2022; Vohlídalová, 2017). Our analysis addresses this gap by showing that the “childcare penalty” is experienced by both men and women—but with notably greater consequences for mothers.

In our full sample, childcare responsibilities account for 4% of the gender gap in research satisfaction, 16% in career satisfaction, 6% in community recognition, 4% in research productivity, and 6% in collaboration. The especially large effect on career satisfaction suggests that caregiving burdens may hinder not just research output, but broader advancement opportunities such as leadership roles, awards, and institutional recognition. Importantly, our analysis by partner employment characteristics reveals that women face the highest caregiving penalties when their partners are employed full-time in non-research-related fields, limiting opportunities for shared caregiving. These findings offer clear implications for institutions seeking to address gender equity: caregiving burdens cannot be tackled without accounting for family structures and external constraints on caregiving flexibility.

**Parental support policies: Solutions or double-edged swords**

Institutional parental support policies are often touted as key tools for reducing gender disparities in academia. Our analysis reveals that institutional support policies for parenting duties have varying impacts on men and women. Existing research has yet to reach a consensus on the effectiveness of parenting support policies in reducing the gender gaps in academia. While some studies revealed the positive side of these policies (Morgan et al., 2021; Feeney et al., 2014), others pointed out the even more disadvantaged situation of women under specific policies (Antecol et al., 2018). Our study adds nuance to this debate, showing that while flexible work schedules and childcare support positively affect women's career trajectories by mitigating the adverse effect of childcare responsibilities, paused tenure clocks and paternity leave may exacerbate the negative effect of childcare responsibilities on their careers. For men, paused tenure policies can mitigate the negative impact of caregiving on academic collaboration, yet paternity leave policies may intensify it.

To contextualize these findings, it is essential to consider that the "parenting penalty" is closely tied to the level of engagement in caregiving activities (Derrick et al., 2022) and the ways in which institutions enforce parental support policies. While such policies have the potential to help mothers balance work and family responsibilities, their effectiveness depends on how they are implemented and enforced. If mothers continue to shoulder a disproportionate share of caregiving duties compared to fathers, these policies may not fully alleviate the structural challenges they face. Given the self-reported nature of our data, the actual gender disparity in childcare responsibilities may be even greater than what is captured in our study. Therefore, the design and execution of institutional parental support policies should prioritize practical mechanisms that meaningfully reduce caregiving burdens, such as expanding childcare services and providing more flexible work arrangements.

Additionally, it is essential to consider how policy enforcement varies across institutions and how organizational norms shape the experiences of women scholars who seek parental accommodations. Specifically, institutions differ in how they apply parental policies, whether they treat them as standardized rights or discretionary benefits, and how they evaluate the professional performance of women after they use such policies (Millward, 2006). As studies showed, the provision of parental support seemed to be a “case-by-case customization” rather than an automatic right (Akram & Pflaeger Young, 2021). Moreover, women who receive parental accommodations may face heightened expectations or implicit biases regarding their job performance (Infanger & Lima, 2019), potentially exacerbating their psychological stress (Antecol et al., 2018). Hence, while parental support policies are an important step toward mitigating gender disparities in academia, their enforcement and institutional culture require careful oversight and evaluation to ensure they do not inadvertently reinforce traditional gender roles or expectations. Instead, these policies should be designed to foster genuine gender equity, providing meaningful and equitable support for all scholars navigating the demands of parenthood.

## Conclusions and limitations

This study offers new empirical evidence on the mechanisms driving gender disparities in academic careers, with a focus on the interplay among childcare responsibilities, gender role beliefs, and institutional support. Despite a shift towards more egalitarian gender role beliefs, women continue to bear a disproportionate share of caregiving duties. This imbalance not only perpetuates existing gender gaps but also places additional psychological and professional strain on women in academia. We have also demonstrated that women's egalitarian beliefs significantly decrease their childcare responsibilities, whereas similar beliefs in men do not produce comparable effects. However, the transformation in caregiving responsibilities is complex and influenced by multiple factors, especially the job contexts of one’s partner.

Our analysis of institutional support policies reveals that their effects are not uniformly positive. Some policies—particularly flexible work arrangements and childcare services—help mitigate the career impacts of caregiving, especially for women. However, others, such as tenure clock extensions and paternity leave, may have unintended negative effects if not carefully designed and equitably implemented. These findings call for a more nuanced approach to institutional policy-making that goes beyond formal provisions to ensure actual behavioral and cultural change.

At the same time, this study has limitations. As with all survey-based research, self-selection bias may influence our results, as academics with strong views on gender and caregiving—particularly women—may have been more likely to participate. Additionally, the U.S.-centric nature of the sample limits generalizability, especially given the limited availability of subsidized childcare and family-friendly policies in many U.S. institutions. Our reliance on self-reported caregiving measures also introduces potential reporting bias and lacks granularity in capturing the types, timing, and emotional dimensions of childcare responsibilities. Future research should employ more detailed, time-use and observational methods to better capture the multifaceted nature of caregiving.

The measurement of academic achievement could also benefit from further refinement. Our indicators—publications, citations, and collaborations—draw heavily from the Web of Science, which may not reflect the full range of academic contributions. Future work should incorporate more holistic metrics and qualitative assessments to better represent the diverse accomplishments of academic parents.

Finally, this study opens the door for future research on the structural factors that influence caregiving and gender equity in academia. For instance, prior studies suggest that women, due to caregiving demands and societal expectations, are more likely to occupy positions with fewer institutional resources (Ceci & Williams, 2011). In future work, we plan to examine how career choices and institutional placements intersect with academics' childcare duties and gender role beliefs, as well as their long-term impact on academics' career trajectories.

## Reference

AAUP. (2023). *The Annual Report on the Economic Status of the Profession, 2022-23*. American Association of University Professors. https://www.aaup.org/report/annual-report-economic-status-profession-2022-23

Akram, S., & Pflaeger Young, Z. (2021). Early Career Researchers' Experiences of Post-Maternity and Parental Leave Provision in UK Politics and International Studies Departments: A Heads of Department and Early Career Researcher Survey. *Political Studies Review*, *19*(1), 58–74. https://doi.org/10.1177/1478929920910363

Antecol, H., Bedard, K., & Stearns, J. (2018). Equal but Inequitable: Who Benefits from Gender-Neutral Tenure Clock Stopping Policies? *American Economic Review*, *108*(9), 2420–2441. https://doi.org/10.1257/aer.20160613

Artis, J. E., & Pavalko, E. K. (2003). Explaining the Decline in Women's Household Labor: Individual Change and Cohort Differences. *Journal of Marriage and Family*, *65*(3), 746–761. https://doi.org/10.1111/j.1741-3737.2003.00746.x

Barbezat, D. A., & Hughes, J. W. (2005). Salary Structure Effects and the Gender Pay Gap in Academia. *Research in Higher Education*, *46*(6), 621–640. https://doi.org/10.1007/s11162-004-4137-1

Baron, R. M., & Kenny, D. A. (1986). The moderator–mediator variable distinction in social psychological research: Conceptual, strategic, and statistical considerations. *Journal of Personality and Social Psychology*, *51*(6), 1173–1182. https://doi.org/10.1037//0022-3514.51.6.1173

Bleske-Rechek, A., & Gunseor, M. M. (2022). Gendered perspectives on sharing the load: Men's and women's attitudes toward family roles and household and childcare tasks. *Evolutionary Behavioral Sciences*, *16*(3), 201–219. https://doi.org/10.1037/ebs0000257.supp

Boehnke, M. (2011). Gender Role Attitudes around the Globe: Egalitarian vs. Traditional Views. *Asian Journal of Social Science*, *39*(1), 57–74. https://doi.org/10.1163/156853111X554438

Bollen, K. A., & Stine, R. (1990). Direct and Indirect Effects: Classical and Bootstrap Estimates of Variability. *Sociological Methodology*, *20*, 115–140. https://doi.org/10.2307/271084

Brambor, T., Clark, W. R., & Golder, M. (2006). Understanding Interaction Models: Improving Empirical Analyses. *Political Analysis*, *14*(1), 63–82. https://doi.org/10.1093/pan/mpi014

Caldarulo, M., Olsen, J., Frandell, A., Islam, S., Johnson, T. P., Feeney, M. K., Michalegko, L., & Welch, E. W. (2022). COVID-19 and gender inequity in science: Consistent harm over time. *PLOS ONE*, *17*(7), e0271089. https://doi.org/10.1371/journal.pone.0271089

Casad, B. J., Franks, J. E., Garasky, C. E., Kittleman, M. M., Roesler, A. C., Hall, D. Y., & Petzel, Z. W. (2021). Gender inequality in academia: Problems and solutions for women faculty in STEM. *Journal of Neuroscience Research*, *99*(1), 13–23. https://doi.org/10.1002/jnr.24631

Casad, B. J., Garasky, C. E., Jancetic, T. R., Brown, A. K., Franks, J. E., & Bach, C. R. (2022). U.S. Women Faculty in the Social Sciences Also Face Gender Inequalities. *Frontiers in Psychology*, *13*. https://www.frontiersin.org/articles/10.3389/fpsyg.2022.792756

Davern, M., Bautista, R., Freese, J., Morgan, S. L., & Smith, T. W. (2021). *General Social Survey 2021 Cross-section*. NORC ed. https://gss.norc.org/content/dam/gss/get-documentation/pdf/codebook/GSS%202021%20Codebook%20R1.pdf

Davies, J. M., Brighton, L. J., Reedy, F., & Bajwah, S. (2022). Maternity provision, contract status, and likelihood of returning to work: Evidence from research intensive universities in the UK. *Gender, Work & Organization*, *29*(5), 1495–1510. https://doi.org/10.1111/gwao.12843

De Luccie, M. F. (1995). Mothers as Gatekeepers: A Model of Maternal Mediators of Father Involvement. *The Journal of Genetic Psychology*, *156*(1), 115–131. https://doi.org/10.1080/00221325.1995.9914811

Derrick, G. E., Chen, P.-Y., van Leeuwen, T., Larivière, V., & Sugimoto, C. R. (2022). The relationship between parenting engagement and academic performance. *Scientific Reports*, *12*(1), 22300. https://doi.org/10.1038/s41598-022-26258-z

Deutsch, F. M., & Yao, B. (2014). Gender differences in faculty attrition in the USA. *Community, Work & Family*, *17*(4), 392–408. https://doi.org/10.1080/13668803.2014.885880

Eagly, A. H., & Wood, W. (2012). Social role theory. In *Handbook of theories of social psychology, Vol. 2* (pp. 458–476). Sage Publications Ltd. https://doi.org/10.4135/9781446249222.n49

Elsevier. (2017). *Gender in the Global Research Landscape*. Elsevier. https://assets.ctfassets.net/zlnfaxb2lcqx/57uxjkQA2aUQSpWayDUd5c/6653475e50db61cfb0f828e291c1e08a/Elsevier-gender-report-2017.pdf

Eren, E. (2022). Never the right time: Maternity planning alongside a science career in academia. *Journal of Gender Studies*, *31*(1), 136–147. https://doi.org/10.1080/09589236.2020.1858765

Feeney, M. K., Bernal, M., & Bowman, L. (2014). Enabling work? Family-friendly policies and academic productivity for men and women scientists. *Science and Public Policy*, *41*(6), 750–764. https://doi.org/10.1093/scipol/scu006

Fisher, R. J. (1993). Social Desirability Bias and the Validity of Indirect Questioning. *Journal of Consumer Research*, *20*(2), 303–315. https://doi.org/10.1086/209351

Haines, E. L., & Stroessner, S. J. (2019). The role prioritization model: How communal men and agentic women can (sometimes) have it all. *Social and Personality Psychology Compass*, *13*(12), e12504. https://doi.org/10.1111/spc3.12504

Hayes, A. F. (2017). *Introduction to Mediation, Moderation, and Conditional Process Analysis, Second Edition: A Regression-Based Approach* (Second Edition). The Guilford Press.

Hosseini, M., & Villanueva, A. C. (2023). A qualitative analysis of the impact of COVID-19 restrictions on gender biases in an Irish University. *PLOS ONE*, *18*(9), e0288467. https://doi.org/10.1371/journal.pone.0288467

Infanger, C., & Lima, M. (2019). Maternity Leave Benefit for Researchers: A Case Study of FAPESP’s Maternity Leave Policy. *International Journal of Environmental Science and Technology*, *11*, 134.

Johnson, D. R., & Creech, J. C. (1983). Ordinal Measures in Multiple Indicator Models: A Simulation Study of Categorization Error. *American Sociological Review*, *48*(3), 398–407. https://doi.org/10.2307/2095231

Kerr, P. S., & Holden, R. R. (1996). Development of the Gender Role Beliefs Scale (GRBS). *Journal of Social Behavior and Personality*, *11*(5), 3–15.

Ley, T. J., & Hamilton, B. H. (2008). The Gender Gap in NIH Grant Applications. *Science*, *322*(5907), 1472–1474. https://doi.org/10.1126/science.1165878

Liu, S., Smith-Doerr, L., Clark, D., & Misra, J. (2025). Faculty Inclusion Before and “After” COVID-19: Investigating the Effects of Caregiving, Health, and Gender. *ADVANCE Journal*. https://doi.org/10.5399/osu/ADVJRNL.6.1.2

Lorber, J. (1994). Daily Bread: Gender and Domestic Labor. In *Paradoxes of Gender* (pp. 172–193). Yale University Press. https://www.jstor.org/stable/j.ctt1bhkntg.12

Lunnemann, P., Jensen, M. H., & Jauffred, L. (2019). Gender bias in Nobel prizes. *Palgrave Communications*, *5*(1), Article 1. https://doi.org/10.1057/s41599-019-0256-3

Madsen, E. B., Nielsen, M. W., Bjørnholm, J., Jagsi, R., & Andersen, J. P. (2022). Author-level data confirm the widening gender gap in publishing rates during COVID-19. *eLife*, *11*, e76559. https://doi.org/10.7554/eLife.76559

Malkinson, T. S., Terhune, D. B., Kollamkulam, M., Guerreiro, M. J., Bassett, D. S., & Makin, T. R. (2023). Gender imbalances in the editorial activities of a selective journal run by academic editors. *PLOS ONE*, *18*(12), e0294805. https://doi.org/10.1371/journal.pone.0294805

Meeussen, L., Veldman, J., & Van Laar, C. (2016). Combining Gender, Work, and Family Identities: The Cross-Over and Spill-Over of Gender Norms into Young Adults’ Work and Family Aspirations. *Frontiers in Psychology*, *7*. https://www.frontiersin.org/articles/10.3389/fpsyg.2016.01781

Meho, L. I. (2021). The gender gap in highly prestigious international research awards, 2001–2020. *Quantitative Science Studies*, *2*(3), 976–989. https://doi.org/10.1162/qss_a_00148

Meteyer, K., & Perry-Jenkins, M. (2010). Father involvement among working-class, dual-earner couples. *Fathering: A Journal of Theory, Research, and Practice about Men as Fathers*, *8*(3), 379–403. https://doi.org/10.3149/fth.0803.379

Millward, L. J. (2006). The transition to motherhood in an organizational context: An interpretative phenomenological analysis. *Journal of Occupational and Organizational Psychology*, *79*(3), 315–333. https://doi.org/10.1348/096317906X110322

Misra, J., Lundquist, J. H., & Templer, A. (2012). Gender, Work Time, and Care Responsibilities Among Faculty1: Gender, Work Time, and Care Responsibilities Among Faculty. *Sociological Forum*, *27*(2), 300–323. https://doi.org/10.1111/j.1573-7861.2012.01319.x

Moore, D. (1995). Role conflict: Not only for women? A comparative analysis of 5 nations. *International Journal of Comparative Sociology*, *36*(1–2), 17–36.

Morgan, A. C., Way, S. F., Hoefer, M. J. D., Larremore, D. B., Galesic, M., & Clauset, A. (2021). The unequal impact of parenthood in academia. *Science Advances*, *7*(9). https://doi.org/10.1126/sciadv.abd1996

Moss-Racusin, C. A., Dovidio, J. F., Brescoll, V. L., Graham, M. J., & Handelsman, J. (2012). Science faculty’s subtle gender biases favor male students. *Proceedings*

*of the National Academy of Sciences*, *109*(41), 16474–16479. https://doi.org/10.1073/pnas.1211286109

Ni, C., Basson, I., Badia, G., Tufenkji, N., Sugimoto, C. R., & Larivière, V. (2025). Gender differences in submission behavior exacerbate publication disparities in elite journals. *eLife*, *12*. https://doi.org/10.7554/eLife.90049.2

Nielsen, M. W., Alegria, S., Börjeson, L., Etzkowitz, H., Falk-Krzesinski, H. J., Joshi, A., Leahey, E., Smith-Doerr, L., Woolley, A. W., & Schiebinger, L. (2017). Gender diversity leads to better science. *Proceedings of the National Academy of Sciences*, *114*(8), 1740–1742. https://doi.org/10.1073/pnas.1700616114

Nielsen, M. W., Pedersen, J. V., & Larregue, J. (2024). Getting ahead in the social sciences: How parenthood and publishing contribute to gender gaps in academic career advancement. *The British Journal of Sociology*, *75*(3), 322–346. https://doi.org/10.1111/1468-4446.13088

Norman, G. (2010). Likert scales, levels of measurement and the “laws” of statistics. *Advances in Health Sciences Education: Theory and Practice*, *15*(5), 625–632. https://doi.org/10.1007/s10459-010-9222-y

O’Laughlin, E. M., & Bischoff, L. G. (2005). Balancing Parenthood and Academia: Work/Family Stress as Influenced by Gender and Tenure Status. *Journal of Family Issues*, *26*(1), 79–106. https://doi.org/10.1177/0192513X04265942

Oliveira, L. D., Reichert, F., Zandonà, E., Soletti, R. C., & Staniscuaski, F. (2021). The 100,000 most influential scientists rank: The underrepresentation of Brazilian women in academia. *Anais Da Academia Brasileira de Ciências*, *93*, e20201952. https://doi.org/10.1590/0001-3765202120201952

Pearl, J. (2022). Direct and Indirect Effects. In H. Geffner, R. Dechter, & J. Y. Halpern (Eds.), *Probabilistic and Causal Inference* (1st ed., pp. 373–392). ACM. https://doi.org/10.1145/3501714.3501736

Percival Carter, E. (2024). Power and the perception of pregnancy in the academy. *Gender, Work & Organization*, *31*(5), 1951–1975. https://doi.org/10.1111/gwao.13001

Prasad, P., & Baron, J. (1996). *Measurement of gender-role attitudes, beliefs, and principles*. https://www.sas.upenn.edu/~baron/papers.htm/pp.htm

Prozesky, H. (2006). Gender differences in the journal publication productivity of South African academic authors. *South African Review of Sociology*, *37*(2), 87–112. https://doi.org/10.1080/21528586.2006.10419149

Sallee, M., Ward, K., & Wolf-Wendel, L. (2016). Can Anyone Have it All? Gendered Views on Parenting and Academic Careers. *Innovative Higher Education*, *41*(3), 187–202. https://doi.org/10.1007/s10755-015-9345-4

Santero, A., & Cannito, M. (2024). Balancing Work, Life, and Care Responsibilities in Academia: The Male Perspective. *Sociologia del lavoro : 168, 1, 2024*, 88–111. https://doi.org/10.3280/SL2024-168005

Sheltzer, J. M., & Smith, J. C. (2014). Elite male faculty in the life sciences employ fewer women. *Proceedings of the National Academy of Sciences*, *111*(28), 10107–10112. https://doi.org/10.1073/pnas.1403334111

Snir, R., Harpaz, I., & Ben-Baruch, D. (2009). Centrality of and Investment in Work and Family Among Israeli High-Tech Workers: A Bicultural Perspective. *Cross-Cultural Research*, *43*(4), 366–385. https://doi.org/10.1177/1069397109336991

Sougou, N. M., Ndiaye, O., Nabil, F., Folayan, M. O., Sarr, S. C., Mbaye, E. M., & Martínez-Pérez, G. Z. (2022). Barriers of West African women scientists in their research and academic careers: A qualitative research. *PLOS ONE*, *17*(3), e0265413. https://doi.org/10.1371/journal.pone.0265413

Spoon, K., LaBerge, N., Wapman, K. H., Zhang, S., Morgan, A. C., Galesic, M., Fosdick, B. K., Larremore, D. B., & Clauset, A. (2023). Gender and retention patterns among U.S. faculty. *Science Advances*, *9*(42), eadi2205. https://doi.org/10.1126/sciadv.adi2205

Stalford, H. (2005). Parenting, Care and Mobility in the Eu: Issues facing migrant scientists. *Innovation: The European Journal of Social Science Research*, *18*(3), 361–380. https://doi.org/10.1080/13511610500186805

Staniscuaski, F., Machado, A. V., Soletti, R. C., Reichert, F., Zandonà, E., Mello-Carpes, P. B., Infanger, C., Ludwig, Z. M. C., & de Oliveira, L. (2023). Bias against parents in science hits women harder. *Humanities and Social Sciences Communications*, *10*(1), 1–9. https://doi.org/10.1057/s41599-023-01722-x

Takahashi, A. M., & Takahashi, S. (2015). Gender promotion differences in economics departments in Japan: A duration analysis. *Journal of Asian Economics*, *41*, 1–19. https://doi.org/10.1016/j.asieco.2015.09.002

Thébaud, S., & Taylor, C. J. (2021). The Specter of Motherhood: Culture and the Production of Gendered Career Aspirations in Science and Engineering. *Gender & Society*, *35*(3), 395–421. https://doi.org/10.1177/08912432211006037

Townsley, N. C., & Broadfoot, K. J. (2008). Care, Career, and Academe: Heeding the Calls of a New Professoriate. *Women's Studies in Communication*, *31*(2), 133–142. https://doi.org/10.1080/07491409.2008.10162525

UIS. (2022). *UIS Statistics*. http://data.uis.unesco.org/#

Ullman, J. B., & Bentler, P. M. (2012). Structural Equation Modeling. In *Handbook of Psychology, Second Edition*. John Wiley & Sons, Ltd. https://doi.org/10.1002/9781118133880.hop202023

UNESCO. (2015). *UNESCO science report: Towards 2030* (F. Schlegel, Ed.). UNESCO Publ.

Vohlídalová, M. (2017). Academic couples, parenthood and women's research careers. *European Educational Research Journal*, *16*(2–3), 166–182. https://doi.org/10.1177/1474904116668883

Westfall, R. L., BoydJr., H. W., & Campbell, D. T. (1957). The Use of Structured Techniques in Motivation Research. *Journal of Marketing*, *22*(2), 134–139. https://doi.org/10.1177/002224295702200202

Wilton, S., & Ross, L. (2017). Flexibility, Sacrifice and Insecurity: A Canadian Study Assessing the Challenges of Balancing Work and Family in Academia. *Journal of Feminist Family Therapy*, *29*(1–2), 66–87. https://doi.org/10.1080/08952833.2016.1272663

Windsor, L. C., & Crawford, K. F. (2020). Best Practices for Normalizing Parents in the Academy: Higher- and Lower-Order Processes and Women and Parents' Success. *PS: Political Science & Politics*, *53*(2), 275–280. https://doi.org/10.1017/S1049096519001938

Witteman, H. O., Hendricks, M., Straus, S., & Tannenbaum, C. (2019). Are gender gaps due to evaluations of the applicant or the science? A natural experiment at a national funding agency. *The Lancet*, *393*(10171), 531–540. https://doi.org/10.1016/S0140-6736(18)32611-4

Xu, Y. J. (2008). Gender Disparity in STEM Disciplines: A Study of Faculty Attrition and Turnover Intentions. *Research in Higher Education*, *49*(7), 607–624. https://doi.org/10.1007/s11162-008-9097-4

Zhang, N., He, G., Shi, D., Zhao, Z., & Li, J. (2022). Does a gender-neutral name associate with the research impact of a scientist? *Journal of Informetrics*, *16*(1), 101251. https://doi.org/10.1016/j.joi.2022.101251

Zheng, X., Yuan, H., & Ni, C. (2022). Meta-Research: How parenthood contributes to gender gaps in academia. *eLife*, *11*, e78909. https://doi.org/10.7554/eLife.78909

**Supplementary Materials**

**Variable, data, and Method Supplemental**
*Includes Tables S1–S5*

**Table S1. Gender composition of the population, surveyed, respondents, and analytical sample.** The gender categorization was estimated using the methods successfully applied in Larivière et al. (2013). This is only used for a rough assessment of the percentage of each gender in the population, sample, responses, and the final analytical sample. Self-identified gender information was used to assign the gender in all other analyses of this study.

| Gender | Population | | Surveyed | | Respondents | | Analytical Sample | |
|---|---|---|---|---|---|---|---|---|
| | N | % | N | % | N | % | N | % |
| ***By gender*** | | | | | | | | |
| Men | 189,046 | 47.3 | 44,725 | 45.10 | 4,126 | 39.9 | 2,540 | 44.6 |
| Women | 113,507 | 28.4 | 29,850 | 30.10 | 4,003 | 38.7 | 3,160 | 55.4 |
| Unknown | 97,121 | 24.3 | 24,594 | 24.80 | 2,204 | 21.3 | - | - |

***Control variables.*** Control variables for statistical modeling analysis in this study include discipline, career stage, race, partner job status, and type according to respondents' self-reported data. The discipline includes arts & humanities, medical sciences, natural science & engineering, social sciences, and interdisciplinary fields (those who chose two or more areas in response). Our analysis categorizes career stages into four levels: trainee (post-doctoral fellows and research associates), early career (assistant professors), middle career (associate professors and senior researchers), and late career (full professors and emeritus professors) (see **Table S2**). Two race categories are classified as white and non-white. We also controlled for the partner job status (employed; self-employed or students;

unemployed, and other) and types (research-oriented; not research-oriented; and other) (see **Table S3**).

**Table S2. Sample distribution by gender, career stage and disciplinary area**

| | Trainee | | Early Career | | Middle Career | | Late Career | | Total | |
|---|---|---|---|---|---|---|---|---|---|---|
| | **N** | **%** | **N** | **%** | **N** | **%** | **N** | **%** | **N** | **%** |
| **Natural Science & Engineering** | | | | | | | | | | |
| Women | 66 | 49.6 | 111 | 53.1 | 317 | 56.2 | 527 | 65.5 | 1021 | 59.7 |
| Men | 67 | 50.4 | 98 | 46.9 | 247 | 43.8 | 278 | 34.5 | 690 | 40.3 |
| **Medical Sciences** | | | | | | | | | | |
| Women | 16 | 29.1 | 84 | 30.9 | 119 | 31.2 | 221 | 50.6 | 440 | 38.4 |
| Men | 39 | 70.9 | 188 | 69.1 | 262 | 68.8 | 216 | 49.4 | 705 | 61.6 |
| **Social Sciences** | | | | | | | | | | |
| Women | 16 | 20.3 | 101 | 30.2 | 221 | 30.9 | 430 | 49.3 | 768 | 38.4 |
| Men | 63 | 79.7 | 233 | 69.8 | 495 | 69.1 | 442 | 50.7 | 1233 | 61.6 |
| **Arts & Humanities** | | | | | | | | | | |
| Women | 2 | 50 | 10 | 27.8 | 41 | 29.7 | 98 | 49 | 151 | 39.9 |
| Men | 2 | 50 | 26 | 72.2 | 97 | 70.3 | 102 | 51 | 227 | 60.1 |
| **Interdisciplinary** | | | | | | | | | | |
| Women | 19 | 33.3 | 30 | 26.5 | 49 | 34 | 61 | 40.7 | 159 | 34.3 |
| Men | 38 | 66.7 | 83 | 73.5 | 95 | 66 | 89 | 59.3 | 305 | 65.7 |

**Table S3. Sample distribution by respondents' gender and their partner's job status and type**

| **Partner's job** | | **Women** | **Men** | **Total** |
|---|---|---|---|---|
| Job status | Employed | 1717 | 2474 | 4191 |
| | Self-employed, student, or out of work and looking for work | 354 | 462 | 816 |
| | Out of work but not looking for work or retired | 341 | 105 | 446 |
| | Other | 122 | 90 | 212 |
| Job type | Research-oriented | 789 | 1104 | 1893 |
| | Non-research- oriented | 1717 | 2005 | 3722 |

| | | | |
|---|---|---|---|
| Other | 28 | 22 | 50 |

***Objective career achievement measures*.** Three indicators were developed to evaluate the objective career achievement of scholars, including Annual relative publication (ARP), average relative citations (ARC), and annual relative coauthors (ARCo), to represent academics' career achievement in publication productivity, citations and research networks, respectively **(See Table S4)**. The normalization for these metrics was based on the domains classified by the National Science Foundation and the year records in the database Web of Science (WOS). For more detailed information, please refer to the researchers' previous study (Zheng et al., 2022).

**1) ARP.** For a respondent $x$, the yearly productivity (YP) is counted through dividing the total number of publications by the number of years between $x$'s earliest and latest publications:

$$YP_x = \frac{TPub}{Y2 - Y1 + 1}$$

where $TPub$ represents the total number of papers of $x$, $Y2$ denotes the year of $x$'s most recent publication and $Y1$ is that of $x$'s earliest publication. The ARP of $x$ is:

$$ARP_x = YP_x \Big/ \frac{1}{n}\sum_{i=1}^{n} YP_i$$

where $YP_i$ is the yearly publication of an academic $i$ in their discipline, and $n$ is the total number of academics in this discipline.

**2) ARC.** For a respondent $x$, we first compute $x$'s baseline citation (BC) as:

$$BC = \frac{1}{j}\sum_{k=1}^{j} Cite_k$$

where $Cite_k$ represents the number of citations received by the $k$th paper in this discipline and year, and $j$ is the total number of papers published in this discipline and year. The ARC of $x$ is:

$$ARC_x = \frac{1}{n}\sum_{i=1}^{n}\frac{Cite_i}{BC_i}$$

where $citation_i$ represents the number of citations received by the $i$th paper, authored by academic $x$, $BC_i$ refers to the baseline citations for the discipline and year corresponding to this paper's publication, and $n$ is the total number of papers published by $x$.

**3) ARCo.** To calculate ARCo, the respondent $x$'s yearly unique coauthors (YUC) is first calculated by:

$$YUC_x = \frac{TCo}{Y2 - Y1 + 1}$$

where $TCo$ is the total number of unique coauthors listed in researcher $x$'s publications, $Y2$ is the year of their most recent publication, and $Y1$ is the year of their first publication. The ARCo of $x$ is:

$$ARCo_x = YUC_x / \frac{1}{n}\sum_{i=1}^{n} YUC_i$$

where $YUC_i$ denotes the number of unique coauthors per year for the $i$th scholar in this discipline, and $n$ is the total number of academics within this discipline.

**Table S4. Descriptive Statistics of childcare responsibilities, Research satisfaction, Career satisfaction, Community recognitions, ARP, ARC and ARCo by Gender**

| Gender | Variable | Mean | Std | N | Margin of Error | Ci_Low | Ci_High |
|---|---|---|---|---|---|---|---|
| Men | Childcare responsibilities | 1.974328 | 1.177702 | 2540 | 0.045801 | 1.928527 | 2.020129 |
| | Research satisfaction | 0.563241 | 0.811245 | 2540 | 0.031549 | 0.531692 | 0.59479 |
| | Career satisfaction | 0.679558 | 0.713582 | 2540 | 0.027751 | 0.651807 | 0.707309 |
| | Community recognitions | 0.745541 | 0.622292 | 2540 | 0.024201 | 0.721340 | 0.769742 |
| | ARP | 2.271846 | 3.030543 | 2540 | 0.117858 | 2.153988 | 2.389704 |
| | ARC | 2.277524 | 3.81489 | 2540 | 0.148362 | 2.129163 | 2.425886 |
| | ARCo | 1.083716 | 1.389866 | 2540 | 0.054052 | 1.029664 | 1.137768 |
| Women | Childcare responsibilities | 2.129263 | 1.154497 | 3160 | 0.040254 | 2.089009 | 2.169516 |
| | Research satisfaction | 0.396448 | 0.901289 | 3160 | 0.031425 | 0.365023 | 0.427873 |
| | Career satisfaction | 0.593274 | 0.783554 | 3160 | 0.027320 | 0.565954 | 0.620594 |
| | Community recognitions | 0.628617 | 0.730607 | 3160 | 0.025474 | 0.603143 | 0.654091 |
| | ARP | 1.813911 | 1.801113 | 3160 | 0.062799 | 1.751112 | 1.87671 |
| | ARC | 2.141185 | 3.139571 | 3160 | 0.109467 | 2.031718 | 2.250652 |
| | ARCo | 0.981357 | 1.1545 | 3160 | 0.040254 | 0.941103 | 1.021611 |

***Egalitarian gender role beliefs.***

The four dimensions of egalitarian gender role beliefs are reflected in the following questions:

- A working mother can establish just as warm and secure a relationship with her children as a mother who doesn't work.
- It is acceptable for me to go to work and my spouse/partner to stay at home and care for the family (for female respondents) / It is acceptable for me to stay at home and care for the family and my spouse/partner to go to work (for male respondents).
- A woman should be prepared to cut down on her paid work for the sake of her family.
- A young child is likely to suffer if both parents work.

**Table S5. Descriptive Statistics of egalitarian gender role beliefs, by item and gender**

| Gender | Variable | Mean | Std | N | Margin of Error | Ci_Low | Ci_High |
|---|---|---|---|---|---|---|---|
| Men | Child suffering | 0.58453 | 1.913265 | 2540 | 0.074407 | 0.510122 | 0.658937 |
| | Mother relationship | 1.421074 | 1.527145 | 2540 | 0.059391 | 1.361683 | 1.480464 |
| | Mother to work | 0.654061 | 1.896609 | 2540 | 0.073759 | 0.580301 | 0.72782 |
| | Women to cut work | -0.802312 | 1.53377 | 2540 | 0.059648 | -0.86196 | -0.742663 |
| | Average | 0.466858 | 0.965398 | 2540 | 0.037544 | 0.429314 | 0.504403 |
| | Child suffering | 1.820325 | 1.516498 | 3160 | 0.052875 | 1.767449 | 1.8732 |
| | Mother relationship | 2.184102 | 1.145908 | 3160 | 0.039954 | 2.144148 | 2.224056 |
| Women | Mother to work | 1.565563 | 1.66083 | 3160 | 0.057908 | 1.507655 | 1.623471 |
| | Women to cut work | -0.381575 | 1.704422 | 3160 | 0.059428 | -0.441003 | -0.322147 |
| | Average | 1.298512 | 0.877616 | 3160 | 0.030600 | 1.267912 | 1.329112 |

**Statistical analysis**

We used several regression analysis techniques to explore the gendered difference in academia, including linear regressions, Structural Equation Modeling (SEM), and moderated linear regressions. Linear regressions were used for comparing gender disparities in gender role beliefs, childcare responsibilities and measures for subjective and objective career achievements. It is also adopted for testing whether firmer egalitarian gender role beliefs can lead to any change in childcare responsibilities among women and men in academics, respectively. SEM was used to test whether child-rearing responsibilities help explain the gender gap in academia as a mediator. A bootstrap sampling procedure with 5,000 iterations was employed to estimate 95% confidence intervals and statistical significance. Moderated linear regressions were used to examine (a) whether academics' egalitarian gender role beliefs can moderate the associations between gender and childcare responsibilities and (b) whether the parental support policies provided by institutions can moderate the associations between child-rearing responsibilities and academic career achievements. Career stage, disciplinary area, race, partner job status, job type and other covariates are controlled. Robust standard errors are clustered by respondents' affiliations.

**Original data of model results**
*Includes Tables S6–S8*

**Table S6.1. Mediation effect analysis results in subjective and objective career achievement measures, total sample.**

| Achievement type | Path | Coefficient | Standard error | Z-value | P-value | CI_low | Ci_high |
|---|---|---|---|---|---|---|---|
| Research satisfaction | Gender --> Childcare | 0.121 | 0.033 | 3.673 | 0.000 | 0.057 | 0.186 |
| | Childcare --> Achievement | -0.083 | 0.021 | -3.897 | 0.000 | -0.125 | -0.041 |
| | Gender --> Achievement | -0.253 | 0.051 | -4.954 | 0.000 | -0.353 | -0.153 |
| Career satisfaction | Gender --> Childcare | 0.121 | 0.033 | 3.673 | 0.000 | 0.057 | 0.186 |
| | Childcare --> Achievement | -0.104 | 0.020 | -5.182 | 0.000 | -0.143 | -0.064 |
| | Gender --> Achievement | -0.066 | 0.047 | -1.405 | 0.160 | -0.157 | 0.026 |
| Community recognition | Gender --> Childcare | 0.121 | 0.033 | 3.673 | 0.000 | 0.057 | 0.186 |
| | Childcare --> Achievement | -0.083 | 0.018 | -4.696 | 0.000 | -0.118 | -0.049 |
| | Gender --> Achievement | -0.169 | 0.042 | -3.995 | 0.000 | -0.252 | -0.086 |
| ARP | Gender --> Childcare | 0.121 | 0.033 | 3.673 | 0.000 | 0.057 | 0.186 |
| | Childcare --> Achievement | -0.139 | 0.026 | -5.396 | 0.000 | -0.189 | -0.088 |
| | Gender --> Achievement | -0.380 | 0.067 | -5.644 | 0.000 | -0.512 | -0.248 |
| ARC | Gender --> Childcare | 0.121 | 0.033 | 3.673 | 0.000 | 0.057 | 0.186 |
| | Childcare --> Achievement | -0.053 | 0.049 | -1.090 | 0.276 | -0.149 | 0.043 |
| | Gender --> Achievement | -0.221 | 0.108 | -2.048 | 0.041 | -0.433 | -0.010 |
| ARCo | Gender --> Childcare | 0.121 | 0.033 | 3.673 | 0.000 | 0.057 | 0.186 |
| | Childcare --> Achievement | -0.058 | 0.016 | -3.681 | 0.000 | -0.089 | -0.027 |
| | Gender --> Achievement | -0.110 | 0.036 | -3.047 | 0.002 | -0.180 | -0.039 |

**Table S6.2. Mediation effect analysis results in subjective and objective career achievement measures, by partner’s job status**

| Partner's job status | Achievement type | Path | Coefficient | Standard error | Z-value | P-value | CI_low | Ci_high |
|---|---|---|---|---|---|---|---|---|
| | Research satisfaction | Gender --> Childcare | 0.126 | 0.041 | 3.058 | 0.002 | 0.045 | 0.207 |
| Employed | | Childcare --> Achievement | -0.079 | 0.025 | -3.199 | 0.001 | -0.128 | -0.031 |
| | | Gender --> Achievement | -0.219 | 0.055 | -3.975 | 0.000 | -0.326 | -0.111 |
| | Career satisfaction | Gender --> Childcare | 0.126 | 0.041 | 3.058 | 0.002 | 0.045 | 0.207 |
| | | Childcare --> Achievement | -0.112 | 0.022 | -5.167 | 0.000 | -0.155 | -0.070 |
| | | Gender --> Achievement | -0.042 | 0.050 | -0.846 | 0.398 | -0.139 | 0.055 |
| | Community recognition | Gender --> Childcare | 0.126 | 0.041 | 3.058 | 0.002 | 0.045 | 0.207 |
| | | Childcare --> Achievement | -0.083 | 0.021 | -3.919 | 0.000 | -0.125 | -0.042 |
| | | Gender --> Achievement | -0.157 | 0.044 | -3.544 | 0.000 | -0.244 | -0.070 |
| | ARP | Gender --> Childcare | 0.126 | 0.041 | 3.058 | 0.002 | 0.045 | 0.207 |
| | | Childcare --> Achievement | -0.136 | 0.034 | -3.987 | 0.000 | -0.203 | -0.069 |
| | | Gender --> Achievement | -0.415 | 0.074 | -5.635 | 0.000 | -0.560 | -0.271 |
| | ARC | Gender --> Childcare | 0.126 | 0.041 | 3.058 | 0.002 | 0.045 | 0.207 |
| | | Childcare --> Achievement | -0.028 | 0.054 | -0.518 | 0.605 | -0.133 | 0.077 |
| | | Gender --> Achievement | -0.282 | 0.119 | -2.380 | 0.017 | -0.514 | -0.050 |
| | ARCo | Gender --> Childcare | 0.126 | 0.041 | 3.058 | 0.002 | 0.045 | 0.207 |
| | | Childcare --> Achievement | -0.052 | 0.020 | -2.614 | 0.009 | -0.090 | -0.013 |
| | | Gender --> Achievement | -0.127 | 0.038 | -3.352 | 0.001 | -0.202 | -0.053 |
| Self-employed, student, or out of work and looking for work | Research satisfaction | Gender --> Childcare | -0.021 | 0.100 | -0.208 | 0.835 | -0.216 | 0.174 |
| | | Childcare --> Achievement | -0.130 | 0.059 | -2.191 | 0.028 | -0.246 | -0.014 |
| | | Gender --> Achievement | -0.221 | 0.135 | -1.640 | 0.101 | -0.484 | 0.043 |
| | Career satisfaction | Gender --> Childcare | -0.021 | 0.100 | -0.208 | 0.835 | -0.216 | 0.174 |
| | | Childcare --> Achievement | -0.102 | 0.054 | -1.903 | 0.057 | -0.208 | 0.003 |
| | | Gender --> Achievement | -0.060 | 0.125 | -0.475 | 0.635 | -0.305 | 0.186 |
| | Community recognition | Gender --> Childcare | -0.021 | 0.100 | -0.208 | 0.835 | -0.216 | 0.174 |
| | | Childcare --> Achievement | -0.099 | 0.050 | -1.981 | 0.048 | -0.197 | -0.001 |
| | | Gender --> Achievement | -0.200 | 0.125 | -1.595 | 0.111 | -0.446 | 0.046 |
| | ARP | Gender --> Childcare | -0.021 | 0.100 | -0.208 | 0.835 | -0.216 | 0.174 |
| | | Childcare --> Achievement | -0.118 | 0.053 | -2.223 | 0.026 | -0.222 | -0.014 |
| | | Gender --> Achievement | -0.253 | 0.139 | -1.822 | 0.068 | -0.524 | 0.019 |
| | ARC | Gender --> Childcare | -0.021 | 0.100 | -0.208 | 0.835 | -0.216 | 0.174 |
| | | Childcare --> Achievement | 0.110 | 0.133 | 0.831 | 0.406 | -0.150 | 0.370 |
| | | Gender --> Achievement | 0.526 | 0.318 | 1.657 | 0.098 | -0.096 | 1.148 |
| | ARCo | Gender --> Childcare | -0.021 | 0.100 | -0.208 | 0.835 | -0.216 | 0.174 |
| | | Childcare --> Achievement | -0.057 | 0.038 | -1.500 | 0.134 | -0.131 | 0.017 |
| | | Gender --> Achievement | 0.062 | 0.107 | 0.582 | 0.561 | -0.148 | 0.273 |
| | Research satisfaction | Gender --> Childcare | 0.232 | 0.125 | 1.861 | 0.063 | -0.012 | 0.476 |
| | | Childcare --> Achievement | -0.045 | 0.079 | -0.567 | 0.571 | -0.199 | 0.110 |
| | | Gender --> Achievement | -0.467 | 0.217 | -2.151 | 0.031 | -0.892 | -0.041 |
| | Career satisfaction | Gender --> Childcare | 0.232 | 0.125 | 1.861 | 0.063 | -0.012 | 0.476 |
| | | Childcare --> Achievement | -0.030 | 0.073 | -0.410 | 0.682 | -0.172 | 0.113 |
| | | Gender --> Achievement | -0.225 | 0.197 | -1.143 | 0.253 | -0.612 | 0.161 |
| | Community recognition | Gender --> Childcare | 0.232 | 0.125 | 1.861 | 0.063 | -0.012 | 0.476 |
| | | Childcare --> Achievement | 0.076 | 0.058 | 1.314 | 0.189 | -0.037 | 0.190 |
| | | Gender --> Achievement | -0.188 | 0.173 | -1.092 | 0.275 | -0.527 | 0.150 |
| Out of work but not looking for work or retired | ARP | Gender --> Childcare | 0.232 | 0.125 | 1.861 | 0.063 | -0.012 | 0.476 |
| | | Childcare --> Achievement | -0.137 | 0.070 | -1.955 | 0.051 | -0.275 | 0.000 |
| | | Gender --> Achievement | -0.176 | 0.249 | -0.706 | 0.480 | -0.665 | 0.313 |
| | ARC | Gender --> Childcare | 0.232 | 0.125 | 1.861 | 0.063 | -0.012 | 0.476 |
| | | Childcare --> Achievement | -0.303 | 0.120 | -2.537 | 0.011 | -0.538 | -0.069 |
| | | Gender --> Achievement | -0.300 | 0.258 | -1.162 | 0.245 | -0.806 | 0.206 |
| | ARCo | Gender --> Childcare | 0.232 | 0.125 | 1.861 | 0.063 | -0.012 | 0.476 |
| | | Childcare --> Achievement | -0.062 | 0.036 | -1.713 | 0.087 | -0.133 | 0.009 |
| | | Gender --> Achievement | -0.061 | 0.128 | -0.480 | 0.631 | -0.312 | 0.189 |

**Table S6.3. Mediation effect analysis results in subjective and objective career achievement measures, by partner's job type**

| Partner's job type | Achievement type | Path | Coefficient | Standard error | Z-value | P-value | CI_low | Ci_high |
|---|---|---|---|---|---|---|---|---|
| | Research satisfaction | Gender --> Childcare | 0.111 | 0.062 | 1.785 | 0.074 | -0.011 | 0.232 |
| | | Childcare --> Achievement | -0.094 | 0.038 | -2.460 | 0.014 | -0.168 | -0.019 |
| | | Gender --> Achievement | -0.213 | 0.087 | -2.452 | 0.014 | -0.383 | -0.043 |
| Research-oriented | Career satisfaction | Gender --> Childcare | 0.111 | 0.062 | 1.785 | 0.074 | -0.011 | 0.232 |
| | | Childcare --> Achievement | -0.121 | 0.035 | -3.469 | 0.001 | -0.189 | -0.052 |
| | | Gender --> Achievement | -0.069 | 0.088 | -0.780 | 0.435 | -0.242 | 0.104 |

| | | | | | | | |
|---|---|---|---|---|---|---|---|
| | Community recognition | Gender --> Childcare | 0.111 | 0.062 | 1.785 | 0.074 | -0.011 | 0.232 |
| | | Childcare --> Achievement | -0.077 | 0.032 | -2.412 | 0.016 | -0.140 | -0.014 |
| | | Gender --> Achievement | -0.167 | 0.063 | -2.655 | 0.008 | -0.289 | -0.044 |
| | ARP | Gender --> Childcare | 0.111 | 0.062 | 1.785 | 0.074 | -0.011 | 0.232 |
| | | Childcare --> Achievement | -0.183 | 0.043 | -4.218 | 0.000 | -0.269 | -0.098 |
| | | Gender --> Achievement | -0.408 | 0.122 | -3.329 | 0.001 | -0.648 | -0.168 |
| | ARC | Gender --> Childcare | 0.111 | 0.062 | 1.785 | 0.074 | -0.011 | 0.232 |
| | | Childcare --> Achievement | 0.102 | 0.115 | 0.885 | 0.376 | -0.124 | 0.327 |
| | | Gender --> Achievement | -0.165 | 0.192 | -0.863 | 0.388 | -0.541 | 0.210 |
| | ARCo | Gender --> Childcare | 0.111 | 0.062 | 1.785 | 0.074 | -0.011 | 0.232 |
| | | Childcare --> Achievement | -0.089 | 0.026 | -3.418 | 0.001 | -0.141 | -0.038 |
| | | Gender --> Achievement | -0.113 | 0.068 | -1.662 | 0.096 | -0.246 | 0.020 |
| Non-research-oriented | Research satisfaction | Gender --> Childcare | 0.106 | 0.043 | 2.465 | 0.014 | 0.022 | 0.190 |
| | | Childcare --> Achievement | -0.077 | 0.026 | -3.024 | 0.002 | -0.127 | -0.027 |
| | | Gender --> Achievement | -0.266 | 0.062 | -4.296 | 0.000 | -0.387 | -0.145 |
| | Career satisfaction | Gender --> Childcare | 0.106 | 0.043 | 2.465 | 0.014 | 0.022 | 0.190 |
| | | Childcare --> Achievement | -0.093 | 0.023 | -4.031 | 0.000 | -0.138 | -0.048 |
| | | Gender --> Achievement | -0.058 | 0.055 | -1.055 | 0.292 | -0.167 | 0.050 |
| | Community recognition | Gender --> Childcare | 0.106 | 0.043 | 2.465 | 0.014 | 0.022 | 0.190 |
| | | Childcare --> Achievement | -0.085 | 0.021 | -4.043 | 0.000 | -0.126 | -0.044 |
| | | Gender --> Achievement | -0.164 | 0.053 | -3.101 | 0.002 | -0.268 | -0.060 |
| | ARP | Gender --> Childcare | 0.106 | 0.043 | 2.465 | 0.014 | 0.022 | 0.190 |
| | | Childcare --> Achievement | -0.110 | 0.034 | -3.248 | 0.001 | -0.177 | -0.044 |
| | | Gender --> Achievement | -0.365 | 0.087 | -4.182 | 0.000 | -0.536 | -0.194 |
| | ARC | Gender --> Childcare | 0.106 | 0.043 | 2.465 | 0.014 | 0.022 | 0.190 |
| | | Childcare --> Achievement | -0.116 | 0.046 | -2.511 | 0.012 | -0.206 | -0.025 |
| | | Gender --> Achievement | -0.267 | 0.131 | -2.044 | 0.041 | -0.523 | -0.011 |
| | ARCo | Gender --> Childcare | 0.106 | 0.043 | 2.465 | 0.014 | 0.022 | 0.190 |
| | | Childcare --> Achievement | -0.040 | 0.021 | -1.861 | 0.063 | -0.082 | 0.002 |
| | | Gender --> Achievement | -0.108 | 0.046 | -2.375 | 0.018 | -0.198 | -0.019 |

**Table S7.1. The moderating effect of egalitarian gender role beliefs between gender and childcare responsibilities, by partner's job status and type**

| Partner's job status/type | Variable | Coefficient | Standard error | T-value | P-value | CI_low | Ci_high |
|---|---|---|---|---|---|---|---|
| Employed | Gender role beliefs | -0.026 | 0.029 | -0.902 | 0.368 | -0.082 | 0.030 |
| | Gender | 0.279 | 0.061 | 4.547 | 0.000 | 0.159 | 0.400 |
| | Gender # Gender role beliefs | -0.093 | 0.040 | -2.333 | 0.020 | -0.172 | -0.015 |
| Self-employed, student, or out of work and looking for work | Gender role beliefs | -0.103 | 0.065 | -1.575 | 0.117 | -0.231 | 0.026 |
| | Gender | 0.331 | 0.137 | 2.409 | 0.017 | 0.060 | 0.601 |
| | Gender # Gender role beliefs | -0.177 | 0.092 | -1.909 | 0.057 | -0.359 | 0.006 |
| Out of work but not looking for work or retired | Gender role beliefs | 0.091 | 0.064 | 1.424 | 0.156 | -0.035 | 0.217 |
| | Gender | 0.605 | 0.241 | 2.504 | 0.013 | 0.128 | 1.081 |
| | Gender # Gender role beliefs | -0.403 | 0.156 | -2.584 | 0.010 | -0.711 | -0.096 |
| Research-oriented | Gender role beliefs | -0.071 | 0.043 | -1.644 | 0.101 | -0.155 | 0.014 |
| | Gender | 0.303 | 0.093 | 3.248 | 0.001 | 0.119 | 0.486 |
| | Gender # Gender role beliefs | -0.094 | 0.060 | -1.586 | 0.114 | -0.212 | 0.023 |
| Non-research-oriented | Gender role beliefs | -0.016 | 0.029 | -0.537 | 0.592 | -0.074 | 0.042 |
| | Gender | 0.281 | 0.059 | 4.742 | 0.000 | 0.165 | 0.398 |
| | Gender # Gender role beliefs | -0.122 | 0.042 | -2.892 | 0.004 | -0.205 | -0.039 |

**Table S7.2. The marginal effect of gender on childcare responsibilities at different levels of egalitarian gender role beliefs, by partner's job status and type**

| Partner's job status/type | Level of gender role beliefs | Coefficient | Standard error | T-value | P-value | CI_low | Ci_high |
|---|---|---|---|---|---|---|---|
| Employed | Mean-1sd | 0.287 | 0.064 | 4.497 | 0.000 | 0.161 | 0.412 |
| | Mean | 0.193 | 0.043 | 4.500 | 0.000 | 0.109 | 0.277 |
| | Mean+1sd | 0.099 | 0.053 | 1.859 | 0.064 | -0.006 | 0.204 |
| Self-employed, student, or out of work and looking for work | Mean-1sd | 0.344 | 0.142 | 2.425 | 0.016 | 0.065 | 0.624 |
| | Mean | 0.167 | 0.105 | 1.582 | 0.115 | -0.041 | 0.374 |
| | Mean+1sd | -0.011 | 0.139 | -0.079 | 0.937 | -0.285 | 0.263 |
| Out of work but not looking for work or retired | Mean-1sd | 0.636 | 0.251 | 2.529 | 0.012 | 0.140 | 1.132 |
| | Mean | 0.230 | 0.148 | 1.551 | 0.123 | -0.062 | 0.523 |
| | Mean+1sd | -0.176 | 0.173 | -1.014 | 0.312 | -0.518 | 0.166 |
| Research-oriented | Mean-1sd | 0.310 | 0.096 | 3.216 | 0.001 | 0.120 | 0.499 |
| | Mean | 0.215 | 0.068 | 3.139 | 0.002 | 0.080 | 0.349 |
| | Mean+1sd | 0.120 | 0.085 | 1.405 | 0.161 | -0.048 | 0.287 |
| Non-research-oriented | Mean-1sd | 0.291 | 0.062 | 4.716 | 0.000 | 0.170 | 0.412 |
| | Mean | 0.168 | 0.043 | 3.951 | 0.000 | 0.085 | 0.252 |

| | | | | | | |
|---|---|---|---|---|---|---|
| Mean+1sd | 0.045 | 0.059 | 0.772 | 0.440 | -0.070 | 0.160 |

**Table S8.1. Change in subjective and objective career achievements with the change in childcare responsibilities, the parental support policies provided by institutions, and their interaction items, by gender**

| | Achievement | Support | Variable | Coefficient | Standard error | T-value | P-value | CI_low | Ci_high |
|---|---|---|---|---|---|---|---|---|---|
| Men | Research Satisfaction | Childcare Support | Childcare responsibilities | -0.037 | 0.031 | -1.207 | 0.228 | -0.098 | 0.023 |
| | | | Childcare Support | 0.303 | 0.183 | 1.655 | 0.099 | -0.057 | 0.664 |
| | | | Child_Res#Childcare Sup | -0.142 | 0.088 | -1.616 | 0.107 | -0.315 | 0.031 |
| | | Flexible Schedule | Childcare responsibilities | -0.063 | 0.033 | -1.911 | 0.057 | -0.129 | 0.002 |
| | | | Flexible Schedule | 0.110 | 0.147 | 0.748 | 0.455 | -0.180 | 0.400 |
| | | | Child_Res#Flexible_Sche | 0.013 | 0.064 | 0.199 | 0.842 | -0.114 | 0.139 |
| | | Paternity Leave | Childcare responsibilities | -0.037 | 0.031 | -1.183 | 0.237 | -0.098 | 0.024 |
| | | | Paternity Leave | 0.430 | 0.160 | 2.684 | 0.008 | 0.115 | 0.746 |
| | | | Childcare Res#Paternity Leave | -0.143 | 0.078 | -1.838 | 0.067 | -0.295 | 0.010 |
| | | Paused Tenure Clock | Childcare responsibilities | -0.059 | 0.032 | -1.840 | 0.066 | -0.123 | 0.004 |
| | | | Paused Tenure Clock | 0.085 | 0.185 | 0.461 | 0.645 | -0.278 | 0.448 |
| | | | Childcare Res#Paused Tenure | -0.008 | 0.087 | -0.097 | 0.923 | -0.180 | 0.163 |
| | Career Satisfaction | Childcare Support | Childcare responsibilities | -0.056 | 0.031 | -1.810 | 0.071 | -0.117 | 0.005 |
| | | | Childcare Support | 0.173 | 0.167 | 1.040 | 0.299 | -0.154 | 0.501 |
| | | | Child_Res#Childcare Sup | -0.038 | 0.082 | -0.461 | 0.645 | -0.198 | 0.123 |
| | | Flexible Schedule | Childcare responsibilities | -0.074 | 0.033 | -2.268 | 0.024 | -0.139 | -0.010 |
| | | | Flexible Schedule | 0.144 | 0.124 | 1.158 | 0.248 | -0.100 | 0.388 |
| | | | Child_Res#Flexible_Sche | 0.047 | 0.057 | 0.829 | 0.407 | -0.065 | 0.159 |
| | | Paternity Leave | Childcare responsibilities | -0.048 | 0.031 | -1.543 | 0.124 | -0.109 | 0.013 |
| | | | Paternity Leave | 0.234 | 0.156 | 1.501 | 0.134 | -0.072 | 0.541 |
| | | | Childcare Res#Paternity Leave | -0.092 | 0.074 | -1.236 | 0.217 | -0.238 | 0.054 |
| | | Paused Tenure Clock | Childcare responsibilities | -0.061 | 0.031 | -1.990 | 0.047 | -0.121 | -0.001 |
| | | | Paused Tenure Clock | 0.088 | 0.184 | 0.479 | 0.632 | -0.273 | 0.449 |
| | | | Childcare Res#Paused Tenure | -0.016 | 0.088 | -0.187 | 0.852 | -0.188 | 0.156 |
| | Community Recognition | Childcare Support | Childcare responsibilities | -0.058 | 0.026 | -2.214 | 0.027 | -0.109 | -0.006 |
| | | | Childcare Support | 0.042 | 0.159 | 0.265 | 0.791 | -0.271 | 0.355 |
| | | | Child_Res#Childcare Sup | -0.035 | 0.074 | -0.474 | 0.636 | -0.180 | 0.110 |
| | | Flexible Schedule | Childcare responsibilities | -0.059 | 0.027 | -2.213 | 0.027 | -0.112 | -0.007 |

| | | | | | | | | |
|---|---|---|---|---|---|---|---|---|
| | | Flexible Schedule | 0.099 | 0.113 | 0.876 | 0.382 | -0.123 | 0.322 |
| | | Child_Res#Flexible_Sche | -0.011 | 0.054 | -0.203 | 0.839 | -0.117 | 0.095 |
| | Paternity Leave | Childcare responsibilities | -0.050 | 0.026 | -1.941 | 0.053 | -0.100 | 0.001 |
| | | Paternity Leave | 0.283 | 0.145 | 1.955 | 0.051 | -0.001 | 0.567 |
| | | Childcare Res#Paternity Leave | -0.078 | 0.063 | -1.239 | 0.216 | -0.203 | 0.046 |
| | Paused Tenure Clock | Childcare responsibilities | -0.063 | 0.026 | -2.400 | 0.017 | -0.115 | -0.011 |
| | | Paused Tenure Clock | 0.060 | 0.171 | 0.349 | 0.727 | -0.276 | 0.396 |
| | | Childcare Res#Paused Tenure | 0.005 | 0.080 | 0.062 | 0.950 | -0.152 | 0.162 |
| ARP | Childcare Support | Childcare responsibilities | -0.160 | 0.054 | -2.991 | 0.003 | -0.266 | -0.055 |
| | | Childcare Support | 0.201 | 0.382 | 0.527 | 0.598 | -0.549 | 0.952 |
| | | Child_Res#Childcare Sup | -0.016 | 0.123 | -0.129 | 0.897 | -0.258 | 0.226 |
| | Flexible Schedule | Childcare responsibilities | -0.188 | 0.058 | -3.224 | 0.001 | -0.302 | -0.073 |
| | | Flexible Schedule | -0.089 | 0.283 | -0.314 | 0.753 | -0.646 | 0.468 |
| | | Child_Res#Flexible_Sche | 0.083 | 0.110 | 0.757 | 0.450 | -0.133 | 0.299 |
| | Paternity Leave | Childcare responsibilities | -0.182 | 0.055 | -3.336 | 0.001 | -0.290 | -0.075 |
| | | Paternity Leave | -0.150 | 0.287 | -0.521 | 0.602 | -0.714 | 0.415 |
| | | Childcare Res#Paternity Leave | 0.109 | 0.101 | 1.079 | 0.281 | -0.090 | 0.308 |
| | Paused Tenure Clock | Childcare responsibilities | -0.221 | 0.048 | -4.569 | 0.000 | -0.316 | -0.126 |
| | | Paused Tenure Clock | -0.537 | 0.357 | -1.503 | 0.134 | -1.239 | 0.165 |
| | | Childcare Res#Paused Tenure | 0.454 | 0.208 | 2.179 | 0.030 | 0.044 | 0.863 |
| ARC | Childcare Support | Childcare responsibilities | 0.008 | 0.111 | 0.069 | 0.945 | -0.211 | 0.227 |
| | | Childcare Support | 0.298 | 0.319 | 0.936 | 0.350 | -0.328 | 0.925 |
| | | Child_Res#Childcare Sup | -0.161 | 0.134 | -1.199 | 0.231 | -0.424 | 0.103 |
| | Flexible Schedule | Childcare responsibilities | 0.035 | 0.121 | 0.287 | 0.774 | -0.203 | 0.272 |
| | | Flexible Schedule | 0.456 | 0.399 | 1.143 | 0.254 | -0.328 | 1.239 |
| | | Child_Res#Flexible_Sche | -0.182 | 0.145 | -1.248 | 0.213 | -0.467 | 0.104 |
| | Paternity Leave | Childcare responsibilities | 0.008 | 0.108 | 0.070 | 0.944 | -0.206 | 0.221 |
| | | Paternity Leave | 0.547 | 0.347 | 1.575 | 0.116 | -0.136 | 1.229 |
| | | Childcare Res#Paternity Leave | -0.159 | 0.142 | -1.125 | 0.261 | -0.437 | 0.119 |
| | Paused Tenure Clock | Childcare responsibilities | 0.004 | 0.104 | 0.038 | 0.969 | -0.200 | 0.208 |
| | | Paused Tenure Clock | 1.081 | 0.617 | 1.751 | 0.081 | -0.132 | 2.295 |
| | | Childcare Res#Paused Tenure | -0.114 | 0.219 | -0.521 | 0.602 | -0.545 | 0.316 |
| ARCo | Childcare Support | Childcare responsibilities | -0.091 | 0.024 | -3.736 | 0.000 | -0.139 | -0.043 |
| | | Childcare Support | -0.012 | 0.138 | -0.090 | 0.928 | -0.283 | 0.258 |

| | | | | | | | | | |
|---|---|---|---|---|---|---|---|---|---|
| | | | Child_Res#Childcare Sup | 0.048 | 0.050 | 0.966 | 0.335 | -0.050 | 0.146 |
| | | Flexible Schedule | Childcare responsibilities | -0.094 | 0.026 | -3.647 | 0.000 | -0.145 | -0.044 |
| | | | Flexible Schedule | 0.028 | 0.132 | 0.215 | 0.830 | -0.231 | 0.288 |
| | | | Child_Res#Flexible_Sche | 0.043 | 0.051 | 0.831 | 0.406 | -0.058 | 0.144 |
| | | Paternity Leave | Childcare responsibilities | -0.086 | 0.023 | -3.665 | 0.000 | -0.132 | -0.040 |
| | | | Paternity Leave | 0.079 | 0.175 | 0.452 | 0.652 | -0.266 | 0.424 |
| | | | Childcare Res#Paternity Leave | 0.020 | 0.056 | 0.363 | 0.717 | -0.090 | 0.130 |
| | | Paused Tenure Clock | Childcare responsibilities | -0.104 | 0.023 | -4.598 | 0.000 | -0.148 | -0.059 |
| | | | Paused Tenure Clock | -0.219 | 0.169 | -1.294 | 0.196 | -0.552 | 0.114 |
| | | | Childcare Res#Paused Tenure | 0.163 | 0.087 | 1.885 | 0.060 | -0.007 | 0.333 |
| Women | Research Satisfaction | Childcare Support | Childcare responsibilities | -0.105 | 0.032 | -3.269 | 0.001 | -0.168 | -0.042 |
| | | | Childcare Support | 0.223 | 0.163 | 1.371 | 0.171 | -0.097 | 0.542 |
| | | | Child_Res#Childcare Sup | -0.061 | 0.069 | -0.888 | 0.375 | -0.197 | 0.075 |
| | | Flexible Schedule | Childcare responsibilities | -0.151 | 0.037 | -4.129 | 0.000 | -0.223 | -0.079 |
| | | | Flexible Schedule | -0.107 | 0.134 | -0.802 | 0.423 | -0.370 | 0.156 |
| | | | Child_Res#Flexible_Sche | 0.088 | 0.055 | 1.611 | 0.108 | -0.019 | 0.195 |
| | | Paternity Leave | Childcare responsibilities | -0.049 | 0.044 | -1.100 | 0.272 | -0.136 | 0.038 |
| | | | Paternity Leave | 0.329 | 0.148 | 2.223 | 0.027 | 0.038 | 0.619 |
| | | | Childcare Res#Paternity Leave | -0.105 | 0.058 | -1.805 | 0.072 | -0.219 | 0.009 |
| | | Paused Tenure Clock | Childcare responsibilities | -0.094 | 0.032 | -2.994 | 0.003 | -0.156 | -0.032 |
| | | | Paused Tenure Clock | 0.231 | 0.149 | 1.552 | 0.121 | -0.061 | 0.524 |
| | | | Childcare Res#Paused Tenure | -0.084 | 0.061 | -1.365 | 0.173 | -0.204 | 0.037 |
| | Career Satisfaction | Childcare Support | Childcare responsibilities | -0.140 | 0.028 | -4.910 | 0.000 | -0.196 | -0.084 |
| | | | Childcare Support | 0.177 | 0.146 | 1.211 | 0.226 | -0.110 | 0.464 |
| | | | Child_Res#Childcare Sup | -0.048 | 0.062 | -0.774 | 0.439 | -0.170 | 0.074 |
| | | Flexible Schedule | Childcare responsibilities | -0.192 | 0.034 | -5.703 | 0.000 | -0.258 | -0.126 |
| | | | Flexible Schedule | -0.078 | 0.124 | -0.628 | 0.530 | -0.322 | 0.166 |
| | | | Child_Res#Flexible_Sche | 0.114 | 0.056 | 2.017 | 0.044 | 0.003 | 0.225 |
| | | Paternity Leave | Childcare responsibilities | -0.125 | 0.039 | -3.211 | 0.001 | -0.202 | -0.049 |
| | | | Paternity Leave | 0.207 | 0.132 | 1.569 | 0.117 | -0.052 | 0.466 |
| | | | Childcare Res#Paternity Leave | -0.030 | 0.050 | -0.588 | 0.557 | -0.128 | 0.069 |
| | | Paused Tenure Clock | Childcare responsibilities | -0.122 | 0.026 | -4.631 | 0.000 | -0.174 | -0.070 |
| | | | Paused Tenure Clock | 0.314 | 0.130 | 2.421 | 0.016 | 0.059 | 0.568 |
| | | | Childcare Res#Paused Tenure | -0.092 | 0.055 | -1.652 | 0.099 | -0.201 | 0.017 |

| | | | | | | | | |
|---|---|---|---|---|---|---|---|---|
| Community Recognition | Childcare Support | Childcare responsibilities | -0.102 | 0.028 | -3.691 | 0.000 | -0.156 | -0.047 |
| | | Childcare Support | 0.090 | 0.128 | 0.700 | 0.484 | -0.162 | 0.341 |
| | | Child_Res#Childcare Sup | 0.005 | 0.054 | 0.092 | 0.927 | -0.101 | 0.111 |
| | Flexible Schedule | Childcare responsibilities | -0.107 | 0.031 | -3.407 | 0.001 | -0.169 | -0.045 |
| | | Flexible Schedule | 0.086 | 0.108 | 0.803 | 0.423 | -0.125 | 0.298 |
| | | Child_Res#Flexible_Sche | 0.025 | 0.046 | 0.551 | 0.582 | -0.064 | 0.115 |
| | Paternity Leave | Childcare responsibilities | -0.064 | 0.041 | -1.551 | 0.122 | -0.145 | 0.017 |
| | | Paternity Leave | 0.213 | 0.135 | 1.578 | 0.115 | -0.052 | 0.478 |
| | | Childcare Res#Paternity Leave | -0.053 | 0.055 | -0.956 | 0.340 | -0.162 | 0.056 |
| | Paused Tenure Clock | Childcare responsibilities | -0.088 | 0.028 | -3.120 | 0.002 | -0.144 | -0.033 |
| | | Paused Tenure Clock | 0.139 | 0.133 | 1.045 | 0.297 | -0.122 | 0.399 |
| | | Childcare Res#Paused Tenure | -0.043 | 0.059 | -0.728 | 0.467 | -0.158 | 0.073 |
| ARP | Childcare Support | Childcare responsibilities | -0.116 | 0.027 | -4.283 | 0.000 | -0.169 | -0.063 |
| | | Childcare Support | -0.139 | 0.154 | -0.901 | 0.368 | -0.443 | 0.164 |
| | | Child_Res#Childcare Sup | 0.087 | 0.064 | 1.373 | 0.171 | -0.038 | 0.212 |
| | Flexible Schedule | Childcare responsibilities | -0.066 | 0.033 | -2.004 | 0.046 | -0.131 | -0.001 |
| | | Flexible Schedule | 0.243 | 0.123 | 1.983 | 0.048 | 0.002 | 0.484 |
| | | Child_Res#Flexible_Sche | -0.070 | 0.051 | -1.387 | 0.166 | -0.170 | 0.029 |
| | Paternity Leave | Childcare responsibilities | -0.048 | 0.043 | -1.114 | 0.266 | -0.131 | 0.036 |
| | | Paternity Leave | 0.302 | 0.146 | 2.065 | 0.040 | 0.015 | 0.589 |
| | | Childcare Res#Paternity Leave | -0.070 | 0.058 | -1.197 | 0.232 | -0.185 | 0.045 |
| | Paused Tenure Clock | Childcare responsibilities | -0.077 | 0.029 | -2.623 | 0.009 | -0.135 | -0.019 |
| | | Paused Tenure Clock | 0.379 | 0.183 | 2.068 | 0.039 | 0.019 | 0.739 |
| | | Childcare Res#Paused Tenure | -0.039 | 0.065 | -0.608 | 0.544 | -0.167 | 0.088 |
| ARC | Childcare Support | Childcare responsibilities | -0.108 | 0.053 | -2.052 | 0.041 | -0.211 | -0.005 |
| | | Childcare Support | -0.205 | 0.213 | -0.962 | 0.337 | -0.623 | 0.214 |
| | | Child_Res#Childcare Sup | 0.120 | 0.090 | 1.329 | 0.185 | -0.057 | 0.297 |
| | Flexible Schedule | Childcare responsibilities | -0.127 | 0.062 | -2.050 | 0.041 | -0.248 | -0.005 |
| | | Flexible Schedule | -0.364 | 0.213 | -1.711 | 0.088 | -0.782 | 0.054 |
| | | Child_Res#Flexible_Sche | 0.093 | 0.094 | 0.991 | 0.322 | -0.092 | 0.278 |
| | Paternity Leave | Childcare responsibilities | -0.087 | 0.087 | -0.996 | 0.320 | -0.257 | 0.084 |
| | | Paternity Leave | 0.048 | 0.258 | 0.186 | 0.853 | -0.459 | 0.555 |
| | | Childcare Res#Paternity Leave | 0.010 | 0.097 | 0.105 | 0.916 | -0.180 | 0.201 |
| | | Childcare responsibilities | -0.075 | 0.051 | -1.479 | 0.140 | -0.175 | 0.025 |

| | | | | | | | | | |
|---|---|---|---|---|---|---|---|---|---|
| | | Paused Tenure Clock | Paused Tenure Clock | 0.318 | 0.235 | 1.353 | 0.177 | -0.144 | 0.779 |
| | | | Childcare Res#Paused Tenure | 0.019 | 0.087 | 0.218 | 0.827 | -0.153 | 0.191 |
| | ARCo | Childcare Support | Childcare responsibilities | -0.044 | 0.023 | -1.883 | 0.060 | -0.089 | 0.002 |
| | | | Childcare Support | -0.105 | 0.095 | -1.108 | 0.269 | -0.292 | 0.082 |
| | | | Child_Res#Childcare Sup | 0.051 | 0.040 | 1.269 | 0.205 | -0.028 | 0.129 |
| | | Flexible Schedule | Childcare responsibilities | -0.018 | 0.027 | -0.678 | 0.498 | -0.072 | 0.035 |
| | | | Flexible Schedule | 0.112 | 0.089 | 1.249 | 0.212 | -0.064 | 0.287 |
| | | | Child_Res#Flexible_Sche | -0.033 | 0.040 | -0.820 | 0.413 | -0.111 | 0.046 |
| | | Paternity Leave | Childcare responsibilities | 0.014 | 0.026 | 0.516 | 0.606 | -0.038 | 0.066 |
| | | | Paternity Leave | 0.281 | 0.089 | 3.150 | 0.002 | 0.106 | 0.456 |
| | | | Childcare Res#Paternity Leave | -0.065 | 0.042 | -1.549 | 0.122 | -0.148 | 0.018 |
| | | Paused Tenure Clock | Childcare responsibilities | -0.022 | 0.027 | -0.802 | 0.423 | -0.074 | 0.031 |
| | | | Paused Tenure Clock | 0.178 | 0.115 | 1.548 | 0.122 | -0.048 | 0.404 |
| | | | Childcare Res#Paused Tenure | -0.028 | 0.048 | -0.592 | 0.554 | -0.123 | 0.066 |

**Table S8.2. Change in subjective and objective career achievements with the change in childcare responsibilities, the parental support policies provided by institutions, and their interaction items, by gender, if partner's job type is employed**

| | Achievement | Support | Variable | Coefficient | Standard error | T-value | P-value | CI_low | Ci_high |
|---|---|---|---|---|---|---|---|---|---|
| Men | Research Satisfaction | Childcare Support | Childcare responsibilities | -0.019 | 0.039 | -0.482 | 0.630 | -0.095 | 0.058 |
| | | | Childcare Support | 0.302 | 0.218 | 1.386 | 0.166 | -0.126 | 0.731 |
| | | | Child_Res#Childcare Sup | -0.123 | 0.100 | -1.227 | 0.221 | -0.319 | 0.074 |
| | | Flexible Schedule | Childcare responsibilities | -0.035 | 0.042 | -0.834 | 0.405 | -0.116 | 0.047 |
| | | | Flexible Schedule | 0.188 | 0.168 | 1.119 | 0.264 | -0.143 | 0.519 |
| | | | Child_Res#Flexible_Sche | -0.020 | 0.075 | -0.262 | 0.793 | -0.167 | 0.128 |
| | | Paternity Leave | Childcare responsibilities | -0.013 | 0.038 | -0.335 | 0.738 | -0.089 | 0.063 |
| | | | Paternity Leave | 0.465 | 0.195 | 2.388 | 0.017 | 0.082 | 0.848 |
| | | | Childcare Res#Paternity Leave | -0.174 | 0.096 | -1.807 | 0.071 | -0.362 | 0.015 |
| | | Paused Tenure Clock | Childcare responsibilities | -0.049 | 0.040 | -1.225 | 0.221 | -0.127 | 0.029 |
| | | | Paused Tenure Clock | -0.042 | 0.211 | -0.197 | 0.844 | -0.457 | 0.374 |
| | | | Childcare Res#Paused Tenure | 0.055 | 0.087 | 0.634 | 0.527 | -0.115 | 0.225 |
| | | Childcare Support | Childcare responsibilities | -0.052 | 0.037 | -1.431 | 0.153 | -0.125 | 0.020 |

| | | | | | | | | |
|---|---|---|---|---|---|---|---|---|
| | | Childcare Support | 0.108 | 0.195 | 0.553 | 0.580 | -0.276 | 0.491 |
| | | Child_Res#Childcare Sup | 0.022 | 0.093 | 0.232 | 0.816 | -0.162 | 0.205 |
| | Flexible Schedule | Childcare responsibilities | -0.063 | 0.041 | -1.538 | 0.125 | -0.143 | 0.017 |
| | | Flexible Schedule | 0.177 | 0.150 | 1.176 | 0.240 | -0.119 | 0.472 |
| | | Child_Res#Flexible_Sche | 0.047 | 0.069 | 0.681 | 0.496 | -0.089 | 0.183 |
| Career Satisfaction | Paternity Leave | Childcare responsibilities | -0.022 | 0.036 | -0.616 | 0.539 | -0.093 | 0.049 |
| | | Paternity Leave | 0.320 | 0.200 | 1.601 | 0.110 | -0.073 | 0.713 |
| | | Childcare Res#Paternity Leave | -0.170 | 0.096 | -1.767 | 0.078 | -0.358 | 0.019 |
| | Paused Tenure Clock | Childcare responsibilities | -0.059 | 0.037 | -1.608 | 0.109 | -0.131 | 0.013 |
| | | Paused Tenure Clock | -0.077 | 0.208 | -0.372 | 0.710 | -0.487 | 0.332 |
| | | Childcare Res#Paused Tenure | 0.073 | 0.092 | 0.798 | 0.425 | -0.107 | 0.254 |
| | Childcare Support | Childcare responsibilities | -0.034 | 0.031 | -1.095 | 0.274 | -0.095 | 0.027 |
| | | Childcare Support | 0.063 | 0.178 | 0.354 | 0.724 | -0.288 | 0.414 |
| | | Child_Res#Childcare Sup | -0.011 | 0.078 | -0.146 | 0.884 | -0.165 | 0.142 |
| | Flexible Schedule | Childcare responsibilities | -0.040 | 0.036 | -1.112 | 0.267 | -0.110 | 0.031 |
| | | Flexible Schedule | 0.035 | 0.144 | 0.244 | 0.808 | -0.248 | 0.318 |
| Community Recognition | | Child_Res#Flexible_Sche | 0.012 | 0.064 | 0.191 | 0.849 | -0.114 | 0.138 |
| | Paternity Leave | Childcare responsibilities | -0.023 | 0.032 | -0.712 | 0.477 | -0.085 | 0.040 |
| | | Paternity Leave | 0.304 | 0.178 | 1.712 | 0.088 | -0.045 | 0.654 |
| | | Childcare Res#Paternity Leave | -0.081 | 0.079 | -1.026 | 0.306 | -0.236 | 0.074 |
| | Paused Tenure Clock | Childcare responsibilities | -0.040 | 0.032 | -1.239 | 0.216 | -0.103 | 0.023 |
| | | Paused Tenure Clock | 0.033 | 0.190 | 0.172 | 0.864 | -0.342 | 0.407 |
| | | Childcare Res#Paused Tenure | 0.034 | 0.085 | 0.393 | 0.695 | -0.134 | 0.201 |
| | Childcare Support | Childcare responsibilities | -0.185 | 0.081 | -2.290 | 0.023 | -0.345 | -0.026 |
| | | Childcare Support | 0.131 | 0.519 | 0.252 | 0.801 | -0.890 | 1.151 |
| | | Child_Res#Childcare Sup | 0.027 | 0.173 | 0.156 | 0.876 | -0.314 | 0.368 |
| | Flexible Schedule | Childcare responsibilities | -0.202 | 0.085 | -2.363 | 0.019 | -0.370 | -0.034 |
| | | Flexible Schedule | -0.039 | 0.401 | -0.097 | 0.923 | -0.827 | 0.749 |
| ARP | | Child_Res#Flexible_Sche | 0.067 | 0.152 | 0.440 | 0.661 | -0.232 | 0.365 |
| | Paternity Leave | Childcare responsibilities | -0.204 | 0.081 | -2.527 | 0.012 | -0.362 | -0.045 |
| | | Paternity Leave | -0.198 | 0.346 | -0.572 | 0.568 | -0.878 | 0.482 |
| | | Childcare Res#Paternity Leave | 0.136 | 0.131 | 1.038 | 0.300 | -0.122 | 0.395 |
| | Paused Tenure Clock | Childcare responsibilities | -0.258 | 0.070 | -3.709 | 0.000 | -0.395 | -0.121 |
| | | Paused Tenure Clock | -0.617 | 0.496 | -1.244 | 0.214 | -1.592 | 0.358 |

| | | | | | | | | | |
|---|---|---|---|---|---|---|---|---|---|
| | | | Childcare Res#Paused Tenure | 0.558 | 0.286 | 1.950 | 0.052 | -0.005 | 1.122 |
| | ARC | Childcare Support | Childcare responsibilities | 0.180 | 0.145 | 1.237 | 0.217 | -0.106 | 0.466 |
| | | | Childcare Support | 0.686 | 0.453 | 1.514 | 0.131 | -0.205 | 1.578 |
| | | | Child_Res#Childcare Sup | -0.338 | 0.190 | -1.780 | 0.076 | -0.710 | 0.035 |
| | | Flexible Schedule | Childcare responsibilities | 0.190 | 0.173 | 1.103 | 0.271 | -0.149 | 0.530 |
| | | | Flexible Schedule | 0.473 | 0.356 | 1.330 | 0.185 | -0.227 | 1.172 |
| | | | Child_Res#Flexible_Sche | -0.236 | 0.179 | -1.321 | 0.188 | -0.588 | 0.116 |
| | | Paternity Leave | Childcare responsibilities | 0.188 | 0.139 | 1.357 | 0.176 | -0.084 | 0.461 |
| | | | Paternity Leave | 1.163 | 0.507 | 2.295 | 0.022 | 0.166 | 2.160 |
| | | | Childcare Res#Paternity Leave | -0.427 | 0.213 | -2.003 | 0.046 | -0.845 | -0.008 |
| | | Paused Tenure Clock | Childcare responsibilities | 0.170 | 0.135 | 1.262 | 0.208 | -0.095 | 0.434 |
| | | | Paused Tenure Clock | 1.595 | 0.717 | 2.225 | 0.027 | 0.185 | 3.004 |
| | | | Childcare Res#Paused Tenure | -0.284 | 0.265 | -1.072 | 0.285 | -0.805 | 0.237 |
| | ARCo | Childcare Support | Childcare responsibilities | -0.091 | 0.034 | -2.650 | 0.008 | -0.159 | -0.024 |
| | | | Childcare Support | -0.037 | 0.175 | -0.210 | 0.834 | -0.382 | 0.308 |
| | | | Child_Res#Childcare Sup | 0.060 | 0.064 | 0.947 | 0.344 | -0.065 | 0.186 |
| | | Flexible Schedule | Childcare responsibilities | -0.092 | 0.036 | -2.552 | 0.011 | -0.162 | -0.021 |
| | | | Flexible Schedule | 0.008 | 0.160 | 0.048 | 0.961 | -0.306 | 0.322 |
| | | | Child_Res#Flexible_Sche | 0.037 | 0.063 | 0.596 | 0.551 | -0.086 | 0.161 |
| | | Paternity Leave | Childcare responsibilities | -0.081 | 0.032 | -2.542 | 0.011 | -0.144 | -0.018 |
| | | | Paternity Leave | 0.106 | 0.198 | 0.536 | 0.592 | -0.282 | 0.494 |
| | | | Childcare Res#Paternity Leave | 0.007 | 0.071 | 0.098 | 0.922 | -0.133 | 0.148 |
| | | Paused Tenure Clock | Childcare responsibilities | -0.104 | 0.029 | -3.548 | 0.000 | -0.162 | -0.047 |
| | | | Paused Tenure Clock | -0.216 | 0.226 | -0.958 | 0.339 | -0.661 | 0.228 |
| | | | Childcare Res#Paused Tenure | 0.173 | 0.117 | 1.477 | 0.141 | -0.057 | 0.403 |
| Women | Research Satisfaction | Childcare Support | Childcare responsibilities | -0.107 | 0.035 | -3.037 | 0.003 | -0.176 | -0.038 |
| | | | Childcare Support | 0.230 | 0.179 | 1.284 | 0.200 | -0.122 | 0.581 |
| | | | Child_Res#Childcare Sup | -0.045 | 0.077 | -0.592 | 0.554 | -0.197 | 0.106 |
| | | Flexible Schedule | Childcare responsibilities | -0.150 | 0.040 | -3.739 | 0.000 | -0.228 | -0.071 |
| | | | Flexible Schedule | -0.112 | 0.153 | -0.734 | 0.463 | -0.412 | 0.188 |
| | | | Child_Res#Flexible_Sche | 0.088 | 0.067 | 1.327 | 0.185 | -0.043 | 0.219 |
| | | Paternity Leave | Childcare responsibilities | -0.057 | 0.050 | -1.132 | 0.258 | -0.156 | 0.042 |
| | | | Paternity Leave | 0.324 | 0.171 | 1.899 | 0.058 | -0.011 | 0.659 |
| | | | Childcare Res#Paternity Leave | -0.086 | 0.068 | -1.268 | 0.206 | -0.219 | 0.047 |

| | | | | | | | | |
|---|---|---|---|---|---|---|---|---|
| | Paused Tenure Clock | Childcare responsibilities | -0.095 | 0.035 | -2.689 | 0.007 | -0.164 | -0.026 |
| | | Paused Tenure Clock | 0.191 | 0.160 | 1.194 | 0.233 | -0.123 | 0.505 |
| | | Childcare Res#Paused Tenure | -0.080 | 0.073 | -1.092 | 0.276 | -0.224 | 0.064 |
| Career Satisfaction | Childcare Support | Childcare responsibilities | -0.163 | 0.032 | -5.072 | 0.000 | -0.226 | -0.100 |
| | | Childcare Support | 0.143 | 0.156 | 0.911 | 0.363 | -0.165 | 0.450 |
| | | Child_Res#Childcare Sup | -0.025 | 0.071 | -0.355 | 0.722 | -0.164 | 0.114 |
| | Flexible Schedule | Childcare responsibilities | -0.196 | 0.037 | -5.250 | 0.000 | -0.269 | -0.122 |
| | | Flexible Schedule | -0.040 | 0.137 | -0.292 | 0.771 | -0.309 | 0.229 |
| | | Child_Res#Flexible_Sche | 0.079 | 0.063 | 1.252 | 0.211 | -0.045 | 0.204 |
| | Paternity Leave | Childcare responsibilities | -0.150 | 0.045 | -3.353 | 0.001 | -0.238 | -0.062 |
| | | Paternity Leave | 0.195 | 0.155 | 1.262 | 0.208 | -0.109 | 0.500 |
| | | Childcare Res#Paternity Leave | -0.018 | 0.059 | -0.306 | 0.760 | -0.133 | 0.097 |
| Community Recognition | Paused Tenure Clock | Childcare responsibilities | -0.141 | 0.030 | -4.672 | 0.000 | -0.201 | -0.082 |
| | | Paused Tenure Clock | 0.275 | 0.139 | 1.982 | 0.048 | 0.002 | 0.547 |
| | | Childcare Res#Paused Tenure | -0.091 | 0.064 | -1.424 | 0.155 | -0.218 | 0.035 |
| | Childcare Support | Childcare responsibilities | -0.113 | 0.032 | -3.581 | 0.000 | -0.175 | -0.051 |
| | | Childcare Support | 0.112 | 0.138 | 0.811 | 0.418 | -0.159 | 0.383 |
| | | Child_Res#Childcare Sup | 0.010 | 0.061 | 0.170 | 0.865 | -0.110 | 0.131 |
| | Flexible Schedule | Childcare responsibilities | -0.112 | 0.037 | -3.050 | 0.002 | -0.185 | -0.040 |
| | | Flexible Schedule | 0.094 | 0.121 | 0.777 | 0.438 | -0.144 | 0.332 |
| | | Child_Res#Flexible_Sche | 0.014 | 0.055 | 0.257 | 0.798 | -0.094 | 0.122 |
| | Paternity Leave | Childcare responsibilities | -0.088 | 0.047 | -1.876 | 0.061 | -0.179 | 0.004 |
| | | Paternity Leave | 0.132 | 0.150 | 0.879 | 0.380 | -0.163 | 0.428 |
| | | Childcare Res#Paternity Leave | -0.034 | 0.062 | -0.548 | 0.584 | -0.157 | 0.088 |
| ARP | Paused Tenure Clock | Childcare responsibilities | -0.103 | 0.033 | -3.135 | 0.002 | -0.167 | -0.038 |
| | | Paused Tenure Clock | 0.074 | 0.145 | 0.511 | 0.610 | -0.211 | 0.360 |
| | | Childcare Res#Paused Tenure | -0.031 | 0.066 | -0.474 | 0.636 | -0.162 | 0.099 |
| | Childcare Support | Childcare responsibilities | -0.100 | 0.034 | -2.974 | 0.003 | -0.167 | -0.034 |
| | | Childcare Support | -0.016 | 0.172 | -0.093 | 0.926 | -0.354 | 0.322 |
| | | Child_Res#Childcare Sup | 0.037 | 0.071 | 0.521 | 0.603 | -0.103 | 0.177 |
| | Flexible Schedule | Childcare responsibilities | -0.055 | 0.039 | -1.429 | 0.154 | -0.132 | 0.021 |
| | | Flexible Schedule | 0.248 | 0.147 | 1.686 | 0.093 | -0.041 | 0.538 |
| | | Child_Res#Flexible_Sche | -0.086 | 0.060 | -1.430 | 0.154 | -0.205 | 0.032 |
| | Paternity Leave | Childcare responsibilities | -0.018 | 0.049 | -0.369 | 0.713 | -0.115 | 0.078 |
| | | Paternity Leave | 0.393 | 0.160 | 2.452 | 0.015 | 0.078 | 0.708 |
| | | Childcare Res#Paternity Leave | -0.105 | 0.065 | -1.609 | 0.108 | -0.234 | 0.023 |
| | Paused Tenure Clock | Childcare responsibilities | -0.063 | 0.037 | -1.716 | 0.087 | -0.135 | 0.009 |
| | | Paused Tenure Clock | 0.456 | 0.209 | 2.182 | 0.030 | 0.045 | 0.867 |
| | | Childcare Res#Paused Tenure | -0.060 | 0.077 | -0.784 | 0.433 | -0.211 | 0.091 |
| ARC | Childcare Support | Childcare responsibilities | -0.140 | 0.056 | -2.487 | 0.013 | -0.250 | -0.029 |
| | | Childcare Support | -0.263 | 0.233 | -1.132 | 0.258 | -0.720 | 0.194 |
| | | Child_Res#Childcare Sup | 0.144 | 0.099 | 1.453 | 0.147 | -0.051 | 0.340 |
| | Flexible Schedule | Childcare responsibilities | -0.171 | 0.058 | -2.931 | 0.004 | -0.286 | -0.056 |
| | | Flexible Schedule | -0.427 | 0.234 | -1.829 | 0.068 | -0.887 | 0.032 |
| | | Child_Res#Flexible_Sche | 0.147 | 0.102 | 1.437 | 0.152 | -0.054 | 0.348 |
| | Paternity Leave | Childcare responsibilities | -0.123 | 0.084 | -1.463 | 0.144 | -0.289 | 0.042 |
| | | Paternity Leave | 0.098 | 0.273 | 0.358 | 0.721 | -0.439 | 0.634 |
| | | Childcare Res#Paternity Leave | 0.036 | 0.100 | 0.363 | 0.717 | -0.160 | 0.233 |
| | Paused Tenure Clock | Childcare responsibilities | -0.114 | 0.056 | -2.018 | 0.044 | -0.224 | -0.003 |
| | | Paused Tenure Clock | 0.204 | 0.271 | 0.754 | 0.451 | -0.328 | 0.736 |
| | | Childcare Res#Paused Tenure | 0.074 | 0.103 | 0.716 | 0.474 | -0.129 | 0.276 |
| ARCo | Childcare Support | Childcare responsibilities | -0.031 | 0.028 | -1.138 | 0.256 | -0.086 | 0.023 |
| | | Childcare Support | 0.020 | 0.113 | 0.178 | 0.859 | -0.201 | 0.241 |
| | | Child_Res#Childcare Sup | 0.001 | 0.049 | 0.019 | 0.985 | -0.096 | 0.098 |
| | Flexible Schedule | Childcare responsibilities | -0.005 | 0.031 | -0.146 | 0.884 | -0.065 | 0.056 |
| | | Flexible Schedule | 0.179 | 0.100 | 1.792 | 0.074 | -0.017 | 0.375 |
| | | Child_Res#Flexible_Sche | -0.063 | 0.043 | -1.468 | 0.143 | -0.147 | 0.021 |
| | Paternity Leave | Childcare responsibilities | 0.019 | 0.030 | 0.652 | 0.515 | -0.039 | 0.078 |
| | | Paternity Leave | 0.307 | 0.102 | 3.005 | 0.003 | 0.106 | 0.507 |
| | | Childcare Res#Paternity Leave | -0.069 | 0.047 | -1.473 | 0.141 | -0.161 | 0.023 |
| | Paused Tenure Clock | Childcare responsibilities | -0.010 | 0.032 | -0.301 | 0.764 | -0.074 | 0.054 |
| | | Paused Tenure Clock | 0.286 | 0.135 | 2.126 | 0.034 | 0.022 | 0.551 |
| | | Childcare Res#Paused Tenure | -0.055 | 0.057 | -0.959 | 0.338 | -0.167 | 0.058 |

**Table S8.3. Change in subjective and objective career achievements with the change in childcare responsibilities, the parental support policies provided by institutions, and their**

**interaction items, by gender, if partner's job type is Self-employed, student, or out of work and looking for work**

| | Achievement | Support | Variable | Coefficient | Standard error | T-value | P-value | CI_low | Ci_high |
|---|---|---|---|---|---|---|---|---|---|
| Men | Research Satisfaction | Childcare Support | Childcare responsibilities | -0.128 | 0.095 | -1.343 | 0.181 | -0.315 | 0.060 |
| | | | Childcare Support | 0.255 | 0.466 | 0.548 | 0.585 | -0.665 | 1.175 |
| | | | Child_Res#Childcare Sup | -0.271 | 0.228 | -1.189 | 0.236 | -0.721 | 0.179 |
| | | Flexible Schedule | Childcare responsibilities | -0.244 | 0.101 | -2.422 | 0.017 | -0.443 | -0.045 |
| | | | Flexible Schedule | -0.567 | 0.370 | -1.532 | 0.127 | -1.299 | 0.164 |
| | | | Child_Res#Flexible_Sche | 0.214 | 0.178 | 1.198 | 0.233 | -0.139 | 0.566 |
| | | Paternity Leave | Childcare responsibilities | -0.161 | 0.098 | -1.642 | 0.103 | -0.354 | 0.033 |
| | | | Paternity Leave | 0.241 | 0.425 | 0.567 | 0.571 | -0.599 | 1.081 |
| | | | Childcare Res#Paternity Leave | -0.075 | 0.194 | -0.387 | 0.699 | -0.459 | 0.309 |
| | | Paused Tenure Clock | Childcare responsibilities | -0.185 | 0.088 | -2.101 | 0.037 | -0.358 | -0.011 |
| | | | Paused Tenure Clock | -0.013 | 0.520 | -0.026 | 0.980 | -1.041 | 1.015 |
| | | | Childcare Res#Paused Tenure | 0.051 | 0.245 | 0.207 | 0.837 | -0.433 | 0.535 |
| | Career Satisfaction | Childcare Support | Childcare responsibilities | -0.123 | 0.086 | -1.434 | 0.153 | -0.294 | 0.047 |
| | | | Childcare Support | 0.302 | 0.443 | 0.683 | 0.496 | -0.572 | 1.177 |
| | | | Child_Res#Childcare Sup | -0.201 | 0.211 | -0.955 | 0.341 | -0.618 | 0.215 |
| | | Flexible Schedule | Childcare responsibilities | -0.216 | 0.092 | -2.352 | 0.020 | -0.397 | -0.035 |
| | | | Flexible Schedule | -0.360 | 0.327 | -1.101 | 0.273 | -1.005 | 0.286 |
| | | | Child_Res#Flexible_Sche | 0.185 | 0.161 | 1.149 | 0.252 | -0.133 | 0.504 |
| | | Paternity Leave | Childcare responsibilities | -0.192 | 0.094 | -2.053 | 0.042 | -0.377 | -0.007 |
| | | | Paternity Leave | 0.015 | 0.331 | 0.047 | 0.963 | -0.639 | 0.669 |
| | | | Childcare Res#Paternity Leave | 0.154 | 0.146 | 1.052 | 0.294 | -0.135 | 0.443 |
| | | Paused Tenure Clock | Childcare responsibilities | -0.157 | 0.080 | -1.973 | 0.050 | -0.314 | 0.000 |
| | | | Paused Tenure Clock | 0.250 | 0.494 | 0.506 | 0.614 | -0.725 | 1.225 |
| | | | Childcare Res#Paused Tenure | -0.047 | 0.213 | -0.219 | 0.827 | -0.467 | 0.374 |
| | Community Recognition | Childcare Support | Childcare responsibilities | -0.081 | 0.071 | -1.144 | 0.254 | -0.221 | 0.059 |
| | | | Childcare Support | 0.331 | 0.391 | 0.847 | 0.398 | -0.441 | 1.103 |
| | | | Child_Res#Childcare Sup | -0.302 | 0.202 | -1.492 | 0.138 | -0.702 | 0.098 |
| | | Flexible Schedule | Childcare responsibilities | -0.133 | 0.090 | -1.488 | 0.139 | -0.310 | 0.044 |
| | | | Flexible Schedule | 0.032 | 0.341 | 0.094 | 0.925 | -0.641 | 0.705 |
| | | | Child_Res#Flexible_Sche | -0.022 | 0.169 | -0.128 | 0.898 | -0.355 | 0.311 |

| | | | | | | | | |
|---|---|---|---|---|---|---|---|---|
| | Paternity Leave | Childcare responsibilities | -0.120 | 0.084 | -1.432 | 0.154 | -0.285 | 0.045 |
| | | Paternity Leave | 0.178 | 0.278 | 0.642 | 0.522 | -0.371 | 0.727 |
| | | Childcare Res#Paternity Leave | -0.086 | 0.143 | -0.604 | 0.546 | -0.368 | 0.196 |
| | Paused Tenure Clock | Childcare responsibilities | -0.125 | 0.077 | -1.618 | 0.108 | -0.277 | 0.027 |
| | | Paused Tenure Clock | 0.248 | 0.400 | 0.620 | 0.536 | -0.542 | 1.038 |
| | | Childcare Res#Paused Tenure | -0.123 | 0.225 | -0.548 | 0.585 | -0.568 | 0.321 |
| ARP | Childcare Support | Childcare responsibilities | -0.015 | 0.081 | -0.181 | 0.856 | -0.176 | 0.146 |
| | | Childcare Support | 0.915 | 0.666 | 1.374 | 0.172 | -0.402 | 2.232 |
| | | Child_Res#Childcare Sup | -0.293 | 0.211 | -1.391 | 0.166 | -0.711 | 0.124 |
| | Flexible Schedule | Childcare responsibilities | -0.066 | 0.093 | -0.705 | 0.482 | -0.251 | 0.119 |
| | | Flexible Schedule | 0.151 | 0.512 | 0.295 | 0.769 | -0.862 | 1.164 |
| | | Child_Res#Flexible_Sche | -0.028 | 0.169 | -0.168 | 0.867 | -0.362 | 0.305 |
| | Paternity Leave | Childcare responsibilities | -0.052 | 0.081 | -0.642 | 0.522 | -0.212 | 0.108 |
| | | Paternity Leave | 0.558 | 0.606 | 0.922 | 0.358 | -0.639 | 1.756 |
| | | Childcare Res#Paternity Leave | -0.080 | 0.193 | -0.415 | 0.679 | -0.462 | 0.302 |
| | Paused Tenure Clock | Childcare responsibilities | -0.105 | 0.100 | -1.050 | 0.296 | -0.302 | 0.092 |
| | | Paused Tenure Clock | -0.810 | 0.433 | -1.873 | 0.063 | -1.666 | 0.045 |
| | | Childcare Res#Paused Tenure | 0.208 | 0.172 | 1.205 | 0.230 | -0.133 | 0.548 |
| | Childcare Support | Childcare responsibilities | -0.054 | 0.095 | -0.573 | 0.567 | -0.241 | 0.133 |
| | | Childcare Support | 0.628 | 0.520 | 1.208 | 0.229 | -0.400 | 1.657 |
| | | Child_Res#Childcare Sup | -0.208 | 0.180 | -1.158 | 0.249 | -0.564 | 0.148 |
| ARC | Flexible Schedule | Childcare responsibilities | -0.165 | 0.097 | -1.695 | 0.092 | -0.357 | 0.027 |
| | | Flexible Schedule | -0.891 | 0.402 | -2.217 | 0.028 | -1.685 | -0.096 |
| | | Child_Res#Flexible_Sche | 0.212 | 0.130 | 1.625 | 0.106 | -0.046 | 0.469 |
| | Paternity Leave | Childcare responsibilities | -0.142 | 0.102 | -1.391 | 0.166 | -0.343 | 0.060 |
| | | Paternity Leave | -0.607 | 0.447 | -1.359 | 0.176 | -1.491 | 0.276 |
| | | Childcare Res#Paternity Leave | 0.180 | 0.150 | 1.197 | 0.233 | -0.117 | 0.476 |
| | Paused Tenure Clock | Childcare responsibilities | -0.128 | 0.086 | -1.477 | 0.142 | -0.298 | 0.043 |
| | | Paused Tenure Clock | -0.775 | 0.440 | -1.760 | 0.081 | -1.644 | 0.095 |
| | | Childcare Res#Paused Tenure | 0.218 | 0.168 | 1.299 | 0.196 | -0.114 | 0.549 |
| ARCo | Childcare Support | Childcare responsibilities | -0.040 | 0.048 | -0.846 | 0.399 | -0.134 | 0.054 |
| | | Childcare Support | 0.261 | 0.253 | 1.031 | 0.304 | -0.240 | 0.762 |
| | | Child_Res#Childcare Sup | -0.047 | 0.094 | -0.499 | 0.619 | -0.233 | 0.139 |
| | Flexible Schedule | Childcare responsibilities | -0.066 | 0.049 | -1.360 | 0.176 | -0.162 | 0.030 |

| | | | | | | | | | |
|---|---|---|---|---|---|---|---|---|---|
| | | | Flexible Schedule | 0.011 | 0.225 | 0.048 | 0.961 | -0.434 | 0.456 |
| | | | Child_Res#Flexible_Sche | 0.065 | 0.085 | 0.758 | 0.450 | -0.104 | 0.233 |
| | | Paternity Leave | Childcare responsibilities | -0.053 | 0.049 | -1.086 | 0.279 | -0.151 | 0.044 |
| | | | Paternity Leave | 0.048 | 0.250 | 0.192 | 0.848 | -0.446 | 0.542 |
| | | | Childcare Res#Paternity Leave | 0.018 | 0.080 | 0.220 | 0.826 | -0.140 | 0.175 |
| | | Paused Tenure Clock | Childcare responsibilities | -0.068 | 0.050 | -1.365 | 0.175 | -0.167 | 0.031 |
| | | | Paused Tenure Clock | -0.514 | 0.185 | -2.787 | 0.006 | -0.879 | -0.149 |
| | | | Childcare Res#Paused Tenure | 0.126 | 0.063 | 1.997 | 0.048 | 0.001 | 0.251 |
| | Research Satisfaction | Childcare Support | Childcare responsibilities | -0.084 | 0.076 | -1.107 | 0.270 | -0.234 | 0.066 |
| | | | Childcare Support | 0.556 | 0.503 | 1.106 | 0.270 | -0.436 | 1.548 |
| | | | Child_Res#Childcare Sup | -0.234 | 0.230 | -1.017 | 0.311 | -0.689 | 0.220 |
| | | Flexible Schedule | Childcare responsibilities | -0.125 | 0.099 | -1.269 | 0.206 | -0.320 | 0.070 |
| | | | Flexible Schedule | 0.081 | 0.388 | 0.209 | 0.835 | -0.684 | 0.846 |
| | | | Child_Res#Flexible_Sche | 0.015 | 0.146 | 0.105 | 0.917 | -0.273 | 0.304 |
| | | Paternity Leave | Childcare responsibilities | 0.004 | 0.119 | 0.030 | 0.976 | -0.230 | 0.237 |
| | | | Paternity Leave | 0.560 | 0.358 | 1.563 | 0.120 | -0.147 | 1.267 |
| | | | Childcare Res#Paternity Leave | -0.193 | 0.140 | -1.377 | 0.170 | -0.469 | 0.084 |
| | | Paused Tenure Clock | Childcare responsibilities | -0.094 | 0.080 | -1.182 | 0.239 | -0.251 | 0.063 |
| | | | Paused Tenure Clock | 0.501 | 0.411 | 1.217 | 0.225 | -0.311 | 1.313 |
| | | | Childcare Res#Paused Tenure | -0.049 | 0.158 | -0.309 | 0.757 | -0.362 | 0.264 |
| Women | Career Satisfaction | Childcare Support | Childcare responsibilities | -0.056 | 0.065 | -0.868 | 0.386 | -0.185 | 0.072 |
| | | | Childcare Support | 0.409 | 0.449 | 0.913 | 0.363 | -0.476 | 1.295 |
| | | | Child_Res#Childcare Sup | -0.174 | 0.166 | -1.049 | 0.296 | -0.503 | 0.154 |
| | | Flexible Schedule | Childcare responsibilities | -0.158 | 0.081 | -1.947 | 0.053 | -0.318 | 0.002 |
| | | | Flexible Schedule | -0.033 | 0.310 | -0.107 | 0.915 | -0.646 | 0.579 |
| | | | Child_Res#Flexible_Sche | 0.162 | 0.124 | 1.306 | 0.193 | -0.083 | 0.407 |
| | | Paternity Leave | Childcare responsibilities | -0.055 | 0.108 | -0.511 | 0.610 | -0.268 | 0.158 |
| | | | Paternity Leave | 0.249 | 0.339 | 0.736 | 0.463 | -0.420 | 0.918 |
| | | | Childcare Res#Paternity Leave | -0.037 | 0.134 | -0.274 | 0.784 | -0.301 | 0.227 |
| | | Paused Tenure Clock | Childcare responsibilities | -0.082 | 0.068 | -1.203 | 0.231 | -0.216 | 0.053 |
| | | | Paused Tenure Clock | 0.288 | 0.353 | 0.815 | 0.416 | -0.409 | 0.985 |
| | | | Childcare Res#Paused Tenure | 0.063 | 0.152 | 0.416 | 0.678 | -0.237 | 0.363 |
| | Community | Childcare Support | Childcare responsibilities | -0.035 | 0.070 | -0.504 | 0.615 | -0.173 | 0.102 |
| | | | Childcare Support | 0.385 | 0.415 | 0.928 | 0.355 | -0.434 | 1.205 |

| | | | | | | | | |
|---|---|---|---|---|---|---|---|---|
| Recognition | | Child_Res#Childcare Sup | -0.122 | 0.141 | -0.862 | 0.390 | -0.401 | 0.157 |
| | Flexible Schedule | Childcare responsibilities | -0.086 | 0.087 | -0.993 | 0.322 | -0.257 | 0.085 |
| | | Flexible Schedule | 0.007 | 0.306 | 0.022 | 0.983 | -0.598 | 0.612 |
| | | Child_Res#Flexible_Sche | 0.072 | 0.113 | 0.636 | 0.526 | -0.151 | 0.294 |
| | Paternity Leave | Childcare responsibilities | 0.120 | 0.112 | 1.064 | 0.289 | -0.102 | 0.342 |
| | | Paternity Leave | 1.015 | 0.343 | 2.954 | 0.004 | 0.337 | 1.693 |
| | | Childcare Res#Paternity Leave | -0.260 | 0.130 | -2.006 | 0.046 | -0.516 | -0.004 |
| | Paused Tenure Clock | Childcare responsibilities | -0.015 | 0.073 | -0.209 | 0.835 | -0.158 | 0.128 |
| | | Paused Tenure Clock | 0.621 | 0.332 | 1.872 | 0.063 | -0.034 | 1.277 |
| | | Childcare Res#Paused Tenure | -0.105 | 0.140 | -0.750 | 0.454 | -0.382 | 0.172 |
| ARP | Childcare Support | Childcare responsibilities | -0.194 | 0.058 | -3.317 | 0.001 | -0.310 | -0.079 |
| | | Childcare Support | -1.247 | 0.279 | -4.474 | 0.000 | -1.797 | -0.696 |
| | | Child_Res#Childcare Sup | 0.527 | 0.123 | 4.299 | 0.000 | 0.285 | 0.770 |
| | Flexible Schedule | Childcare responsibilities | -0.124 | 0.058 | -2.117 | 0.036 | -0.239 | -0.008 |
| | | Flexible Schedule | 0.244 | 0.260 | 0.936 | 0.351 | -0.270 | 0.758 |
| | | Child_Res#Flexible_Sche | 0.024 | 0.104 | 0.231 | 0.818 | -0.182 | 0.230 |
| | Paternity Leave | Childcare responsibilities | -0.043 | 0.092 | -0.472 | 0.638 | -0.225 | 0.138 |
| | | Paternity Leave | 0.472 | 0.318 | 1.487 | 0.139 | -0.155 | 1.100 |
| | | Childcare Res#Paternity Leave | -0.104 | 0.118 | -0.880 | 0.380 | -0.338 | 0.130 |
| | Paused Tenure Clock | Childcare responsibilities | -0.112 | 0.067 | -1.674 | 0.096 | -0.244 | 0.020 |
| | | Paused Tenure Clock | 0.144 | 0.294 | 0.489 | 0.625 | -0.436 | 0.724 |
| | | Childcare Res#Paused Tenure | 0.012 | 0.110 | 0.108 | 0.914 | -0.206 | 0.230 |
| ARC | Childcare Support | Childcare responsibilities | 0.058 | 0.179 | 0.323 | 0.747 | -0.296 | 0.411 |
| | | Childcare Support | 0.026 | 0.755 | 0.034 | 0.973 | -1.465 | 1.516 |
| | | Child_Res#Childcare Sup | 0.028 | 0.302 | 0.093 | 0.926 | -0.569 | 0.625 |
| | Flexible Schedule | Childcare responsibilities | 0.160 | 0.261 | 0.613 | 0.541 | -0.355 | 0.674 |
| | | Flexible Schedule | -0.084 | 0.606 | -0.138 | 0.890 | -1.281 | 1.114 |
| | | Child_Res#Flexible_Sche | -0.207 | 0.279 | -0.744 | 0.458 | -0.758 | 0.343 |
| | Paternity Leave | Childcare responsibilities | 0.213 | 0.314 | 0.679 | 0.498 | -0.407 | 0.834 |
| | | Paternity Leave | 0.241 | 0.854 | 0.282 | 0.778 | -1.446 | 1.928 |
| | | Childcare Res#Paternity Leave | -0.272 | 0.338 | -0.802 | 0.424 | -0.940 | 0.397 |
| | Paused Tenure Clock | Childcare responsibilities | 0.129 | 0.193 | 0.667 | 0.506 | -0.253 | 0.511 |
| | | Paused Tenure Clock | 0.931 | 0.825 | 1.128 | 0.261 | -0.698 | 2.559 |
| | | Childcare Res#Paused Tenure | -0.247 | 0.263 | -0.939 | 0.349 | -0.767 | 0.272 |

| | | | | | | | | | |
|---|---|---|---|---|---|---|---|---|---|
| | ARCo | Childcare Support | Childcare responsibilities | -0.114 | 0.069 | -1.660 | 0.099 | -0.250 | 0.022 |
| | | | Childcare Support | -1.209 | 0.262 | -4.622 | 0.000 | -1.726 | -0.693 |
| | | | Child_Res#Childcare Sup | 0.474 | 0.108 | 4.389 | 0.000 | 0.261 | 0.687 |
| | | Flexible Schedule | Childcare responsibilities | -0.113 | 0.084 | -1.336 | 0.183 | -0.279 | 0.054 |
| | | | Flexible Schedule | -0.306 | 0.342 | -0.894 | 0.373 | -0.982 | 0.370 |
| | | | Child_Res#Flexible_Sche | 0.142 | 0.141 | 1.001 | 0.318 | -0.138 | 0.421 |
| | | Paternity Leave | Childcare responsibilities | 0.027 | 0.087 | 0.309 | 0.757 | -0.144 | 0.198 |
| | | | Paternity Leave | 0.331 | 0.233 | 1.420 | 0.158 | -0.130 | 0.792 |
| | | | Childcare Res#Paternity Leave | -0.111 | 0.106 | -1.047 | 0.297 | -0.319 | 0.098 |
| | | Paused Tenure Clock | Childcare responsibilities | -0.072 | 0.077 | -0.944 | 0.346 | -0.224 | 0.079 |
| | | | Paused Tenure Clock | -0.388 | 0.291 | -1.335 | 0.184 | -0.963 | 0.186 |
| | | | Childcare Res#Paused Tenure | 0.114 | 0.098 | 1.171 | 0.243 | -0.079 | 0.307 |

**Table S8.4. Change in subjective and objective career achievements with the change in childcare responsibilities, the parental support policies provided by institutions, and their interaction items, by gender, if partner's job type is Out of work but not looking for work or retired**

| | Achievement | Support | Variable | Coefficient | Standard error | T-value | P-value | CI_low | Ci_high |
|---|---|---|---|---|---|---|---|---|---|
| Men | Research Satisfaction | Childcare Support | Childcare responsibilities | -0.004 | 0.089 | -0.043 | 0.966 | -0.180 | 0.173 |
| | | | Childcare Support | 0.323 | 0.615 | 0.526 | 0.600 | -0.890 | 1.537 |
| | | | Child_Res#Childcare Sup | 0.088 | 0.334 | 0.262 | 0.794 | -0.572 | 0.747 |
| | | Flexible Schedule | Childcare responsibilities | -0.010 | 0.115 | -0.091 | 0.928 | -0.237 | 0.216 |
| | | | Flexible Schedule | 0.131 | 0.418 | 0.313 | 0.754 | -0.694 | 0.956 |
| | | | Child_Res#Flexible_Sche | 0.099 | 0.195 | 0.505 | 0.614 | -0.287 | 0.484 |
| | | Paternity Leave | Childcare responsibilities | -0.028 | 0.090 | -0.306 | 0.760 | -0.206 | 0.151 |
| | | | Paternity Leave | -0.087 | 0.407 | -0.213 | 0.832 | -0.891 | 0.718 |
| | | | Childcare Res#Paternity Leave | 0.445 | 0.179 | 2.484 | 0.014 | 0.091 | 0.798 |
| | | Paused Tenure Clock | Childcare responsibilities | 0.044 | 0.091 | 0.486 | 0.628 | -0.136 | 0.225 |
| | | | Paused Tenure Clock | 1.010 | 0.412 | 2.452 | 0.015 | 0.196 | 1.823 |
| | | | Childcare Res#Paused Tenure | -0.473 | 0.224 | -2.114 | 0.036 | -0.915 | -0.031 |
| | | Childcare Support | Childcare responsibilities | 0.025 | 0.071 | 0.346 | 0.730 | -0.116 | 0.166 |

| | | | | | | | | |
|---|---|---|---|---|---|---|---|---|
| Career Satisfaction | | Childcare Support | 0.422 | 0.541 | 0.779 | 0.437 | -0.647 | 1.491 |
| | | Child_Res#Childcare Sup | -0.097 | 0.313 | -0.310 | 0.757 | -0.716 | 0.522 |
| | Flexible Schedule | Childcare responsibilities | 0.029 | 0.101 | 0.286 | 0.775 | -0.170 | 0.227 |
| | | Flexible Schedule | 0.268 | 0.365 | 0.734 | 0.464 | -0.453 | 0.990 |
| | | Child_Res#Flexible_Sche | -0.043 | 0.183 | -0.235 | 0.815 | -0.405 | 0.319 |
| | Paternity Leave | Childcare responsibilities | -0.022 | 0.078 | -0.279 | 0.780 | -0.176 | 0.132 |
| | | Paternity Leave | -0.419 | 0.407 | -1.028 | 0.305 | -1.224 | 0.386 |
| | | Childcare Res#Paternity Leave | 0.448 | 0.149 | 3.009 | 0.003 | 0.154 | 0.742 |
| | Paused Tenure Clock | Childcare responsibilities | 0.052 | 0.072 | 0.725 | 0.469 | -0.090 | 0.195 |
| | | Paused Tenure Clock | 0.901 | 0.423 | 2.128 | 0.035 | 0.065 | 1.737 |
| | | Childcare Res#Paused Tenure | -0.504 | 0.242 | -2.086 | 0.039 | -0.982 | -0.027 |
| Community Recognition | Childcare Support | Childcare responsibilities | 0.001 | 0.066 | 0.010 | 0.992 | -0.130 | 0.131 |
| | | Childcare Support | -0.416 | 0.476 | -0.874 | 0.383 | -1.357 | 0.524 |
| | | Child_Res#Childcare Sup | 0.259 | 0.158 | 1.637 | 0.104 | -0.053 | 0.572 |
| | Flexible Schedule | Childcare responsibilities | 0.065 | 0.074 | 0.873 | 0.384 | -0.082 | 0.212 |
| | | Flexible Schedule | 0.412 | 0.279 | 1.477 | 0.142 | -0.139 | 0.962 |
| | | Child_Res#Flexible_Sche | -0.140 | 0.147 | -0.947 | 0.345 | -0.431 | 0.152 |
| | Paternity Leave | Childcare responsibilities | 0.014 | 0.064 | 0.215 | 0.830 | -0.113 | 0.141 |
| | | Paternity Leave | -0.053 | 0.442 | -0.119 | 0.905 | -0.926 | 0.821 |
| | | Childcare Res#Paternity Leave | 0.196 | 0.167 | 1.175 | 0.242 | -0.133 | 0.525 |
| | Paused Tenure Clock | Childcare responsibilities | 0.026 | 0.064 | 0.397 | 0.692 | -0.101 | 0.153 |
| | | Paused Tenure Clock | -0.056 | 0.381 | -0.147 | 0.883 | -0.808 | 0.696 |
| | | Childcare Res#Paused Tenure | 0.048 | 0.205 | 0.236 | 0.813 | -0.356 | 0.452 |
| ARP | Childcare Support | Childcare responsibilities | -0.090 | 0.087 | -1.029 | 0.305 | -0.262 | 0.083 |
| | | Childcare Support | -0.344 | 0.352 | -0.978 | 0.330 | -1.040 | 0.351 |
| | | Child_Res#Childcare Sup | 0.135 | 0.144 | 0.939 | 0.349 | -0.149 | 0.420 |
| | Flexible Schedule | Childcare responsibilities | -0.082 | 0.089 | -0.927 | 0.356 | -0.257 | 0.093 |
| | | Flexible Schedule | -0.213 | 0.395 | -0.540 | 0.590 | -0.994 | 0.567 |
| | | Child_Res#Flexible_Sche | 0.020 | 0.175 | 0.112 | 0.911 | -0.326 | 0.365 |
| | Paternity Leave | Childcare responsibilities | -0.087 | 0.084 | -1.045 | 0.298 | -0.253 | 0.078 |
| | | Paternity Leave | -0.510 | 0.316 | -1.613 | 0.109 | -1.136 | 0.115 |
| | | Childcare Res#Paternity Leave | 0.119 | 0.153 | 0.780 | 0.437 | -0.183 | 0.422 |
| | Paused Tenure Clock | Childcare responsibilities | -0.090 | 0.083 | -1.090 | 0.278 | -0.254 | 0.074 |
| | | Paused Tenure Clock | -0.238 | 0.315 | -0.757 | 0.450 | -0.861 | 0.384 |

| | | | | | | | | | |
|---|---|---|---|---|---|---|---|---|---|
| | | | Childcare Res#Paused Tenure | 0.209 | 0.201 | 1.038 | 0.301 | -0.189 | 0.606 |
| | ARC | Childcare Support | Childcare responsibilities | -0.264 | 0.121 | -2.186 | 0.030 | -0.502 | -0.025 |
| | | | Childcare Support | -1.003 | 0.757 | -1.324 | 0.188 | -2.499 | 0.494 |
| | | | Child_Res#Childcare Sup | 0.295 | 0.316 | 0.932 | 0.353 | -0.330 | 0.920 |
| | | Flexible Schedule | Childcare responsibilities | -0.191 | 0.138 | -1.387 | 0.168 | -0.464 | 0.081 |
| | | | Flexible Schedule | 0.067 | 0.443 | 0.152 | 0.879 | -0.809 | 0.944 |
| | | | Child_Res#Flexible_Sche | -0.198 | 0.192 | -1.027 | 0.306 | -0.578 | 0.183 |
| | | Paternity Leave | Childcare responsibilities | -0.211 | 0.122 | -1.722 | 0.087 | -0.452 | 0.031 |
| | | | Paternity Leave | 0.431 | 0.470 | 0.916 | 0.361 | -0.499 | 1.360 |
| | | | Childcare Res#Paternity Leave | -0.205 | 0.232 | -0.885 | 0.378 | -0.663 | 0.253 |
| | | Paused Tenure Clock | Childcare responsibilities | -0.203 | 0.096 | -2.103 | 0.037 | -0.393 | -0.012 |
| | | | Paused Tenure Clock | 1.246 | 1.230 | 1.013 | 0.313 | -1.186 | 3.678 |
| | | | Childcare Res#Paused Tenure | -0.267 | 0.567 | -0.471 | 0.639 | -1.388 | 0.854 |
| | ARCo | Childcare Support | Childcare responsibilities | -0.076 | 0.043 | -1.776 | 0.078 | -0.161 | 0.009 |
| | | | Childcare Support | -0.143 | 0.226 | -0.633 | 0.528 | -0.589 | 0.304 |
| | | | Child_Res#Childcare Sup | 0.088 | 0.100 | 0.873 | 0.384 | -0.111 | 0.286 |
| | | Flexible Schedule | Childcare responsibilities | -0.039 | 0.043 | -0.896 | 0.372 | -0.124 | 0.047 |
| | | | Flexible Schedule | 0.326 | 0.308 | 1.060 | 0.291 | -0.282 | 0.935 |
| | | | Child_Res#Flexible_Sche | -0.100 | 0.116 | -0.862 | 0.390 | -0.329 | 0.129 |
| | | Paternity Leave | Childcare responsibilities | -0.075 | 0.043 | -1.721 | 0.087 | -0.161 | 0.011 |
| | | | Paternity Leave | 0.008 | 0.267 | 0.029 | 0.977 | -0.519 | 0.535 |
| | | | Childcare Res#Paternity Leave | 0.101 | 0.130 | 0.780 | 0.437 | -0.156 | 0.358 |
| | | Paused Tenure Clock | Childcare responsibilities | -0.078 | 0.040 | -1.922 | 0.057 | -0.158 | 0.002 |
| | | | Paused Tenure Clock | -0.173 | 0.220 | -0.784 | 0.435 | -0.608 | 0.263 |
| | | | Childcare Res#Paused Tenure | 0.162 | 0.109 | 1.489 | 0.139 | -0.053 | 0.378 |
| Women | Research Satisfaction | Childcare Support | Childcare responsibilities | -0.153 | 0.175 | -0.876 | 0.384 | -0.502 | 0.196 |
| | | | Childcare Support | -1.190 | 1.136 | -1.048 | 0.298 | -3.455 | 1.075 |
| | | | Child_Res#Childcare Sup | 0.705 | 0.691 | 1.020 | 0.311 | -0.672 | 2.081 |
| | | Flexible Schedule | Childcare responsibilities | -0.379 | 0.237 | -1.601 | 0.114 | -0.851 | 0.093 |
| | | | Flexible Schedule | -1.156 | 0.658 | -1.757 | 0.083 | -2.469 | 0.156 |
| | | | Child_Res#Flexible_Sche | 0.638 | 0.318 | 2.006 | 0.049 | 0.004 | 1.273 |
| | | Paternity Leave | Childcare responsibilities | -0.222 | 0.224 | -0.991 | 0.325 | -0.669 | 0.225 |
| | | | Paternity Leave | -0.950 | 0.806 | -1.179 | 0.242 | -2.558 | 0.657 |
| | | | Childcare Res#Paternity Leave | 0.068 | 0.416 | 0.164 | 0.870 | -0.762 | 0.899 |

| | | | | | | | | |
|---|---|---|---|---|---|---|---|---|
| | Paused Tenure Clock | Childcare responsibilities | -0.089 | 0.177 | -0.503 | 0.616 | -0.442 | 0.264 |
| | | Paused Tenure Clock | 0.317 | 1.083 | 0.293 | 0.771 | -1.843 | 2.477 |
| | | Childcare Res#Paused Tenure | -0.586 | 0.629 | -0.932 | 0.355 | -1.841 | 0.668 |
| | Childcare Support | Childcare responsibilities | 0.084 | 0.155 | 0.543 | 0.589 | -0.225 | 0.394 |
| | | Childcare Support | 0.299 | 0.899 | 0.333 | 0.740 | -1.493 | 2.091 |
| | | Child_Res#Childcare Sup | 0.088 | 0.495 | 0.178 | 0.859 | -0.899 | 1.075 |
| Career Satisfaction | Flexible Schedule | Childcare responsibilities | -0.247 | 0.195 | -1.267 | 0.209 | -0.635 | 0.141 |
| | | Flexible Schedule | -1.329 | 0.682 | -1.949 | 0.055 | -2.689 | 0.031 |
| | | Child_Res#Flexible_Sche | 0.691 | 0.283 | 2.439 | 0.017 | 0.126 | 1.256 |
| | Paternity Leave | Childcare responsibilities | 0.007 | 0.222 | 0.031 | 0.975 | -0.435 | 0.449 |
| | | Paternity Leave | -0.368 | 0.681 | -0.540 | 0.591 | -1.725 | 0.990 |
| | | Childcare Res#Paternity Leave | 0.007 | 0.342 | 0.022 | 0.983 | -0.673 | 0.688 |
| | Paused Tenure Clock | Childcare responsibilities | 0.117 | 0.157 | 0.748 | 0.457 | -0.195 | 0.429 |
| | | Paused Tenure Clock | 1.101 | 0.828 | 1.330 | 0.188 | -0.549 | 2.750 |
| | | Childcare Res#Paused Tenure | -0.633 | 0.465 | -1.363 | 0.177 | -1.560 | 0.293 |
| | Childcare Support | Childcare responsibilities | 0.107 | 0.125 | 0.853 | 0.396 | -0.143 | 0.357 |
| | | Childcare Support | -2.297 | 0.939 | -2.446 | 0.017 | -4.169 | -0.425 |
| | | Child_Res#Childcare Sup | 1.110 | 0.530 | 2.094 | 0.040 | 0.053 | 2.167 |
| Community Recognition | Flexible Schedule | Childcare responsibilities | 0.166 | 0.134 | 1.238 | 0.220 | -0.102 | 0.435 |
| | | Flexible Schedule | -0.157 | 0.624 | -0.252 | 0.802 | -1.400 | 1.086 |
| | | Child_Res#Flexible_Sche | 0.125 | 0.273 | 0.456 | 0.650 | -0.420 | 0.670 |
| | Paternity Leave | Childcare responsibilities | 0.288 | 0.188 | 1.537 | 0.129 | -0.086 | 0.662 |
| | | Paternity Leave | 0.015 | 0.698 | 0.022 | 0.982 | -1.376 | 1.407 |
| | | Childcare Res#Paternity Leave | -0.247 | 0.273 | -0.904 | 0.369 | -0.792 | 0.298 |
| | Paused Tenure Clock | Childcare responsibilities | 0.232 | 0.144 | 1.611 | 0.112 | -0.055 | 0.520 |
| | | Paused Tenure Clock | 0.206 | 0.839 | 0.246 | 0.806 | -1.466 | 1.879 |
| | | Childcare Res#Paused Tenure | -0.378 | 0.377 | -1.001 | 0.320 | -1.129 | 0.374 |
| | Childcare Support | Childcare responsibilities | -0.236 | 0.223 | -1.058 | 0.294 | -0.682 | 0.209 |
| | | Childcare Support | 2.011 | 0.985 | 2.041 | 0.045 | 0.045 | 3.977 |
| | | Child_Res#Childcare Sup | -1.189 | 0.467 | -2.547 | 0.013 | -2.120 | -0.257 |
| ARP | Flexible Schedule | Childcare responsibilities | -0.472 | 0.258 | -1.829 | 0.072 | -0.987 | 0.043 |
| | | Flexible Schedule | -1.105 | 0.836 | -1.322 | 0.191 | -2.773 | 0.563 |
| | | Child_Res#Flexible_Sche | 0.497 | 0.374 | 1.328 | 0.189 | -0.250 | 1.245 |
| | Paternity Leave | Childcare responsibilities | -0.541 | 0.398 | -1.359 | 0.179 | -1.335 | 0.254 |
| | | Paternity Leave | -1.581 | 1.140 | -1.386 | 0.170 | -3.858 | 0.695 |
| | | Childcare Res#Paternity Leave | 0.396 | 0.492 | 0.805 | 0.424 | -0.586 | 1.379 |
| | Paused Tenure Clock | Childcare responsibilities | -0.284 | 0.215 | -1.324 | 0.190 | -0.713 | 0.144 |
| | | Paused Tenure Clock | -0.020 | 0.708 | -0.028 | 0.977 | -1.432 | 1.392 |
| | | Childcare Res#Paused Tenure | -0.233 | 0.343 | -0.679 | 0.499 | -0.917 | 0.451 |
| | Childcare Support | Childcare responsibilities | -0.060 | 0.340 | -0.175 | 0.862 | -0.739 | 0.620 |
| | | Childcare Support | 0.983 | 1.219 | 0.806 | 0.423 | -1.451 | 3.417 |
| | | Child_Res#Childcare Sup | -0.595 | 0.577 | -1.031 | 0.306 | -1.746 | 0.556 |
| | Flexible Schedule | Childcare responsibilities | -0.043 | 0.460 | -0.094 | 0.925 | -0.961 | 0.874 |
| | | Flexible Schedule | 0.106 | 0.690 | 0.154 | 0.878 | -1.270 | 1.483 |
| ARC | | Child_Res#Flexible_Sche | -0.118 | 0.392 | -0.301 | 0.765 | -0.900 | 0.664 |
| | Paternity Leave | Childcare responsibilities | -0.188 | 0.365 | -0.515 | 0.609 | -0.916 | 0.540 |
| | | Paternity Leave | -0.907 | 0.724 | -1.253 | 0.214 | -2.352 | 0.538 |
| | | Childcare Res#Paternity Leave | 0.075 | 0.279 | 0.269 | 0.789 | -0.482 | 0.632 |
| | Paused Tenure Clock | Childcare responsibilities | -0.134 | 0.340 | -0.394 | 0.695 | -0.814 | 0.545 |
| | | Paused Tenure Clock | -0.725 | 0.882 | -0.822 | 0.414 | -2.486 | 1.036 |
| | | Childcare Res#Paused Tenure | 0.580 | 0.412 | 1.406 | 0.164 | -0.243 | 1.402 |
| | Childcare Support | Childcare responsibilities | -0.054 | 0.101 | -0.530 | 0.598 | -0.255 | 0.148 |
| | | Childcare Support | 0.826 | 0.425 | 1.942 | 0.056 | -0.023 | 1.675 |
| | | Child_Res#Childcare Sup | -0.437 | 0.274 | -1.596 | 0.115 | -0.983 | 0.109 |
| | Flexible Schedule | Childcare responsibilities | -0.153 | 0.109 | -1.409 | 0.164 | -0.369 | 0.064 |
| | | Flexible Schedule | -0.165 | 0.466 | -0.355 | 0.723 | -1.095 | 0.764 |
| ARCo | | Child_Res#Flexible_Sche | 0.202 | 0.201 | 1.004 | 0.319 | -0.199 | 0.603 |
| | Paternity Leave | Childcare responsibilities | -0.059 | 0.154 | -0.379 | 0.706 | -0.367 | 0.250 |
| | | Paternity Leave | -0.011 | 0.466 | -0.023 | 0.982 | -0.940 | 0.919 |
| | | Childcare Res#Paternity Leave | -0.087 | 0.212 | -0.410 | 0.683 | -0.511 | 0.337 |
| | Paused Tenure Clock | Childcare responsibilities | -0.080 | 0.100 | -0.794 | 0.430 | -0.280 | 0.121 |
| | | Paused Tenure Clock | -0.099 | 0.370 | -0.269 | 0.789 | -0.838 | 0.639 |
| | | Childcare Res#Paused Tenure | -0.157 | 0.182 | -0.864 | 0.391 | -0.521 | 0.206 |

**Table S8.5. Change in subjective and objective career achievements with the change in childcare responsibilities, the parental support policies provided by institutions, and their interaction items, by gender, if partner's job type is Research-oriented**

| | Achieve ment | Support | Variable | Coeffic ient | Standard error | T-value | P-value | CI_low | Ci_high |
|---|---|---|---|---|---|---|---|---|---|
| Men | Research Satisfact ion | Childcare Support | Childcare responsibilities | -0.047 | 0.055 | -0.847 | 0.398 | -0.155 | 0.062 |
| | | | Childcare Support | 0.150 | 0.324 | 0.463 | 0.644 | -0.489 | 0.789 |
| | | | Child_Res#Childcare Sup | -0.104 | 0.159 | -0.658 | 0.511 | -0.417 | 0.208 |
| | | Flexible Schedule | Childcare responsibilities | -0.044 | 0.066 | -0.674 | 0.501 | -0.174 | 0.085 |
| | | | Flexible Schedule | 0.220 | 0.272 | 0.808 | 0.420 | -0.316 | 0.756 |
| | | | Child_Res#Flexible_S che | -0.069 | 0.120 | -0.575 | 0.566 | -0.305 | 0.167 |
| | | Paternity Leave | Childcare responsibilities | 0.002 | 0.058 | 0.036 | 0.972 | -0.111 | 0.115 |
| | | | Paternity Leave | 0.570 | 0.300 | 1.901 | 0.059 | -0.021 | 1.161 |
| | | | Childcare Res#Paternity Leave | -0.321 | 0.130 | -2.478 | 0.014 | -0.577 | -0.066 |
| | | Paused Tenure Clock | Childcare responsibilities | -0.052 | 0.056 | -0.934 | 0.351 | -0.162 | 0.058 |
| | | | Paused Tenure Clock | 0.243 | 0.318 | 0.765 | 0.445 | -0.383 | 0.869 |
| | | | Childcare Res#Paused Tenure | -0.097 | 0.139 | -0.701 | 0.484 | -0.371 | 0.176 |
| | Career Satisfact ion | Childcare Support | Childcare responsibilities | -0.101 | 0.056 | -1.812 | 0.071 | -0.210 | 0.009 |
| | | | Childcare Support | -0.155 | 0.270 | -0.573 | 0.567 | -0.686 | 0.377 |
| | | | Child_Res#Childcare Sup | 0.092 | 0.117 | 0.790 | 0.430 | -0.138 | 0.322 |
| | | Flexible Schedule | Childcare responsibilities | -0.067 | 0.062 | -1.070 | 0.286 | -0.190 | 0.056 |
| | | | Flexible Schedule | 0.257 | 0.244 | 1.052 | 0.294 | -0.224 | 0.738 |
| | | | Child_Res#Flexible_S che | -0.035 | 0.112 | -0.310 | 0.757 | -0.255 | 0.186 |
| | | Paternity Leave | Childcare responsibilities | -0.054 | 0.057 | -0.943 | 0.347 | -0.166 | 0.058 |
| | | | Paternity Leave | 0.245 | 0.274 | 0.895 | 0.372 | -0.295 | 0.786 |
| | | | Childcare Res#Paternity Leave | -0.132 | 0.122 | -1.076 | 0.283 | -0.373 | 0.109 |
| | | Paused Tenure Clock | Childcare responsibilities | -0.070 | 0.053 | -1.310 | 0.191 | -0.175 | 0.035 |
| | | | Paused Tenure Clock | 0.160 | 0.295 | 0.543 | 0.588 | -0.420 | 0.740 |
| | | | Childcare Res#Paused Tenure | -0.078 | 0.125 | -0.627 | 0.531 | -0.323 | 0.167 |
| | Commu nity Recognit ion | Childcare Support | Childcare responsibilities | -0.062 | 0.046 | -1.324 | 0.187 | -0.153 | 0.030 |
| | | | Childcare Support | -0.099 | 0.225 | -0.441 | 0.659 | -0.541 | 0.343 |
| | | | Child_Res#Childcare Sup | 0.020 | 0.092 | 0.211 | 0.833 | -0.162 | 0.201 |
| | | Flexible Schedule | Childcare responsibilities | 0.011 | 0.045 | 0.240 | 0.810 | -0.078 | 0.100 |

| | | | | | | | | |
|---|---|---|---|---|---|---|---|---|
| | | Flexible Schedule | 0.522 | 0.206 | 2.531 | 0.012 | 0.116 | 0.927 |
| | | Child_Res#Flexible_Sche | -0.216 | 0.096 | -2.243 | 0.026 | -0.405 | -0.026 |
| | Paternity Leave | Childcare responsibilities | -0.022 | 0.043 | -0.504 | 0.615 | -0.106 | 0.063 |
| | | Paternity Leave | 0.339 | 0.243 | 1.392 | 0.165 | -0.141 | 0.818 |
| | | Childcare Res#Paternity Leave | -0.163 | 0.109 | -1.492 | 0.137 | -0.378 | 0.052 |
| | Paused Tenure Clock | Childcare responsibilities | -0.054 | 0.045 | -1.193 | 0.234 | -0.142 | 0.035 |
| | | Paused Tenure Clock | 0.061 | 0.250 | 0.245 | 0.806 | -0.431 | 0.553 |
| | | Childcare Res#Paused Tenure | -0.018 | 0.121 | -0.145 | 0.885 | -0.257 | 0.221 |
| ARP | Childcare Support | Childcare responsibilities | -0.237 | 0.075 | -3.147 | 0.002 | -0.385 | -0.089 |
| | | Childcare Support | 0.500 | 0.911 | 0.549 | 0.583 | -1.294 | 2.294 |
| | | Child_Res#Childcare Sup | -0.071 | 0.286 | -0.250 | 0.803 | -0.634 | 0.491 |
| | Flexible Schedule | Childcare responsibilities | -0.271 | 0.114 | -2.385 | 0.018 | -0.495 | -0.047 |
| | | Flexible Schedule | 0.133 | 0.557 | 0.239 | 0.812 | -0.965 | 1.231 |
| | | Child_Res#Flexible_Sche | 0.060 | 0.200 | 0.301 | 0.764 | -0.334 | 0.455 |
| | Paternity Leave | Childcare responsibilities | -0.303 | 0.098 | -3.095 | 0.002 | -0.496 | -0.110 |
| | | Paternity Leave | -0.494 | 0.552 | -0.895 | 0.372 | -1.582 | 0.594 |
| | | Childcare Res#Paternity Leave | 0.215 | 0.198 | 1.087 | 0.278 | -0.175 | 0.605 |
| | Paused Tenure Clock | Childcare responsibilities | -0.304 | 0.096 | -3.154 | 0.002 | -0.494 | -0.114 |
| | | Paused Tenure Clock | -0.455 | 0.503 | -0.904 | 0.367 | -1.447 | 0.537 |
| | | Childcare Res#Paused Tenure | 0.290 | 0.178 | 1.631 | 0.104 | -0.060 | 0.641 |
| ARC | Childcare Support | Childcare responsibilities | 0.421 | 0.302 | 1.394 | 0.165 | -0.174 | 1.017 |
| | | Childcare Support | 1.192 | 0.690 | 1.728 | 0.085 | -0.168 | 2.551 |
| | | Child_Res#Childcare Sup | -0.518 | 0.347 | -1.494 | 0.137 | -1.201 | 0.165 |
| | Flexible Schedule | Childcare responsibilities | 0.499 | 0.346 | 1.443 | 0.150 | -0.183 | 1.181 |
| | | Flexible Schedule | 1.174 | 0.725 | 1.620 | 0.107 | -0.254 | 2.603 |
| | | Child_Res#Flexible_Sche | -0.605 | 0.346 | -1.749 | 0.082 | -1.286 | 0.077 |
| | Paternity Leave | Childcare responsibilities | 0.468 | 0.303 | 1.546 | 0.124 | -0.129 | 1.064 |
| | | Paternity Leave | 1.549 | 0.889 | 1.743 | 0.083 | -0.203 | 3.301 |
| | | Childcare Res#Paternity Leave | -0.727 | 0.392 | -1.856 | 0.065 | -1.498 | 0.045 |
| | Paused Tenure Clock | Childcare responsibilities | 0.406 | 0.275 | 1.478 | 0.141 | -0.135 | 0.948 |
| | | Paused Tenure Clock | 1.476 | 1.205 | 1.225 | 0.222 | -0.898 | 3.850 |
| | | Childcare Res#Paused Tenure | -0.553 | 0.455 | -1.215 | 0.226 | -1.451 | 0.344 |
| ARCo | Childcare Support | Childcare responsibilities | -0.143 | 0.047 | -3.062 | 0.002 | -0.235 | -0.051 |
| | | Childcare Support | -0.270 | 0.293 | -0.924 | 0.356 | -0.847 | 0.306 |

| | | | | | | | | | |
|---|---|---|---|---|---|---|---|---|---|
| | | | Child_Res#Childcare Sup | 0.125 | 0.099 | 1.262 | 0.208 | -0.070 | 0.320 |
| | | Flexible Schedule | Childcare responsibilities | -0.096 | 0.042 | -2.261 | 0.025 | -0.179 | -0.012 |
| | | | Flexible Schedule | 0.511 | 0.274 | 1.867 | 0.063 | -0.028 | 1.051 |
| | | | Child_Res#Flexible_Sche | -0.039 | 0.095 | -0.412 | 0.680 | -0.225 | 0.147 |
| | | Paternity Leave | Childcare responsibilities | -0.128 | 0.040 | -3.160 | 0.002 | -0.207 | -0.048 |
| | | | Paternity Leave | -0.026 | 0.307 | -0.084 | 0.933 | -0.630 | 0.579 |
| | | | Childcare Res#Paternity Leave | 0.045 | 0.096 | 0.470 | 0.639 | -0.145 | 0.235 |
| | | Paused Tenure Clock | Childcare responsibilities | -0.134 | 0.041 | -3.244 | 0.001 | -0.215 | -0.053 |
| | | | Paused Tenure Clock | -0.147 | 0.282 | -0.523 | 0.602 | -0.703 | 0.408 |
| | | | Childcare Res#Paused Tenure | 0.105 | 0.098 | 1.081 | 0.281 | -0.087 | 0.298 |
| Women | Research Satisfaction | Childcare Support | Childcare responsibilities | -0.127 | 0.057 | -2.233 | 0.026 | -0.239 | -0.015 |
| | | | Childcare Support | -0.036 | 0.305 | -0.118 | 0.906 | -0.636 | 0.564 |
| | | | Child_Res#Childcare Sup | 0.009 | 0.120 | 0.074 | 0.941 | -0.227 | 0.245 |
| | | Flexible Schedule | Childcare responsibilities | -0.204 | 0.061 | -3.332 | 0.001 | -0.325 | -0.083 |
| | | | Flexible Schedule | -0.398 | 0.247 | -1.612 | 0.108 | -0.884 | 0.088 |
| | | | Child_Res#Flexible_Sche | 0.192 | 0.098 | 1.952 | 0.052 | -0.002 | 0.386 |
| | | Paternity Leave | Childcare responsibilities | -0.004 | 0.069 | -0.065 | 0.948 | -0.141 | 0.132 |
| | | | Paternity Leave | 0.494 | 0.232 | 2.128 | 0.034 | 0.037 | 0.952 |
| | | | Childcare Res#Paternity Leave | -0.192 | 0.097 | -1.983 | 0.048 | -0.383 | -0.001 |
| | | Paused Tenure Clock | Childcare responsibilities | -0.097 | 0.051 | -1.899 | 0.059 | -0.198 | 0.004 |
| | | | Paused Tenure Clock | 0.260 | 0.254 | 1.022 | 0.308 | -0.240 | 0.760 |
| | | | Childcare Res#Paused Tenure | -0.097 | 0.102 | -0.947 | 0.344 | -0.297 | 0.104 |
| | Career Satisfaction | Childcare Support | Childcare responsibilities | -0.171 | 0.052 | -3.294 | 0.001 | -0.273 | -0.069 |
| | | | Childcare Support | -0.159 | 0.276 | -0.576 | 0.565 | -0.702 | 0.384 |
| | | | Child_Res#Childcare Sup | 0.073 | 0.107 | 0.680 | 0.497 | -0.138 | 0.284 |
| | | Flexible Schedule | Childcare responsibilities | -0.241 | 0.062 | -3.919 | 0.000 | -0.363 | -0.120 |
| | | | Flexible Schedule | -0.360 | 0.229 | -1.577 | 0.116 | -0.810 | 0.090 |
| | | | Child_Res#Flexible_Sche | 0.222 | 0.103 | 2.154 | 0.032 | 0.019 | 0.424 |
| | | Paternity Leave | Childcare responsibilities | -0.106 | 0.070 | -1.512 | 0.132 | -0.245 | 0.032 |
| | | | Paternity Leave | 0.258 | 0.225 | 1.146 | 0.253 | -0.185 | 0.702 |
| | | | Childcare Res#Paternity Leave | -0.072 | 0.093 | -0.774 | 0.439 | -0.256 | 0.111 |
| | | Paused Tenure Clock | Childcare responsibilities | -0.156 | 0.047 | -3.297 | 0.001 | -0.249 | -0.063 |
| | | | Paused Tenure Clock | 0.122 | 0.220 | 0.556 | 0.579 | -0.310 | 0.554 |
| | | | Childcare Res#Paused Tenure | 0.028 | 0.087 | 0.325 | 0.746 | -0.143 | 0.200 |

| | | | | | | | | |
|---|---|---|---|---|---|---|---|---|
| Community Recognition | Childcare Support | Childcare responsibilities | -0.098 | 0.048 | -2.023 | 0.044 | -0.193 | -0.003 |
| | | Childcare Support | -0.208 | 0.251 | -0.831 | 0.407 | -0.702 | 0.285 |
| | | Child_Res#Childcare Sup | 0.102 | 0.103 | 0.990 | 0.323 | -0.101 | 0.306 |
| | Flexible Schedule | Childcare responsibilities | -0.101 | 0.056 | -1.789 | 0.075 | -0.212 | 0.010 |
| | | Flexible Schedule | -0.082 | 0.187 | -0.438 | 0.662 | -0.449 | 0.286 |
| | | Child_Res#Flexible_Sche | 0.071 | 0.078 | 0.908 | 0.364 | -0.083 | 0.225 |
| | Paternity Leave | Childcare responsibilities | -0.012 | 0.067 | -0.173 | 0.862 | -0.143 | 0.120 |
| | | Paternity Leave | 0.354 | 0.192 | 1.844 | 0.066 | -0.024 | 0.732 |
| | | Childcare Res#Paternity Leave | -0.094 | 0.081 | -1.157 | 0.248 | -0.254 | 0.066 |
| | Paused Tenure Clock | Childcare responsibilities | -0.088 | 0.052 | -1.683 | 0.094 | -0.190 | 0.015 |
| | | Paused Tenure Clock | -0.089 | 0.211 | -0.419 | 0.675 | -0.504 | 0.327 |
| | | Childcare Res#Paused Tenure | 0.051 | 0.092 | 0.554 | 0.580 | -0.131 | 0.233 |
| ARP | Childcare Support | Childcare responsibilities | -0.123 | 0.052 | -2.352 | 0.019 | -0.226 | -0.020 |
| | | Childcare Support | -0.107 | 0.241 | -0.444 | 0.657 | -0.581 | 0.367 |
| | | Child_Res#Childcare Sup | 0.022 | 0.112 | 0.198 | 0.843 | -0.198 | 0.242 |
| | Flexible Schedule | Childcare responsibilities | -0.091 | 0.059 | -1.547 | 0.123 | -0.206 | 0.025 |
| | | Flexible Schedule | 0.284 | 0.233 | 1.220 | 0.224 | -0.174 | 0.742 |
| | | Child_Res#Flexible_Sche | -0.050 | 0.098 | -0.509 | 0.611 | -0.243 | 0.143 |
| | Paternity Leave | Childcare responsibilities | -0.052 | 0.077 | -0.668 | 0.505 | -0.204 | 0.101 |
| | | Paternity Leave | 0.452 | 0.256 | 1.767 | 0.078 | -0.052 | 0.955 |
| | | Childcare Res#Paternity Leave | -0.094 | 0.107 | -0.879 | 0.380 | -0.305 | 0.117 |
| | Paused Tenure Clock | Childcare responsibilities | -0.083 | 0.048 | -1.754 | 0.081 | -0.177 | 0.010 |
| | | Paused Tenure Clock | 0.612 | 0.335 | 1.827 | 0.069 | -0.047 | 1.272 |
| | | Childcare Res#Paused Tenure | -0.070 | 0.125 | -0.559 | 0.577 | -0.316 | 0.176 |
| ARC | Childcare Support | Childcare responsibilities | -0.053 | 0.107 | -0.494 | 0.622 | -0.265 | 0.159 |
| | | Childcare Support | -0.476 | 0.390 | -1.221 | 0.223 | -1.245 | 0.292 |
| | | Child_Res#Childcare Sup | 0.170 | 0.158 | 1.076 | 0.283 | -0.141 | 0.482 |
| | Flexible Schedule | Childcare responsibilities | -0.001 | 0.116 | -0.008 | 0.994 | -0.230 | 0.228 |
| | | Flexible Schedule | 0.009 | 0.433 | 0.021 | 0.983 | -0.843 | 0.862 |
| | | Child_Res#Flexible_Sche | -0.039 | 0.196 | -0.201 | 0.841 | -0.425 | 0.347 |
| | Paternity Leave | Childcare responsibilities | 0.118 | 0.166 | 0.709 | 0.479 | -0.209 | 0.444 |
| | | Paternity Leave | 0.539 | 0.496 | 1.086 | 0.278 | -0.438 | 1.515 |
| | | Childcare Res#Paternity Leave | -0.210 | 0.189 | -1.107 | 0.269 | -0.583 | 0.163 |
| | | Childcare responsibilities | 0.007 | 0.097 | 0.069 | 0.945 | -0.183 | 0.197 |

| | | | | | | | | | |
|---|---|---|---|---|---|---|---|---|---|
| | | Paused Tenure Clock | Paused Tenure Clock | 0.609 | 0.489 | 1.247 | 0.214 | -0.353 | 1.571 |
| | | | Childcare Res#Paused Tenure | -0.003 | 0.176 | -0.016 | 0.987 | -0.349 | 0.343 |
| | ARCo | Childcare Support | Childcare responsibilities | -0.088 | 0.038 | -2.336 | 0.020 | -0.162 | -0.014 |
| | | | Childcare Support | -0.159 | 0.157 | -1.013 | 0.312 | -0.468 | 0.150 |
| | | | Child_Res#Childcare Sup | 0.058 | 0.073 | 0.802 | 0.423 | -0.085 | 0.202 |
| | | Flexible Schedule | Childcare responsibilities | -0.064 | 0.042 | -1.550 | 0.122 | -0.146 | 0.017 |
| | | | Flexible Schedule | 0.112 | 0.172 | 0.653 | 0.514 | -0.226 | 0.451 |
| | | | Child_Res#Flexible_Sche | -0.017 | 0.066 | -0.262 | 0.793 | -0.148 | 0.113 |
| | | Paternity Leave | Childcare responsibilities | 0.010 | 0.048 | 0.206 | 0.837 | -0.085 | 0.105 |
| | | | Paternity Leave | 0.476 | 0.162 | 2.947 | 0.004 | 0.158 | 0.794 |
| | | | Childcare Res#Paternity Leave | -0.126 | 0.068 | -1.841 | 0.067 | -0.260 | 0.009 |
| | | Paused Tenure Clock | Childcare responsibilities | -0.063 | 0.038 | -1.653 | 0.100 | -0.138 | 0.012 |
| | | | Paused Tenure Clock | 0.244 | 0.198 | 1.230 | 0.220 | -0.146 | 0.634 |
| | | | Childcare Res#Paused Tenure | -0.017 | 0.070 | -0.246 | 0.806 | -0.155 | 0.121 |

**Table S8.6. Change in subjective and objective career achievements with the change in childcare responsibilities, the parental support policies provided by institutions, and their interaction items, by gender, if partner's job type is Non-research-oriented**

| | Achievement | Support | Variable | Coefficient | Standard error | T-value | P-value | CI_low | Ci_high |
|---|---|---|---|---|---|---|---|---|---|
| Men | Research Satisfaction | Childcare Support | Childcare responsibilities | -0.040 | 0.040 | -1.001 | 0.318 | -0.119 | 0.039 |
| | | | Childcare Support | 0.372 | 0.219 | 1.699 | 0.090 | -0.058 | 0.803 |
| | | | Child_Res#Childcare Sup | -0.159 | 0.106 | -1.507 | 0.133 | -0.367 | 0.048 |
| | | Flexible Schedule | Childcare responsibilities | -0.080 | 0.041 | -1.928 | 0.055 | -0.161 | 0.002 |
| | | | Flexible Schedule | 0.054 | 0.179 | 0.298 | 0.766 | -0.299 | 0.406 |
| | | | Child_Res#Flexible_Sche | 0.054 | 0.081 | 0.673 | 0.501 | -0.104 | 0.213 |
| | | Paternity Leave | Childcare responsibilities | -0.060 | 0.038 | -1.566 | 0.118 | -0.136 | 0.015 |
| | | | Paternity Leave | 0.306 | 0.189 | 1.621 | 0.106 | -0.065 | 0.678 |
| | | | Childcare Res#Paternity Leave | -0.011 | 0.093 | -0.118 | 0.906 | -0.193 | 0.171 |
| | | Paused Tenure Clock | Childcare responsibilities | -0.070 | 0.039 | -1.773 | 0.077 | -0.147 | 0.008 |
| | | | Paused Tenure Clock | -0.025 | 0.218 | -0.112 | 0.910 | -0.453 | 0.404 |
| | | | Childcare Res#Paused Tenure | 0.051 | 0.099 | 0.515 | 0.607 | -0.144 | 0.246 |
| | | Childcare Support | Childcare responsibilities | -0.043 | 0.039 | -1.093 | 0.275 | -0.121 | 0.034 |

| | | | | | | | | |
|---|---|---|---|---|---|---|---|---|
| Career Satisfaction | | Childcare Support | 0.335 | 0.197 | 1.699 | 0.090 | -0.053 | 0.722 |
| | | Child_Res#Childcare Sup | -0.103 | 0.105 | -0.978 | 0.329 | -0.310 | 0.104 |
| | Flexible Schedule | Childcare responsibilities | -0.081 | 0.040 | -2.021 | 0.044 | -0.161 | -0.002 |
| | | Flexible Schedule | 0.099 | 0.154 | 0.641 | 0.522 | -0.204 | 0.401 |
| | | Child_Res#Flexible_Sche | 0.081 | 0.068 | 1.192 | 0.234 | -0.053 | 0.215 |
| | Paternity Leave | Childcare responsibilities | -0.052 | 0.038 | -1.386 | 0.166 | -0.126 | 0.022 |
| | | Paternity Leave | 0.187 | 0.190 | 0.981 | 0.327 | -0.187 | 0.561 |
| | | Childcare Res#Paternity Leave | -0.043 | 0.094 | -0.456 | 0.648 | -0.227 | 0.142 |
| | Paused Tenure Clock | Childcare responsibilities | -0.063 | 0.037 | -1.697 | 0.090 | -0.135 | 0.010 |
| | | Paused Tenure Clock | 0.014 | 0.209 | 0.066 | 0.947 | -0.398 | 0.425 |
| | | Childcare Res#Paused Tenure | 0.037 | 0.096 | 0.384 | 0.701 | -0.152 | 0.226 |
| Community Recognition | Childcare Support | Childcare responsibilities | -0.060 | 0.034 | -1.805 | 0.072 | -0.126 | 0.005 |
| | | Childcare Support | 0.110 | 0.213 | 0.514 | 0.608 | -0.310 | 0.529 |
| | | Child_Res#Childcare Sup | -0.068 | 0.106 | -0.641 | 0.522 | -0.276 | 0.141 |
| | Flexible Schedule | Childcare responsibilities | -0.097 | 0.035 | -2.808 | 0.005 | -0.166 | -0.029 |
| | | Flexible Schedule | -0.094 | 0.133 | -0.705 | 0.481 | -0.356 | 0.168 |
| | | Child_Res#Flexible_Sche | 0.093 | 0.062 | 1.501 | 0.134 | -0.029 | 0.214 |
| | Paternity Leave | Childcare responsibilities | -0.070 | 0.032 | -2.152 | 0.032 | -0.133 | -0.006 |
| | | Paternity Leave | 0.193 | 0.175 | 1.102 | 0.271 | -0.151 | 0.538 |
| | | Childcare Res#Paternity Leave | 0.006 | 0.077 | 0.084 | 0.933 | -0.146 | 0.159 |
| | Paused Tenure Clock | Childcare responsibilities | -0.073 | 0.032 | -2.262 | 0.024 | -0.136 | -0.010 |
| | | Paused Tenure Clock | 0.037 | 0.207 | 0.178 | 0.859 | -0.371 | 0.445 |
| | | Childcare Res#Paused Tenure | 0.029 | 0.096 | 0.302 | 0.763 | -0.160 | 0.218 |
| ARP | Childcare Support | Childcare responsibilities | -0.124 | 0.072 | -1.725 | 0.085 | -0.266 | 0.017 |
| | | Childcare Support | 0.045 | 0.321 | 0.139 | 0.890 | -0.586 | 0.675 |
| | | Child_Res#Childcare Sup | 0.009 | 0.126 | 0.068 | 0.946 | -0.239 | 0.256 |
| | Flexible Schedule | Childcare responsibilities | -0.143 | 0.070 | -2.060 | 0.040 | -0.280 | -0.007 |
| | | Flexible Schedule | -0.149 | 0.316 | -0.470 | 0.639 | -0.770 | 0.473 |
| | | Child_Res#Flexible_Sche | 0.069 | 0.124 | 0.554 | 0.580 | -0.175 | 0.312 |
| | Paternity Leave | Childcare responsibilities | -0.131 | 0.069 | -1.890 | 0.060 | -0.267 | 0.005 |
| | | Paternity Leave | 0.035 | 0.267 | 0.130 | 0.897 | -0.490 | 0.559 |
| | | Childcare Res#Paternity Leave | 0.062 | 0.101 | 0.609 | 0.543 | -0.137 | 0.260 |
| | Paused Tenure Clock | Childcare responsibilities | -0.191 | 0.056 | -3.385 | 0.001 | -0.302 | -0.080 |
| | | Paused Tenure Clock | -0.644 | 0.493 | -1.308 | 0.192 | -1.613 | 0.324 |

| | | | | | | | | | |
|---|---|---|---|---|---|---|---|---|---|
| | | | Childcare Res#Paused Tenure | 0.594 | 0.331 | 1.798 | 0.073 | -0.056 | 1.245 |
| | ARC | Childcare Support | Childcare responsibilities | -0.164 | 0.082 | -1.986 | 0.048 | -0.326 | -0.002 |
| | | | Childcare Support | -0.038 | 0.327 | -0.118 | 0.906 | -0.682 | 0.605 |
| | | | Child_Res#Childcare Sup | -0.040 | 0.114 | -0.348 | 0.728 | -0.263 | 0.184 |
| | | Flexible Schedule | Childcare responsibilities | -0.170 | 0.080 | -2.131 | 0.034 | -0.328 | -0.013 |
| | | | Flexible Schedule | 0.164 | 0.505 | 0.324 | 0.746 | -0.830 | 1.157 |
| | | | Child_Res#Flexible_Sche | 0.004 | 0.146 | 0.029 | 0.977 | -0.283 | 0.291 |
| | | Paternity Leave | Childcare responsibilities | -0.190 | 0.082 | -2.311 | 0.021 | -0.352 | -0.028 |
| | | | Paternity Leave | 0.042 | 0.317 | 0.134 | 0.894 | -0.581 | 0.666 |
| | | | Childcare Res#Paternity Leave | 0.154 | 0.116 | 1.332 | 0.184 | -0.074 | 0.382 |
| | | Paused Tenure Clock | Childcare responsibilities | -0.175 | 0.081 | -2.152 | 0.032 | -0.335 | -0.015 |
| | | | Paused Tenure Clock | 0.833 | 0.651 | 1.280 | 0.201 | -0.447 | 2.112 |
| | | | Childcare Res#Paused Tenure | 0.140 | 0.187 | 0.747 | 0.456 | -0.228 | 0.508 |
| | ARCo | Childcare Support | Childcare responsibilities | -0.069 | 0.030 | -2.311 | 0.021 | -0.128 | -0.010 |
| | | | Childcare Support | 0.131 | 0.164 | 0.799 | 0.425 | -0.192 | 0.454 |
| | | | Child_Res#Childcare Sup | 0.005 | 0.060 | 0.091 | 0.928 | -0.113 | 0.124 |
| | | Flexible Schedule | Childcare responsibilities | -0.086 | 0.031 | -2.773 | 0.006 | -0.147 | -0.025 |
| | | | Flexible Schedule | -0.148 | 0.134 | -1.101 | 0.272 | -0.411 | 0.116 |
| | | | Child_Res#Flexible_Sche | 0.060 | 0.056 | 1.067 | 0.287 | -0.051 | 0.171 |
| | | Paternity Leave | Childcare responsibilities | -0.070 | 0.028 | -2.476 | 0.014 | -0.125 | -0.014 |
| | | | Paternity Leave | 0.128 | 0.160 | 0.801 | 0.423 | -0.187 | 0.443 |
| | | | Childcare Res#Paternity Leave | 0.020 | 0.062 | 0.323 | 0.747 | -0.101 | 0.141 |
| | | Paused Tenure Clock | Childcare responsibilities | -0.093 | 0.025 | -3.675 | 0.000 | -0.142 | -0.043 |
| | | | Paused Tenure Clock | -0.268 | 0.222 | -1.205 | 0.229 | -0.705 | 0.169 |
| | | | Childcare Res#Paused Tenure | 0.209 | 0.130 | 1.608 | 0.109 | -0.047 | 0.466 |
| Women | Research Satisfaction | Childcare Support | Childcare responsibilities | -0.093 | 0.042 | -2.216 | 0.027 | -0.175 | -0.010 |
| | | | Childcare Support | 0.357 | 0.205 | 1.738 | 0.083 | -0.047 | 0.761 |
| | | | Child_Res#Childcare Sup | -0.098 | 0.096 | -1.024 | 0.306 | -0.286 | 0.090 |
| | | Flexible Schedule | Childcare responsibilities | -0.119 | 0.044 | -2.689 | 0.007 | -0.206 | -0.032 |
| | | | Flexible Schedule | 0.045 | 0.151 | 0.299 | 0.765 | -0.251 | 0.341 |
| | | | Child_Res#Flexible_Sche | 0.023 | 0.065 | 0.348 | 0.728 | -0.106 | 0.151 |
| | | Paternity Leave | Childcare responsibilities | -0.071 | 0.060 | -1.197 | 0.232 | -0.188 | 0.046 |
| | | | Paternity Leave | 0.245 | 0.184 | 1.334 | 0.183 | -0.116 | 0.607 |
| | | | Childcare Res#Paternity Leave | -0.057 | 0.074 | -0.774 | 0.439 | -0.203 | 0.088 |

| | | | | | | | | |
|---|---|---|---|---|---|---|---|---|
| | Paused Tenure Clock | Childcare responsibilities | -0.095 | 0.041 | -2.293 | 0.022 | -0.176 | -0.014 |
| | | Paused Tenure Clock | 0.189 | 0.178 | 1.063 | 0.288 | -0.160 | 0.538 |
| | | Childcare Res#Paused Tenure | -0.062 | 0.075 | -0.829 | 0.407 | -0.210 | 0.085 |
| Career Satisfaction | Childcare Support | Childcare responsibilities | -0.126 | 0.035 | -3.553 | 0.000 | -0.195 | -0.056 |
| | | Childcare Support | 0.362 | 0.176 | 2.057 | 0.040 | 0.016 | 0.707 |
| | | Child_Res#Childcare Sup | -0.115 | 0.083 | -1.388 | 0.166 | -0.277 | 0.048 |
| | Flexible Schedule | Childcare responsibilities | -0.170 | 0.040 | -4.256 | 0.000 | -0.249 | -0.092 |
| | | Flexible Schedule | 0.033 | 0.142 | 0.234 | 0.815 | -0.246 | 0.313 |
| | | Child_Res#Flexible_Sche | 0.066 | 0.063 | 1.041 | 0.299 | -0.058 | 0.190 |
| | Paternity Leave | Childcare responsibilities | -0.146 | 0.052 | -2.817 | 0.005 | -0.247 | -0.044 |
| | | Paternity Leave | 0.160 | 0.166 | 0.965 | 0.335 | -0.166 | 0.485 |
| | | Childcare Res#Paternity Leave | 0.010 | 0.067 | 0.145 | 0.885 | -0.122 | 0.141 |
| Community Recognition | Paused Tenure Clock | Childcare responsibilities | -0.108 | 0.033 | -3.232 | 0.001 | -0.173 | -0.042 |
| | | Paused Tenure Clock | 0.413 | 0.153 | 2.689 | 0.007 | 0.111 | 0.714 |
| | | Childcare Res#Paused Tenure | -0.152 | 0.068 | -2.230 | 0.026 | -0.287 | -0.018 |
| | Childcare Support | Childcare responsibilities | -0.095 | 0.034 | -2.819 | 0.005 | -0.162 | -0.029 |
| | | Childcare Support | 0.261 | 0.146 | 1.791 | 0.074 | -0.025 | 0.547 |
| | | Child_Res#Childcare Sup | -0.059 | 0.064 | -0.929 | 0.353 | -0.184 | 0.066 |
| | Flexible Schedule | Childcare responsibilities | -0.106 | 0.039 | -2.720 | 0.007 | -0.182 | -0.029 |
| | | Flexible Schedule | 0.152 | 0.131 | 1.166 | 0.244 | -0.105 | 0.409 |
| | | Child_Res#Flexible_Sche | 0.007 | 0.058 | 0.122 | 0.903 | -0.108 | 0.122 |
| | Paternity Leave | Childcare responsibilities | -0.084 | 0.053 | -1.599 | 0.111 | -0.188 | 0.019 |
| | | Paternity Leave | 0.147 | 0.172 | 0.854 | 0.394 | -0.191 | 0.485 |
| | | Childcare Res#Paternity Leave | -0.032 | 0.073 | -0.432 | 0.666 | -0.175 | 0.112 |
| ARP | Paused Tenure Clock | Childcare responsibilities | -0.081 | 0.033 | -2.463 | 0.014 | -0.146 | -0.016 |
| | | Paused Tenure Clock | 0.262 | 0.161 | 1.620 | 0.106 | -0.056 | 0.579 |
| | | Childcare Res#Paused Tenure | -0.102 | 0.072 | -1.413 | 0.159 | -0.243 | 0.040 |
| | Childcare Support | Childcare responsibilities | -0.115 | 0.031 | -3.728 | 0.000 | -0.176 | -0.055 |
| | | Childcare Support | -0.142 | 0.211 | -0.674 | 0.500 | -0.556 | 0.272 |
| | | Child_Res#Childcare Sup | 0.127 | 0.082 | 1.538 | 0.125 | -0.035 | 0.289 |
| | Flexible Schedule | Childcare responsibilities | -0.061 | 0.039 | -1.583 | 0.114 | -0.137 | 0.015 |
| | | Flexible Schedule | 0.196 | 0.145 | 1.357 | 0.176 | -0.088 | 0.481 |
| | | Child_Res#Flexible_Sche | -0.074 | 0.057 | -1.281 | 0.201 | -0.187 | 0.039 |
| | Paternity Leave | Childcare responsibilities | -0.046 | 0.044 | -1.045 | 0.297 | -0.133 | 0.041 |
| | | Paternity Leave | 0.239 | 0.174 | 1.375 | 0.170 | -0.103 | 0.581 |
| | | Childcare Res#Paternity Leave | -0.067 | 0.064 | -1.040 | 0.299 | -0.194 | 0.060 |
| | Paused Tenure Clock | Childcare responsibilities | -0.075 | 0.033 | -2.268 | 0.024 | -0.139 | -0.010 |
| | | Paused Tenure Clock | 0.285 | 0.195 | 1.462 | 0.145 | -0.098 | 0.669 |
| | | Childcare Res#Paused Tenure | -0.044 | 0.071 | -0.625 | 0.532 | -0.184 | 0.095 |
| | Childcare Support | Childcare responsibilities | -0.107 | 0.050 | -2.122 | 0.034 | -0.205 | -0.008 |
| | | Childcare Support | -0.054 | 0.275 | -0.196 | 0.845 | -0.596 | 0.488 |
| | | Child_Res#Childcare Sup | 0.096 | 0.111 | 0.866 | 0.387 | -0.122 | 0.313 |
| ARC | Flexible Schedule | Childcare responsibilities | -0.152 | 0.064 | -2.367 | 0.018 | -0.278 | -0.026 |
| | | Flexible Schedule | -0.450 | 0.231 | -1.944 | 0.053 | -0.905 | 0.005 |
| | | Child_Res#Flexible_Sche | 0.142 | 0.092 | 1.533 | 0.126 | -0.040 | 0.323 |
| | Paternity Leave | Childcare responsibilities | -0.163 | 0.090 | -1.799 | 0.073 | -0.341 | 0.015 |
| | | Paternity Leave | -0.144 | 0.299 | -0.480 | 0.631 | -0.733 | 0.445 |
| | | Childcare Res#Paternity Leave | 0.128 | 0.110 | 1.166 | 0.244 | -0.088 | 0.343 |
| | Paused Tenure Clock | Childcare responsibilities | -0.092 | 0.057 | -1.626 | 0.105 | -0.203 | 0.019 |
| | | Paused Tenure Clock | 0.059 | 0.228 | 0.259 | 0.796 | -0.389 | 0.507 |
| | | Childcare Res#Paused Tenure | 0.027 | 0.083 | 0.319 | 0.750 | -0.137 | 0.190 |
| | Childcare Support | Childcare responsibilities | -0.021 | 0.032 | -0.659 | 0.510 | -0.084 | 0.042 |
| | | Childcare Support | -0.071 | 0.130 | -0.548 | 0.584 | -0.326 | 0.184 |
| | | Child_Res#Childcare Sup | 0.049 | 0.051 | 0.958 | 0.338 | -0.051 | 0.149 |
| ARCo | Flexible Schedule | Childcare responsibilities | 0.008 | 0.038 | 0.201 | 0.841 | -0.067 | 0.082 |
| | | Flexible Schedule | 0.118 | 0.104 | 1.137 | 0.256 | -0.086 | 0.321 |
| | | Child_Res#Flexible_Sche | -0.047 | 0.050 | -0.951 | 0.342 | -0.145 | 0.050 |
| | Paternity Leave | Childcare responsibilities | 0.017 | 0.027 | 0.609 | 0.543 | -0.037 | 0.070 |
| | | Paternity Leave | 0.187 | 0.111 | 1.683 | 0.093 | -0.031 | 0.405 |
| | | Childcare Res#Paternity Leave | -0.038 | 0.053 | -0.717 | 0.474 | -0.141 | 0.066 |
| | Paused Tenure Clock | Childcare responsibilities | 0.000 | 0.035 | 0.006 | 0.995 | -0.069 | 0.069 |
| | | Paused Tenure Clock | 0.148 | 0.132 | 1.123 | 0.262 | -0.111 | 0.408 |
| | | Childcare Res#Paused Tenure | -0.042 | 0.063 | -0.660 | 0.510 | -0.167 | 0.083 |

**Survey questions used by the study**

With which gender do you most identify?

- Female
- Male
- Other, please specify
- Prefer not to disclose

Please specify your ethnicity:

- White
- Black or African American
- American Indian or Native American
- Asian or Pacific Islander
- Hispanic or Latino
- Other, please specify
- Prefer not to disclose

What is your current marital status? This includes marriage or domestic partnership with the opposite or same sex. In this context, a domestic partnership includes those who are officially registered or who have lived together for more than 2 years.

- Never married or never taken part in a domestic partnership
- Never married but took part in a domestic partnership
- Married, or in a domestic partnership
- Separated

- Divorced
- Widowed
- Not list, please specify

What is/are your general area(s) of study or research? Check all that apply.

- Arts & Humanities
- Medical Sciences
- Natural Science & Engineering
- Social Sciences
- Other, please specify

What is your current rank or role? Check all that apply. - Student (Bachelor, Master, or Doctoral)

- Student (Bachelor, Master, or Doctoral)
- Post-doctoral fellow
- Lecturer (teaching graduate or undergraduate courses)
- Technician or technician assistant (e.g., statistician, laboratory assistant)
- Research associate (at a public or private institution)
- Senior researcher (at a public or private institution)
- Assistant professor
- Associate professor
- Full professor
- Emeritus professor
- Other, please specify

How many children (of all ages) do you have, including step-, adopted, and biological children?

- 0
- 1
- 2
- 3
- 4
- 5
- 6 or more

Did/Does someone (partners, spouses, other relatives or nannies) help you take care of your child(ren) before they turned school age (age 6 and above)?

- Not at all
- Slightly
- Moderately
- Very
- Extremely

What is/was the employment status (for most of the time) of your current or most recent spouse/partner during your marriage or partnership?

- Employed for wages
- Military
- Self-employed

- Student
- Out of work and looking for work
- Out of work but not looking for work
- Retired

Does/Did your current or most recent spouse/partner's primary job duties include conducting research?

- Yes
- No
- I am not sure
- Not applicable

I am satisfied with the progress I have made towards meeting my research achievement goals.

- Strongly disagree
- Disagree
- Somewhat disagree
- Neither agree nor disagree
- Somewhat agree
- Agree
- Strongly agree
- Not applicable

I am satisfied with the progress I have made towards meeting my career achievement goals.

- Strongly disagree
- Disagree
- Somewhat disagree
- Neither agree nor disagree
- Somewhat agree
- Agree
- Strongly agree
- Not applicable

I have been recognized for my contributions to scholarly communities.

- Strongly disagree
- Disagree
- Somewhat disagree
- Neither agree nor disagree
- Somewhat agree
- Agree
- Strongly agree
- Not applicable

To what extent do you agree with the following statements? -A working mother can establish just as warm and secure a relationship with her children as a mother who doesn’t work

- Strongly disagree
- Disagree

- Somewhat disagree
- Neither agree nor disagree
- Somewhat agree
- Agree
- Strongly agree

To what extent do you agree with the following statements? -It is acceptable for me to go to work and my spouse/partner to stay at home and care for the family (for female respondents) / It is acceptable for me to stay at home and care for the family and my spouse/partner to go to work (For male respondents)

- Strongly disagree
- Disagree
- Somewhat disagree
- Neither agree nor disagree
- Somewhat agree
- Agree
- Strongly agree

To what extent do you agree with the following statements? -A woman should be prepared to cut down on her paid work for the sake of her family

- Strongly disagree
- Disagree
- Somewhat disagree

- Neither agree nor disagree
- Somewhat agree
- Agree
- Strongly agree

To what extent do you agree with the following statements? -A young child is likely to suffer if both parents work

- Strongly disagree
- Disagree
- Somewhat disagree
- Neither agree nor disagree
- Somewhat agree
- Agree
- Strongly agree

Note: Only questions pertinent to analyses in this research are displayed.